\documentclass[aps,prb,twocolumn,english,showpacs,superscriptaddress,amssymb,amsfonts,nofootinbib]{revtex4-2}
\usepackage{graphicx}
\usepackage{amsmath}
\usepackage{braket}
\usepackage{tikz}
\usepackage{float}
\usetikzlibrary{positioning}
\usepackage{soul,xcolor}
\usepackage[export]{adjustbox}
\usepackage{placeins}
\usepackage[scr=boondox]{mathalpha}
\usepackage{ulem}
\usepackage{ragged2e}
\usepackage{amsthm}
\usepackage{tabularx}
\usepackage{multirow}

\usepackage[linktocpage]{hyperref}
\hypersetup{colorlinks=true,citecolor=blue,linkcolor=blue, urlcolor=blue, breaklinks=true}

\newcommand{\Rom}[1]{\uppercase\expandafter{\romannumeral #1\relax}}

\newtheorem*{assumption*}{\assumptionnumber}
\providecommand{\assumptionnumber}{}


\theoremstyle{definition}

\theoremstyle{theorem}

\theoremstyle{corollary}

\theoremstyle{lemma}

\theoremstyle{Proposition}

\theoremstyle{definition}

\makeatletter

\newcommand{\Rmnum}[1]{\expandafter\@slowromancap\romannumeral #1@}
\makeatother

\newcommand{\bea}{\begin{eqnarray}}
\newcommand{\eea}{\end{eqnarray}}
\newcommand{\bct}{\begin{center}}
\newcommand{\ect}{\end{center}}
\newcommand{\bpm}{\begin{pmatrix}}
\newcommand{\epm}{\end{pmatrix}}
\newcommand{\bal}{\begin{aligned}}
\newcommand{\eal}{\end{aligned}}

\newcommand{\ovl}[1]{\mathrel{\overline{\raisebox{-0.15ex}{$#1$}}}}

\begin{document}
    \title{Exact Diagonalization Study on Avalanches in Many-Body Localized Constrained Spin Chains}
    \author{Shuangyuan Lu}
    \affiliation{Department of Physics, The Ohio State University, Columbus, OH 43210, USA}
	
    \date{\today}
	
    \begin{abstract}
    Avalanches are believed to be the mechanism behind the transition from many-body localization to the thermal phase. We utilize spin chains with constraints to study the physics of quantum avalanches by exact diagonalization of disordered systems coupled to a thermal bath. Single-spin observables are used to characterize localization and quantify the influence of the thermal bath on disordered spin chains. Constraints on the Hilbert space confine the dynamics to a subspace, effectively reducing the Hilbert space dimension and enabling the study of larger systems with limited computational resources. We study the PXP model, and in addition, we construct spin chains with constraints by searching for constraints with a genetic algorithm to reach larger system sizes. We define quantities to measure the strength of avalanches and use these to compare different models. We find that avalanches are more pronounced in models with constraints compared to those without constraints. Results from exact diagonalization are compared with those from studying the Lindblad master equation. We also identify models that exhibit no thermal phases and find stable ergodicity breaking.   

    \end{abstract}

\maketitle


\section{Introduction}
The thermalized phase is a general state of matter for many-body physics and has been studied for centuries. Over time, states evolve into thermal equilibrium, and initial information becomes increasingly encoded in nonlocal degrees of freedom, making it inaccessible to local observables, a property called thermalization. The absence of thermalization motivates the study of Anderson localization (AL) \cite{Anderson_1958}, many body localization (MBL) \cite{Abanin_2019, Nandkishore_2015,Basko_2006} and scar states \cite{Serbyn_2021, Moudgalya_2022, Papic_2021}. MBL is special because it remains stable with interactions (unlike AL) and all states break ergodicity (unlike scar states). There are many numerical studies \cite{Oganesyan_2007,Lim_2016, Luitz_2015, Kj_ll_2014, _nidari__2008, Pal_2010, Serbyn_2015, Khemani_2017a} and theories \cite{Basko_2006, Vosk_2015} of MBL phase.

After decades of study, the nature of the transition to MBL and even the question of the existence of MBL in large system sizes remain open. The avalanche mechanism has been proposed as the reason MBL does not exist in dimensions higher than 1D, supported by convincing scaling arguments \cite{De_Roeck_2017}. A rare thermal bath can cause the collapse of a large, well localized system, and this doubt extends to one dimension as well \cite{De_Roeck_2016, Thiery_2017, Thiery_2018, Luitz_2017, Luitz_2020, Khemani_2017b, Goremykina_2019, Morningstar_2020, Ha_2023, Abanin_2021, Kiefer_Emmanouilidis_2021, Sels_2021, Sels_2022, Sels_2023}. Recent studies have pushed MBL simulations to larger sizes and shown a drift of the transition point toward the stronger disorder value \cite{Sels_2021, _untajs_2020a, _untajs_2020b, Sierant_2021}. Even though this tendency may not be generalized deep in the MBL phase \cite{Abanin_2021, Sierant_2020_a, Panda_2020}, reasonable doubt about the stability of MBL is raised based on these calculations, along with results from Gorini–Kossakowski–Sudarshan–Lindblad (GKSL) equation \cite{Morningstar_2022, Sels_2022}, also referred to as Lindblad master equation, and other numerical results \cite{Sels_2023}. It is claimed in \cite{Sels_2023, Sels_2022} that MBL is not stable in thermodynamic limit even in one dimension. 

The interpretation of simulation results is complex and varies across the literature. Supporters of the MBL phase argue for a transition point at stronger disorder strengths \cite{Crowley_2022, Doggen_2021, Sierant_2020_a}. Conversely, those who support the instability of MBL contend that the picture of local integrals of motion (LIOM) \cite{Serbyn_2013, Huse_2014, Chandran_2015} only applies to small systems and will eventually exhibit larger-than-exponential tails of operators acting on large spin clusters, leading to the breakdown of MBL in large systems. Operator growth calculations support this argument in various MBL systems \cite{Weisse_2024} The picture of MBL breakdown by quantum avalanching is becoming increasingly consistent and convincing. Numerical \cite{Thiery_2017, Thiery_2018, Tu_2023, Morningstar_2020, Ha_2023, Goihl_2019, Sels_2021, Sels_2022, Szo_dra_2024} studies and experiments \cite{L_onard_2023} on avalanches face the challenge of insufficiently large system sizes. A sufficiently large thermal bath coupled to a sufficiently large localized spin chain is beyond the scope of previous works.

In this paper, we numerically study the physics of avalanches in MBL. We know in integrable systems like free fermions, AL is stable and will not have avalanches. Recent studies show that quasi-periodic models have more stable MBL phases than models with random fields. \cite{Tu_2023} So avalanches in systems closer to integrable systems or with quasi-periodic disorders will require a much larger system even if they exist. Few works so far point to the opposite direction, examining which kinds of systems are prone to avalanches. Engineering new models with less avalanche stability will help increase numerical and even theoretical understanding of avalanches and the MBL phase itself. In this work, we focus on interacting spin chains and aim to find models in which the phenomena of avalanches are easier to observe.

\begin{table*}[ht]
    \centering
    \renewcommand{\arraystretch}{1.3}
    \begin{tabular*}{\linewidth}{@{\extracolsep{\fill}} cccccc}
    \hline
    Model  & Constraints & Phenomena  & Asymptotic Dimension & Avalanches & Methods \\
    \hline
    Heisenberg & None & Thermal $\rightarrow$ MBL  & Exponential & Less Obvious & ED, GKSL Equation\\
    PXP  & weak & Thermal $\rightarrow$ MBL  &Exponential & \multirow{2}{*}{\scalebox{1}[2]{$\downarrow$}} & ED, GKSL Equation\\
    \Rom{1} & \multirow{3}{*}{\scalebox{1}[3]{$\downarrow$}} & Thermal $\rightarrow$ MBL  & Exponential & & ED\\
    \Rom{2} & & Thermal $\rightarrow$ MBL & Exponential & More Obvious  & ED\\
    \Rom{3} & & MBL  & Exponential  & None &ED\\
    \Rom{4} & Strong & MBL   &Polynomial & None &ED \\
    \hline
    \end{tabular*}
    \caption{Summary of the properties of the models studied in this paper.}
    \label{tab:summary}
\end{table*}

New ideas for studying MBL have been proposed in constrained systems \cite{Herviou2021, Chen_2018, Sierant_2021, Royen_2024}. Constraints confine the dynamics to a subspace within the full Hilbert space. In systems of Rydberg atoms, energy blockages effectively constrain the Hilbert space, realizing famous examples such as the PXP model \cite{Lin_2019, Bernien_2017, Turner_2018a, Turner_2018b}. The dimension of the Hilbert space in models with constraints can scale more slowly than the spin-$\frac12$ chains, making larger size numerics possible. In constrained systems, the transition from thermalization to MBL can occur, with the study of avalanches offering valuable insights. Larger system sizes not only provide more data points but also reveal more long-range behavior of localization, where the breakdown of LIOMs should occur based on the operator growth argument \cite{Weisse_2024}. Previous numerical results in the PXP model indicate that MBL in this model is less stable than the random-field Heisenberg model \cite{Sierant_2021}.

In this paper, we study avalanches in MBL within constrained systems. The PXP model is our starting point, and we also design other models to reach larger system sizes. We use a genetic algorithm (GA) \cite{Katoch_2020} to search through a  large set of different constraints for PXP-type of models that minimize the growth exponent of the Hilbert space dimension. We study spin chains coupled to a thermal bath directly by exact diagonalization (ED). An algorithm called the polynomially filtered exact diagonalization method (POLFED) proposed in Ref. \cite{Sierant_2020_b} reduces the memory cost of the ED algorithm and makes studying larger system sizes possible. We will study how a small finite thermal bath breaks MBL for large disordered spin chains in parameter regions where the spin chains exhibit localization within numerically accessible sizes. 

We study six models in this paper. To provide a clear overview, we summarize their key properties in Table~\ref{tab:summary}. For each model, we list the strength of its constraints, the observed localization phenomena, the asymptotic dimension of its Hilbert space, the visibility of avalanche effects, and the methods used in our analysis. Detailed discussions of these models are presented in the following sections.

In the second section, we begin by presenting results of the PXP model under the MBL thermalization transition. Although this has been discussed in Ref. \cite{Chen_2018,Sierant_2021}, we review the set up and propose new ways of using local observables to characterize the localization. This will be useful in Section 3, where we continue to explore the physics of avalanches in the PXP model. We define a new quantity called ``the avalanche ratio" to characterize the strength of avalanches. In Section 4 and 5, we propose general constrained models and show numerical results of their MBL avalanches. In Section 5, we also study models with a stable localization phase due to  strong constraints. In Section 6, we compare the results of the Lindblad master equation with our ED results.

\section{MBL transition of the PXP model}\label{sec:PXP}
The PXP model is known for its scar states and is commonly used to study weak ergodicity breaking \cite{Bernien_2017, Lin_2019, Moudgalya_2022, Papic_2021, Serbyn_2021, Turner_2018a, Turner_2018b}. However, the PXP model is also valuable for another reason. By introducing a random magnetic field in $z$ direction, we can investigate MBL induced by this random field. Specifically, the Hamiltonian of the disordered PXP model on an open spin chain of size $L$ is 
\begin{equation}\label{eq.PXP}
\begin{aligned}
&\hat{H}_{\text{PXP}} = \frac{1}{2} \sum_{i=1}^L h_i \sigma^z_i \\
+ \frac{1}{2} &\left( \sum_{i = 2}^{L-1} J_i P^z_{i-1} \sigma^x_i P^z_{i+1} + J_1\sigma^x_1 P^z_2 + J_L P^z_{L-1} \sigma^x_L \right)
\end{aligned}
\end{equation}

where $\sigma^x, \sigma^z$ are the Pauli matrices, and $P^z = \frac{1}{2} (\mathbb{I} + \sigma^z)$ is the projection operator onto the $| 0\rangle$ state. The random field $h_i$ is uniformly distributed within the range [-h, h]. The interaction strength $J_i$ is site-independent, with $J_i = J$  in this section when avalanches are not being studied. 

This Hamiltonian preserves a Krylov subspace \cite{Moudgalya_2022, Moudgalya_2021, Herviou2021} of the Hilbert space. The subspace is defined by the constraint that no neighboring spins are in the $|1\rangle$ state at the same time. 
\bea
(\mathbb{I} - P_{i}^z) (\mathbb{I} - P^z_{i+1}) |\psi\rangle = 0 \quad \quad \forall i = 1 \cdots L-1
\eea
Thus, we project the Hamiltonian into this subspace and  focus solely on behavior of Hamiltonian and states within this subspace. In this subspace, every $s^z$ basis state is connected to the all-$|0\rangle$ state by the Hamiltonian, meaning there are no further subspaces formed by a subset of its $s^z$ basis states within it, as illustrated by graphs in \cite{Turner_2018a}. The size of the Hilbert space is given by the $L+2$-th number of Fibonacci series and scales as $D \sim \left( \frac{1 + \sqrt{5}}{2} \right) ^L$. We utilize the slower scaling of the Hilbert space dimension to investigate larger system sizes. Focusing on high-temperature physics, we calculate the $N = 100$ eigenstates with energy closest to $\left(E_{max} + E_{min}\right) / 2$. 

\begin{figure}[h]
    \centering
    \includegraphics[width=0.48\columnwidth]{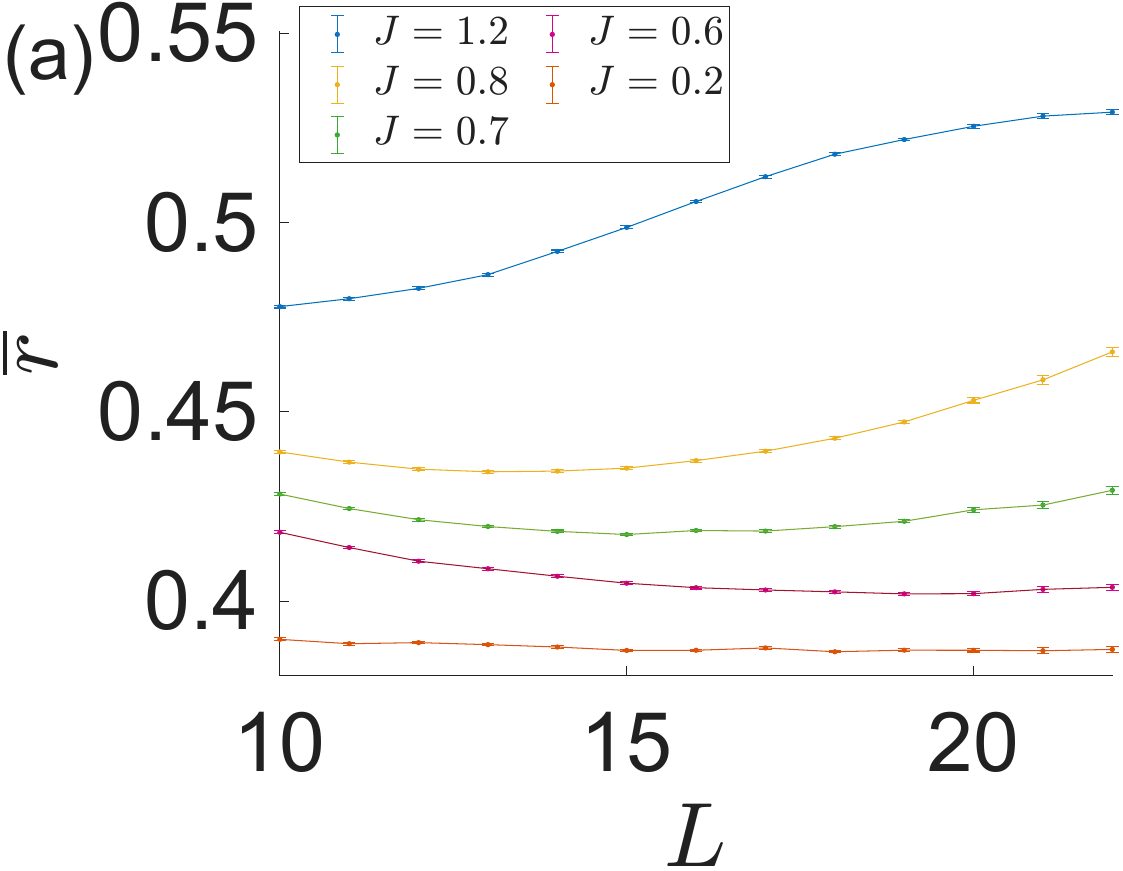}
    \includegraphics[width=0.48\columnwidth]{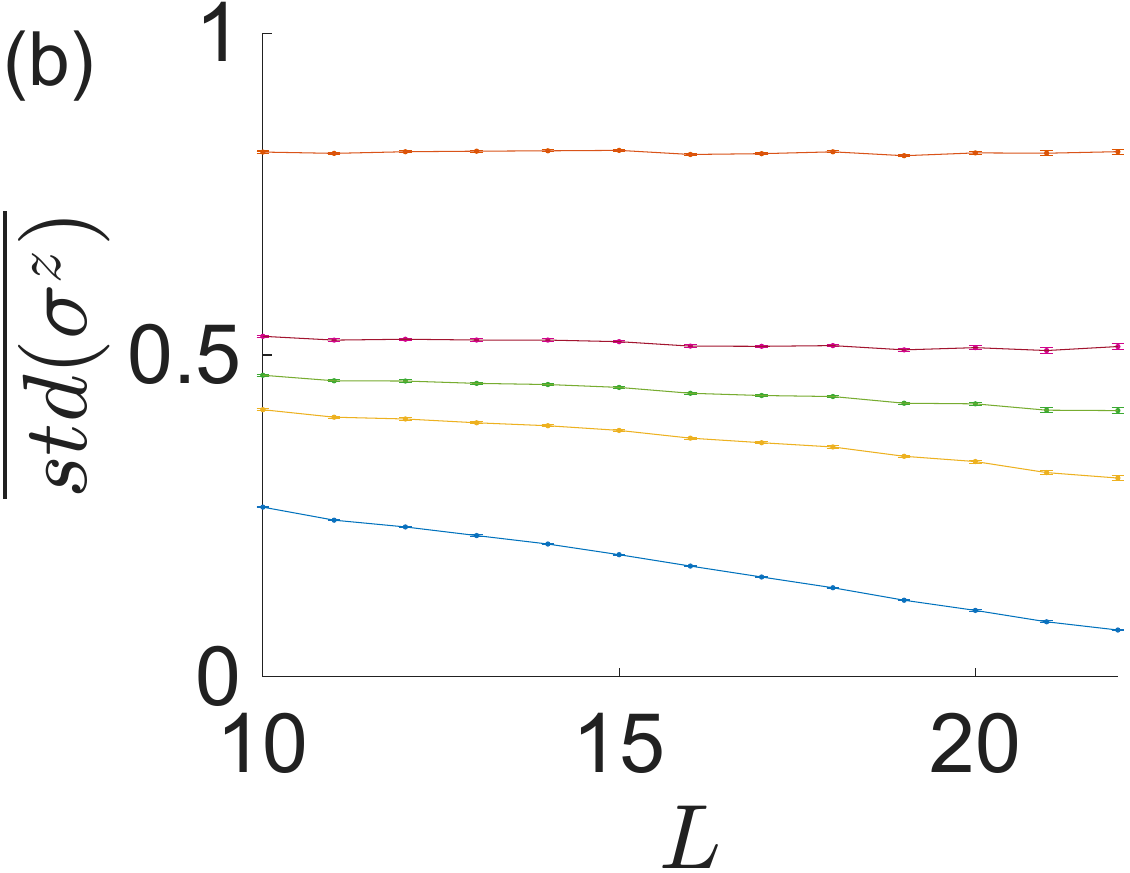}    
    \caption{Two different methods to observe the transition. (a) Average gap ratio versus system size $L$ for different coupling strengths $J$. (b) Average standard deviation of $\sigma^z_L$ versus $L$ for different $J$. These two figures share the same legend. The transition occurs around $J = 0.6$. }
    \label{fig:level_L}
\end{figure}

With increasing disorder, this model undergoes a transition from the thermal phase to the MBL phase, similar to spin-$\frac12$ models like the random-field Heisenberg model. It is evident that when $J=0$, the system is localized. To observe the transition explicitly, we fix $h = 1$ and vary the interaction strength $J$. It is common to use energy level statistics to distinguish whether the system is in the thermal or MBL phase. Let $\{E_n\}$ be the set of sorted eigenenergies and define the gap ratio $r =\text{min}\{\frac{E_n - E_{n-1}}{E_{n-1} - E_{n-2}}, \frac{E_{n-1} - E_{n-2}}{E_n - E_{n-1}} \}$. The average of this ratio, $\ovl{r}$, is approximately $0.529$ in the thermal phase and  $0.386$ in the localized phase \cite{Oganesyan_2007, Atas_2013}. In Fig. \ref{fig:level_L} (a), we show $\ovl{r}$ as a function of system length $L$ for different values of $J = 0.2, 0.6, 0.7, 0.8, 1.2$. When $J$ is large enough, $\ovl{r}$ increases with the system size and saturates around $0.53$. Conversely, When $J$ is small, the behavior is reversed and $\ovl{r}$ decreases to  $0.39$. This demonstrates the transition between the thermal phase and the MBL phase. The transition point varies with system sizes, but for $L = 22$, the $\ovl{r}$ stops increasing when $J$ is less than about $0.6$.

Next, we use additional methods to demonstrate this transition, which will be useful later when we study avalanches. Our challenge is to investigate MBL when the disorder strength, and consequently the degree of localization, varies across different sites. We aim to use information from a single spin at site $L$ to distinguish between the MBL phase and the thermal phase, thereby determining the localization condition at this site. In this section, we first demonstrate that this approach is effective in diagnosing MBL in disordered systems. Similar ideas are discussed in \cite{Colmenarez_2024}. Since the average strength of disorder is the same at all sites, the discussion below in this section applies to measurements taken at any site.

The first method involves using the standard deviation of $\sigma^z_L$. We can calculate expectation value $s^z$ of the last spin of the chain for each eigenstate, $\langle \sigma^z_L \rangle_i = \langle \psi_i | \sigma^z_L|\psi_i \rangle$, with $i$ labelling the $i$-th eigenstates. The distribution of this observable provides information about MBL. To be explicit, the standard deviation of $\langle \sigma^z_L \rangle_i$ is defined as:
\bea
\text{std}(\sigma^z) = \sqrt{\frac{1}{N}\sum_i \langle \sigma^z_L \rangle_i^2 - \left (\frac{1}{N} \sum_i \langle \sigma^z_L \rangle_i \right )^2}
\eea
where $N$ is the number of eigenstates we compute. For convenience, we omit the label $L$ since we always calculate the last spin of the chain. In this paper, we choose $N = 100$, which is much smaller than the total Hilbert space dimension when $L$ is large. The average of this standard deviation over different configurations of disorder, $\ovl{\text{std}(\sigma^z)}$, provides a diagnostic for MBL.

\begin{figure}[h]
    \centering
    \includegraphics[width=0.48\linewidth]{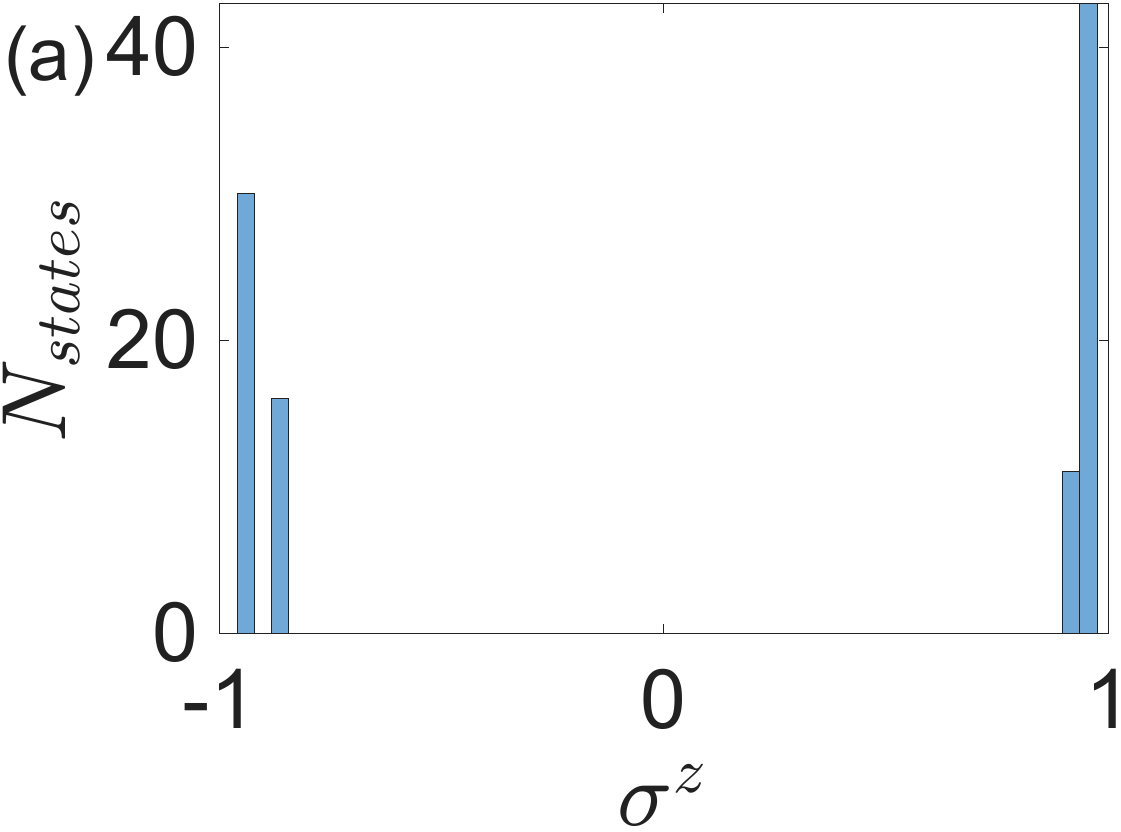}
    \includegraphics[width=0.48\linewidth]{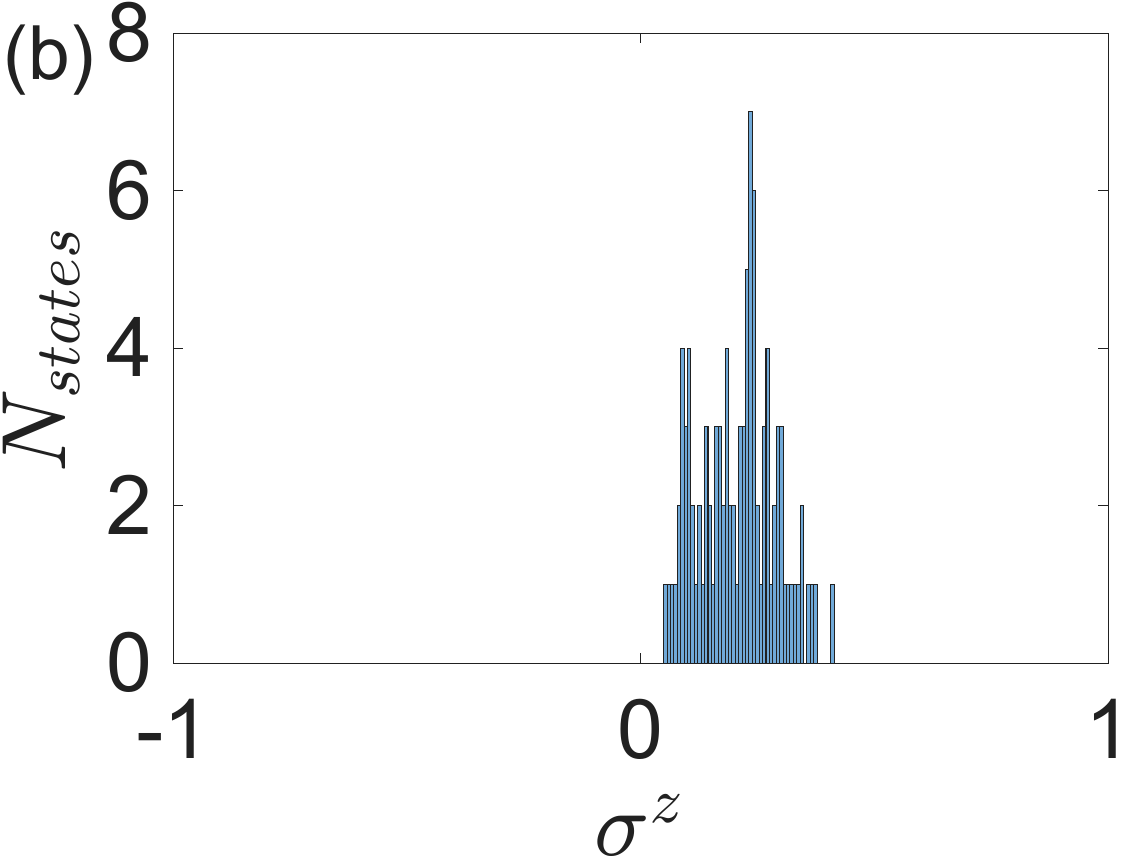} 
    \caption{The distribution of $\sigma^z_L$ for $100$ eigenstates in a specific disorder configuration. (a) Deep in the MBL phase ($J = 0.2$, $L = 22$), spins are aligned along the $z$-axis, leading to a large standard deviation. (b) Deep in the thermal phase ($J = 1.2$, $L = 22$), observables are distributed around a central value, with the standard deviation decreasing to zero in the thermodynamic limit.}
    \label{fig:sz_hist}
\end{figure}

If the system is in the thermal phase, the standard deviation decreases with system size and approaches zero in the thermodynamic limit. This is a direct consequence of the eigenstate thermalization hypothesis (ETH). ETH has been tested in many thermal systems \cite{Kim_2014, Deutsch_2018}. According to the ETH, in the thermal phase, any local observable should only be a function of energy. In the MBL phase, the behavior is completely different. Since the spin is localized, it only couples to a finite number of other spins. Therefore, the distribution of $\langle \sigma^z_L \rangle_i$ resembles that of a finite system, and the standard deviation remains finite and does not change with system size. Two examples of the distribution of $s^z$ are shown in Fig. \ref{fig:sz_hist} for $L = 22$ with $J = 0.2$ and $J = 1.2$, respectively. The numerical results of $\ovl{\text{std}(\sigma^z)}$ versus system length $L$ for different interaction strengths $J$ are shown in Figure \ref{fig:level_L} (b). We observe that the behavior matches our description above and the results of the gap ratio. 

The second method we propose to describe localization involves using the number of clusters of measurements.  We calculate the expectation values of $\sigma_L^z$ and $\sigma_L^x$ and examine whether the distribution of $p_i = (\langle \sigma^z_L \rangle_i, \langle \sigma^x_L \rangle_i)$ forms clusters on the 2-D plane. In the  thermal phase, measurement values are close for states with similar energy, according to the ETH, forming a single cluster. In the MBL phase, the measurement values resemble those of a finite system and form a finite number of clusters. This method is more complex, but its virtue lies in its generality. It does not rely on a specific measurement like $s^z$ and can utilize different measurements together. We do not include $s^y$ because the expectation value is always zero in our examples, as the Hamiltonian is real in the $s^z$ basis.  

\begin{figure}[h]
    \centering
    \includegraphics[width=0.48\linewidth]{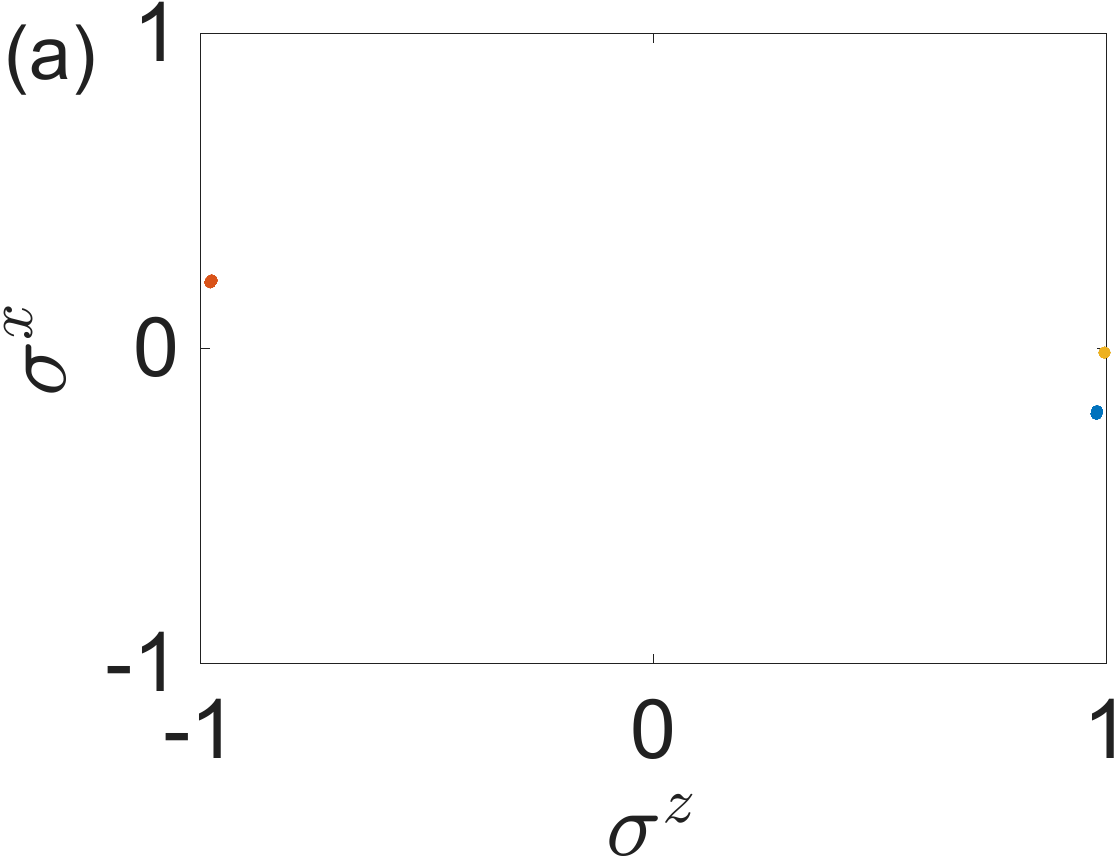}
    \includegraphics[width=0.48\linewidth]{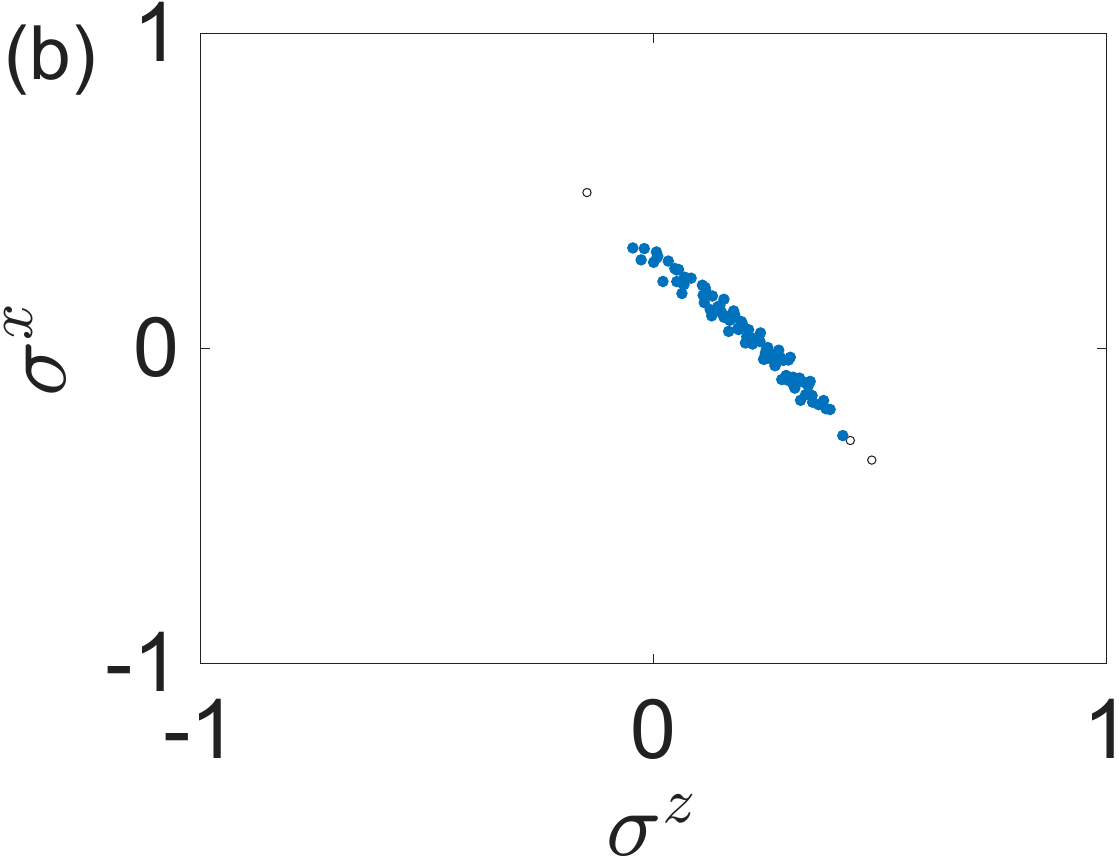} 
    \caption{The distribution of $\sigma^z$ and $\sigma^x$ for $100$ eigenstates in a specific disorder configuration with clusters identified by the DBSCAN algorithm. (a) Deep in the MBL phase ($J = 0.2, L = 22$), data points form several clusters, labeled by different colors. (b) Deep in the thermal phase ($J = 1.2, L = 22$), data points form a single cluster. Black empty circles represent noise points.}
    \label{fig:sz_sx}
\end{figure}

In Fig. \ref{fig:sz_sx}, we show two examples of the distribution of point positions $p_i$.  It is easy to see that in the localized phase, point positions tend to form many clusters because different states do not talk to each other. When the system is deep in thermal phase, points will form a single cluster. 

\begin{figure}[h]
    \centering
    \includegraphics[width=0.5\linewidth]{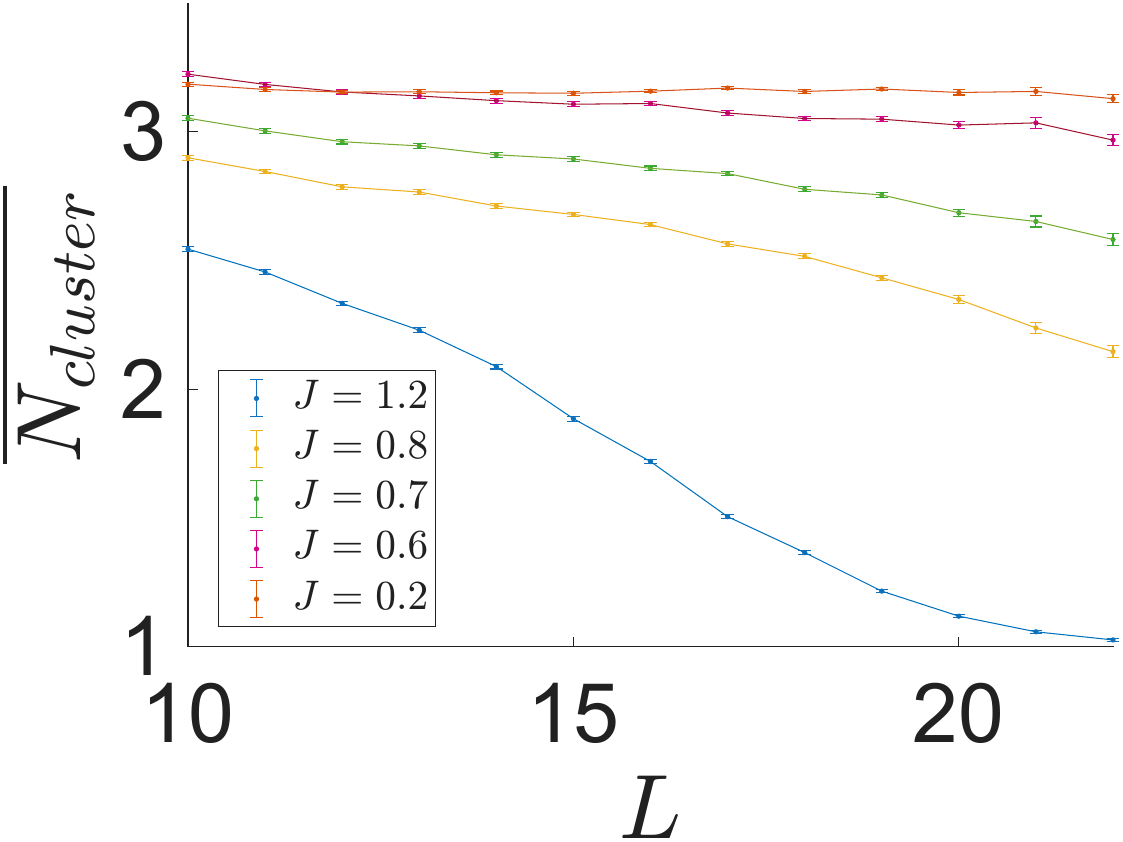}
    \caption{The number of clusters identified by the DBSCAN algorithm versus system size $L$ for different coupling strengths $J$. }
    \label{fig:n_cluster}
\end{figure}

The definition of clusters is technical. Specifically, for the algorithm, we use Density-Based Spatial Clustering of Applications with Noise (DBSCAN) \cite{Ester_1996, Schubert_2017} to identify clusters for each case. This algorithm sets two parameters: the minimum number of points $minPts$ that a cluster must have and the maximum distance $\epsilon$ that two points are identified as neighbors. We set $minPts = 10$ and $\epsilon = 0.1$. For more details on this algorithm, please refer to Appendix \ref{dbscan}. For each realization of disorder, we obtain the number of clusters $N_{cluster}$. The average number of clusters $\ovl{N_{cluster}}$ over disorder samples is calculated and used as a metric to distinguish the strength of localization. Our numerical results of  $\ovl{N_{cluster}}$ versus system size $L$ are shown in Fig. \ref{fig:n_cluster}. The localized phase exhibits a finite (greater than 1) number of clusters, while the cluster number decreases to 1 asymptotically in the thermal phase. 

The standard deviation method can be easily generalized to study the transition region and determine the transition point. We need larger system sizes to determine whether the standard deviation approaches zero in the long-range limit. However, the method of counting clusters is more complex. Larger sizes are needed to accurately identify whether the system is in the thermal phase or the MBL phase. The number of states $N$ needs to increase accordingly,  $\epsilon$ needs to decrease, and $minPts$ needs to increase appropriately. However, the goal of this paper is not to pinpoint the transition point. We make use of the standard deviation and the cluster number to quantify the influence on the disordered systems. We focus more on the changes in these quantities rather than their absolute values. 

\section{Avalanches of the PXP model} \label{sec:PXP_avalanche}
\begin{figure}[h]
    \centering
    \includegraphics[width=\linewidth]{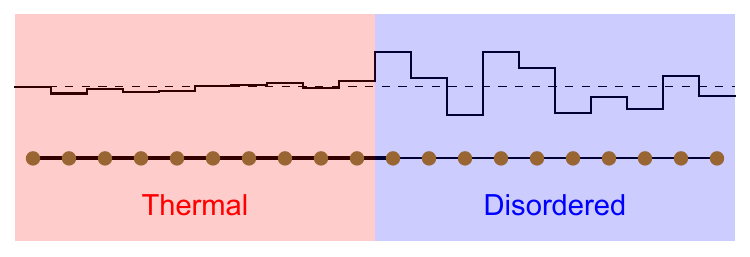}
    \caption{Basic setup of the spin chain to study avalanches. the lattice is divided into two parts: the disordered region with strong disorder and weak interactions, shown in blue, and the bath region with weak disorder (for generality) and strong interactions, shown in red.}
    \label{fig:lattice}
\end{figure}

To study avalanches, we introduce a thermal bath to the disordered spin chain. As shown in Fig. \ref{fig:lattice}, the lattice now consists of two parts: thermal bath region and (strong) disordered region. The thermal bath is deep in the thermal phase, while the disordered region can be in either phase. We are primarily interested in the disordered region when it is in the MBL phase or the transition region. To achieve this, we set different values for $J_i, h_i$ for sites $i \leq L_{thermal}$ (thermal bath), and $i > L_{thermal}$ (disordered region). 
\bea
&J_i = 
\begin{cases}
J_{thermal} & i \leq L_{thermal} \\
J_{disordered} &  i > L_{thermal}
\end{cases}\\
h_i \in & 
\begin{cases}
[-h_{thermal}, h_{thermal}] & i \leq L_{thermal} \\
[-h_{disordered}, h_{disordered}] &  i > L_{thermal}
\end{cases}
\eea
The distribution of $h_i$ on both sides follows a uniform distribution. On the bath side, $J_{thermal} > h_{thermal}$ ensuring it is in the thermal phase. On the disordered side, $J_{disordered} < h_{disordered}$ for most cases, placing it in the localized phase. To match their energy scales, we set $J_{thermal} = h_{disordered} = 1$. This ensures that all states are influenced by the bath and the energy gap of the bath is minimized, optimizing the influence of the bath on the disordered region. We set $h_{thermal} = 0.2$ in all the models discussed in this paper. We fix the size of the thermal bath and observe how its influence changes as the size of the disordered region increases. In Appendix. \ref{append.random_matrix}, we also use a random matrix as thermal bath to benchmark the effectiveness of spin chain thermal bath.

\begin{figure}[h]
    \centering
    \includegraphics[width=0.48\columnwidth]{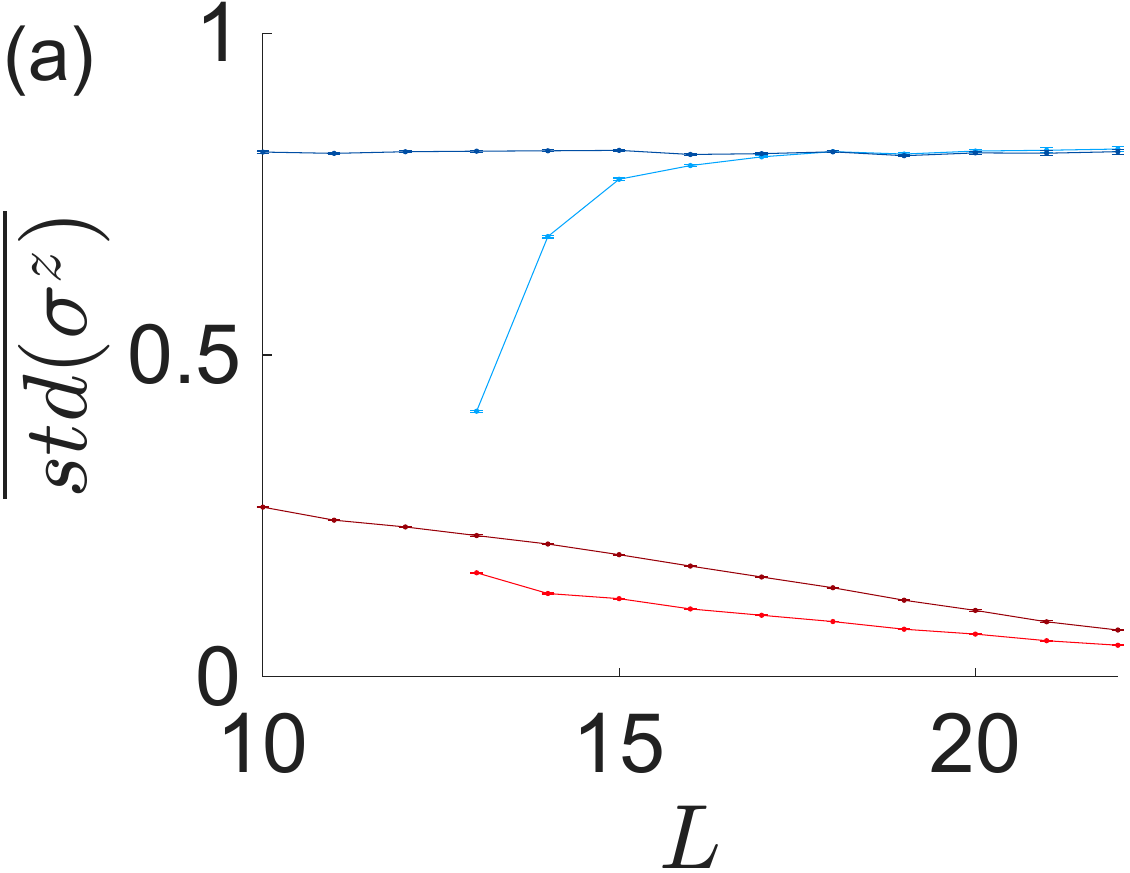}
    \includegraphics[width=0.48\columnwidth]{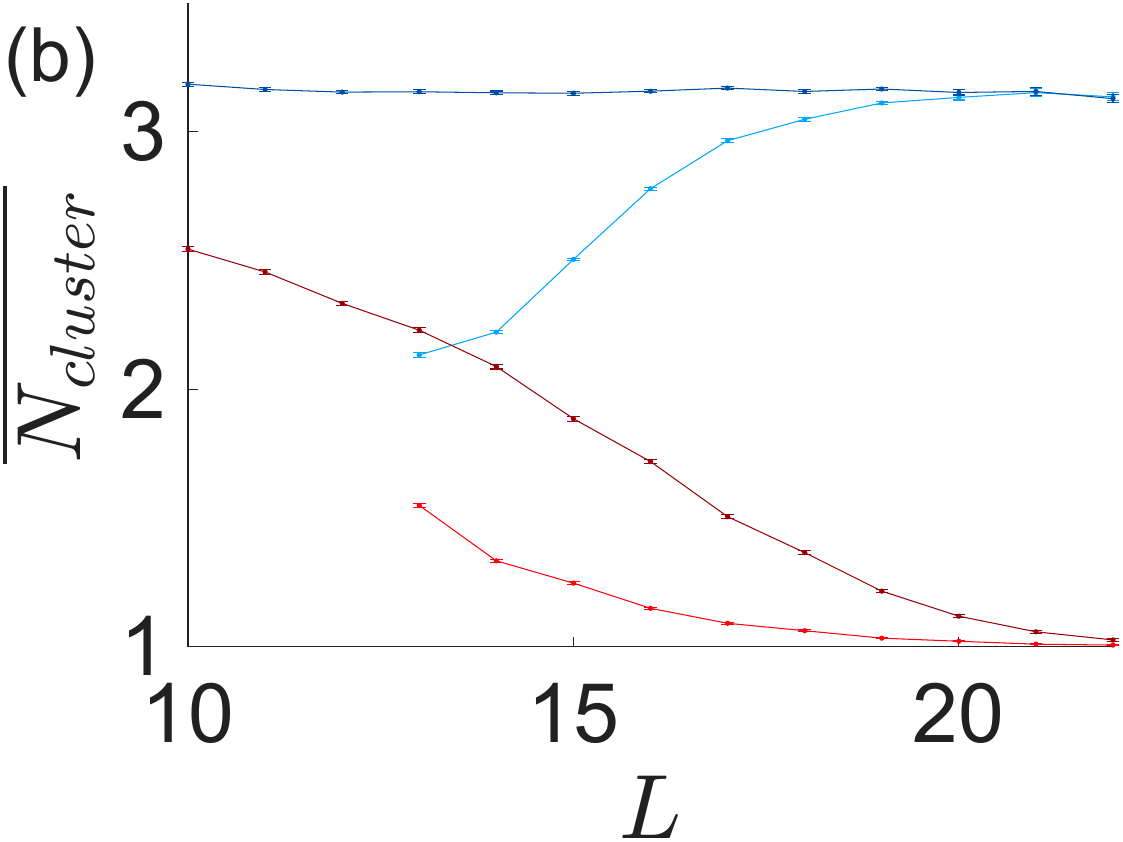}
    \caption{(a) The standard deviation and (b) the number of clusters versus $L$ for systems with and without the thermal bath. Deep in the MBL phase ($J = 0.2$, in blue), the bath influences only a few nearby sites. Deep in the thermal phase ($J = 1.2$, in red), the standard deviation decreases to zero, and the number of clusters reduces to one in the thermodynamic limit. Darker colors represent data without a bath, while lighter colors represent data with a bath ($L_{thermal} = 12$).}
    \label{fig:sz_L_loc_the}
\end{figure}

To gain an initial understanding of the system, we begin with extreme cases. If the disordered region is deep in the MBL phase, the spins far from the bath are independent of the spins coupled to the bath and are therefore not influenced. Beyond a certain length, the spin behavior is the same as if there were no thermal bath. Conversely, in the deep thermal regime, the heat bath combines with the system to form a larger thermal system, and the spins at the end of the chain behave qualitatively similar to the system without the thermal bath. We illustrate this behavior in Fig. \ref{fig:sz_L_loc_the}. Specifically, we plot $\ovl{\text{std}(\sigma^z)}$ as a function of system length $L$. When $J=0.2$ (deeply localized), both cases (with or without bath) approach a finite value, and the two sets of data almost coincide. This indicates that the value is not influenced by the bath when $L$ is sufficiently large. when $J=1.2$ (deeply thermal), both cases (with or without bath) show a decrease in the value. Although the bath causes the standard deviation to decrease more rapidly, the two sets of data are qualitatively similar and approach zero as $L$ becomes very large. The same analysis also applies to the data of the cluster number, as shown in Fig. \ref{fig:sz_L_loc_the} (b).

\begin{figure}[h]
    \centering
    \includegraphics[width=0.48\columnwidth]{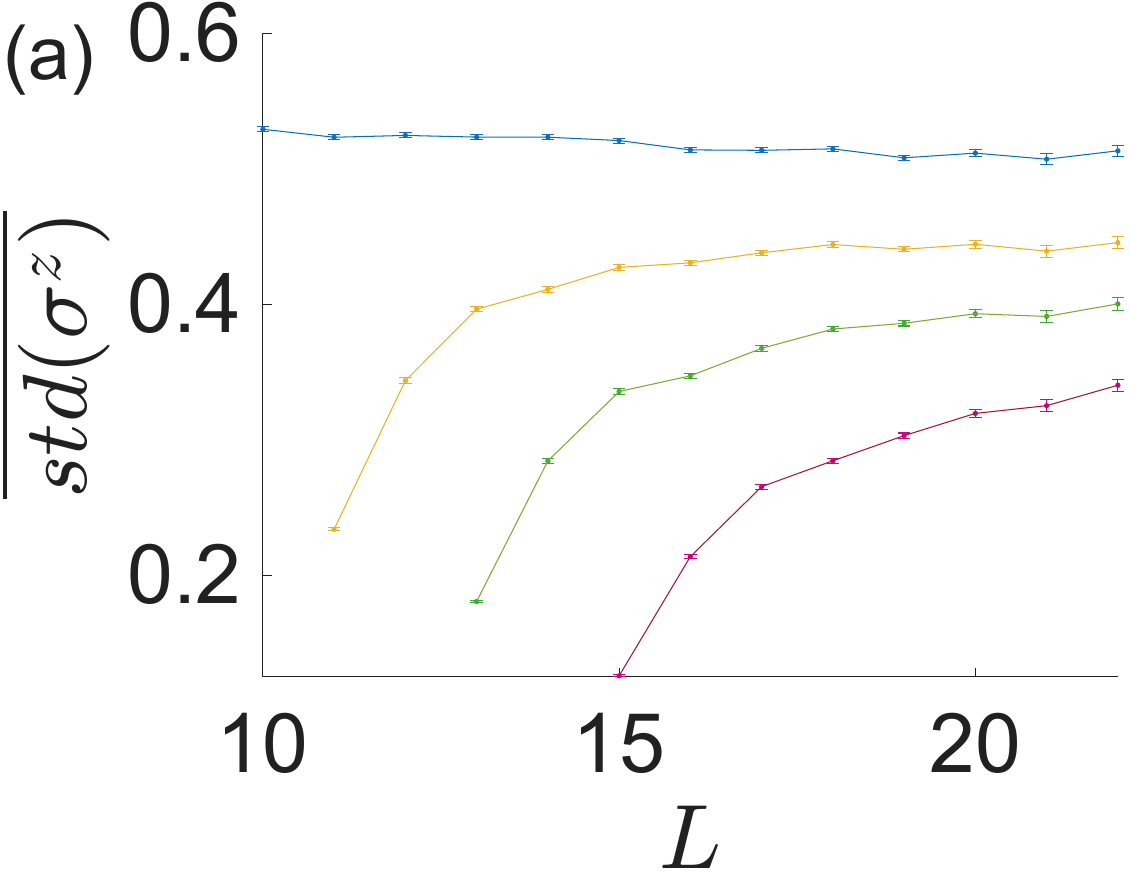}
    \includegraphics[width=0.48\columnwidth]{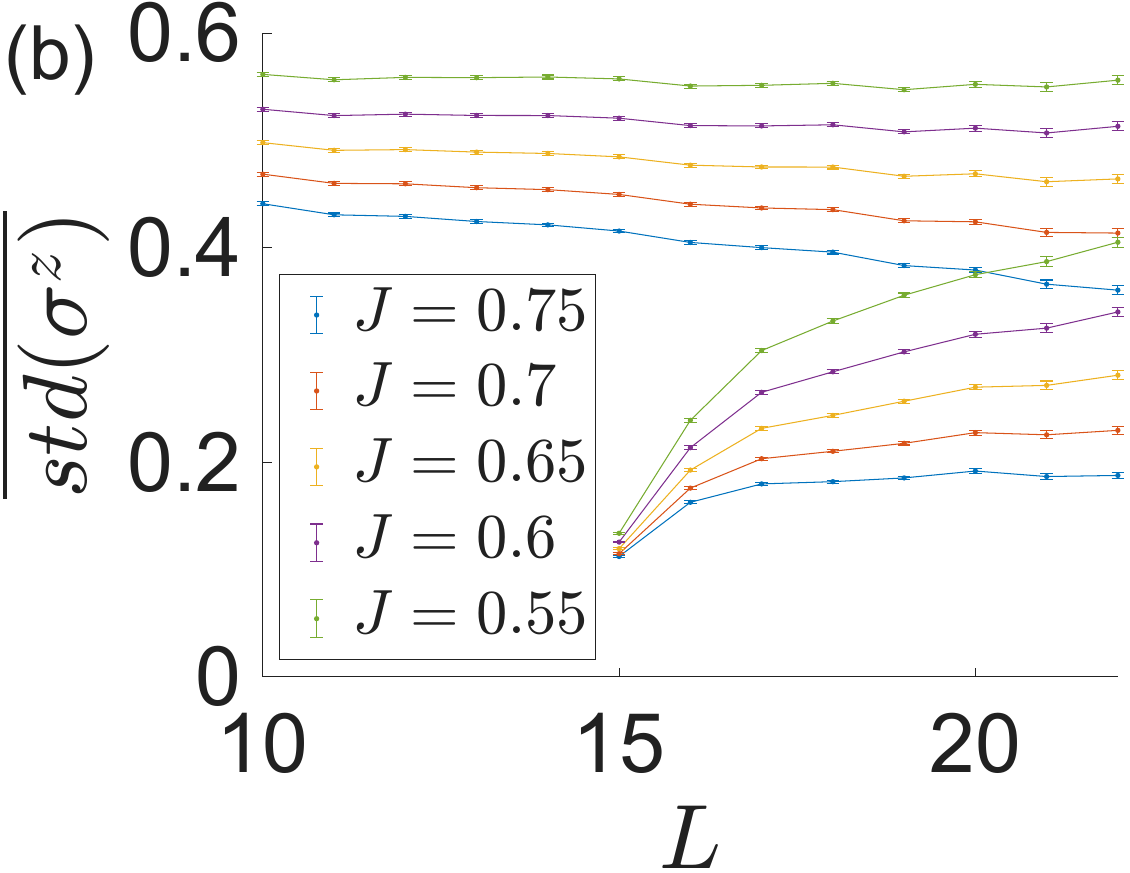}
    \caption{Comparison of $\ovl{std(\sigma^z)}$ with and without the thermal bath. (a) Fixing $J = 0.6$, the blue line represents the case without the bath, while the other three lines correspond to the cases with baths of sizes $L_{thermal} = 10, 12, 14$. The bath size can be identified by $L$ starting from $L_{thermal} + 1$. (b) $L_{thermal} = 14$ is fixed while $J$ varies. The lower curves starting from $L= 15$ correspond to cases with the bath. Although the data stop at $L = 22$, the difference between the cases with and without the bath will persist for larger system sizes in some instances.}
    \label{fig:sz_L_6}
\end{figure}

The situation becomes more complex when $J$ is close to the transition point, as the thermal bath has a greater influence on the system. To predict the influence of the bath on a very large localized lattice, we need to observe how the influence of the bath varies with system sizes $L$ and $L_{thermal}$. To examine this, we calculate $\ovl{std(\sigma^z)}$ for $J = 0.6$ both with ($L_{thermal} = 10, 12, 14$) and without the thermal bath, as shown in Fig. \ref{fig:sz_L_6} (a). We use different values of $L_{thermal}$, with $L$ starting from $L_{thermal} + 1$. This allows us to see that the two standard deviations differ over a long system size. Both the standard deviations, with and without the thermal bath, change very slowly when the system size is large, and the difference between them also changes slowly. We also present $\ovl{std(\sigma^z)}$ for various values of $J$ while fixing $L_{thermal} = 14$ in Fig. \ref{fig:sz_L_6}. 

\begin{figure}[h]
    \centering
    \includegraphics[width=0.48\linewidth]{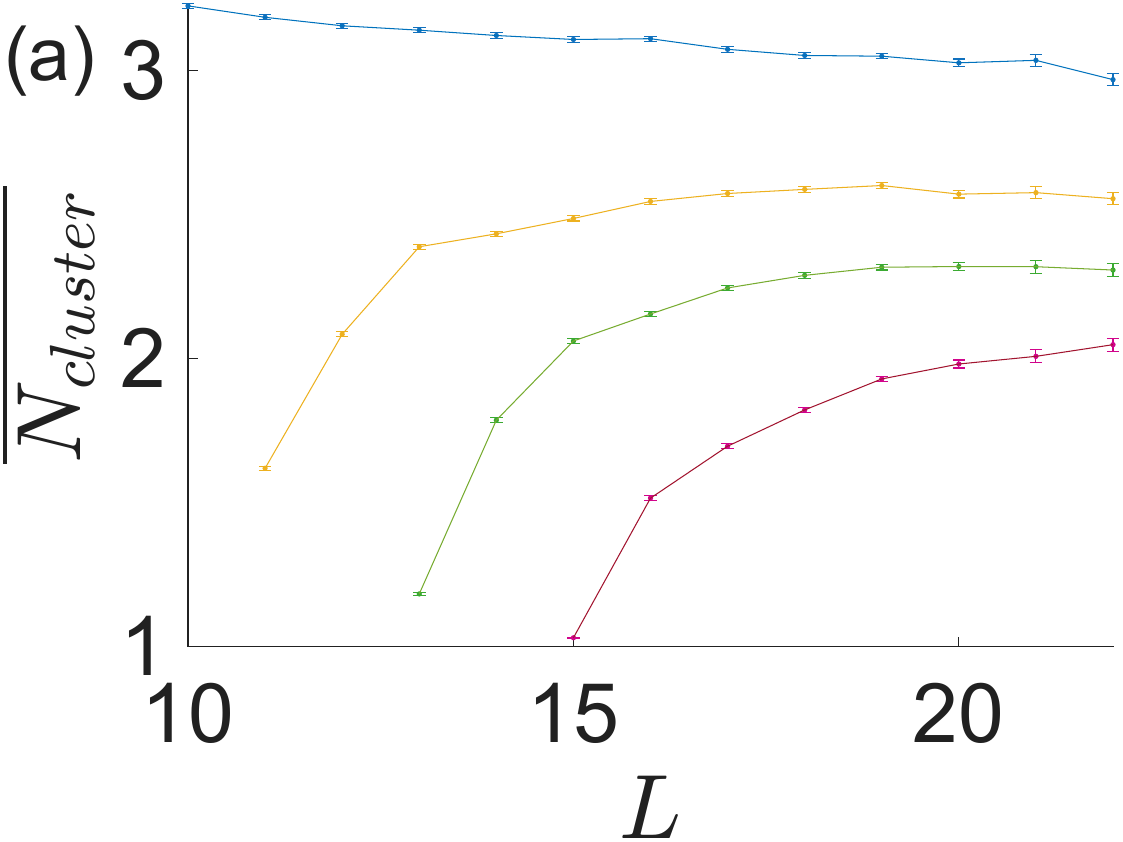}
    \includegraphics[width=0.48\linewidth]{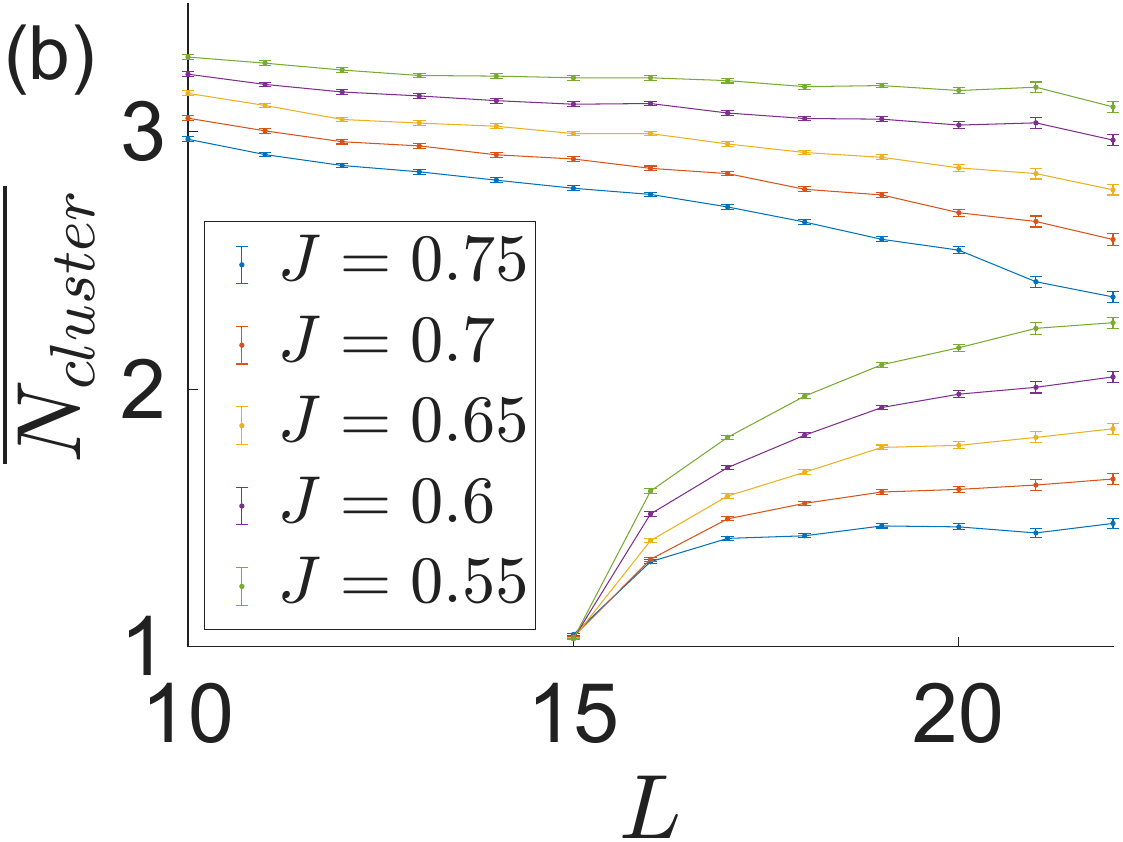}
    \caption{The number of clusters for the same parameters as in Fig. \ref{fig:sz_L_6}. (a) $J = 0.6$ with different bath sizes. (b) $L_{thermal} = 14$ with different coupling strengths $J$.}
    \label{fig:PXP_n_cluster}
\end{figure}

Another method to observe the influence of the bath on the disordered region is through the cluster number we have defined. We calculate the cluster numbers for different system sizes $L$ and $L_{thermal}$, including the case without a thermal bath ($L_{thermal}$ = 0). In Fig. \ref{fig:PXP_n_cluster} (a), we show the case of $J = 0.6$ and $L_{thermal} = 0, 12, 14, 16$. The case without a thermal bath exhibits a stable, large cluster number, while cases with thermal baths show the cluster numbers rising from around $1$ and saturating below the case without the thermal bath. This indicates that the disordered region is significantly influenced by the thermal bath and remains in a state between being thermal and being localized. Fig. \ref{fig:PXP_n_cluster} (b) shows the same quantity for different values of $J$.

\begin{figure}[h]
    \centering
    \includegraphics[width=0.48\linewidth]{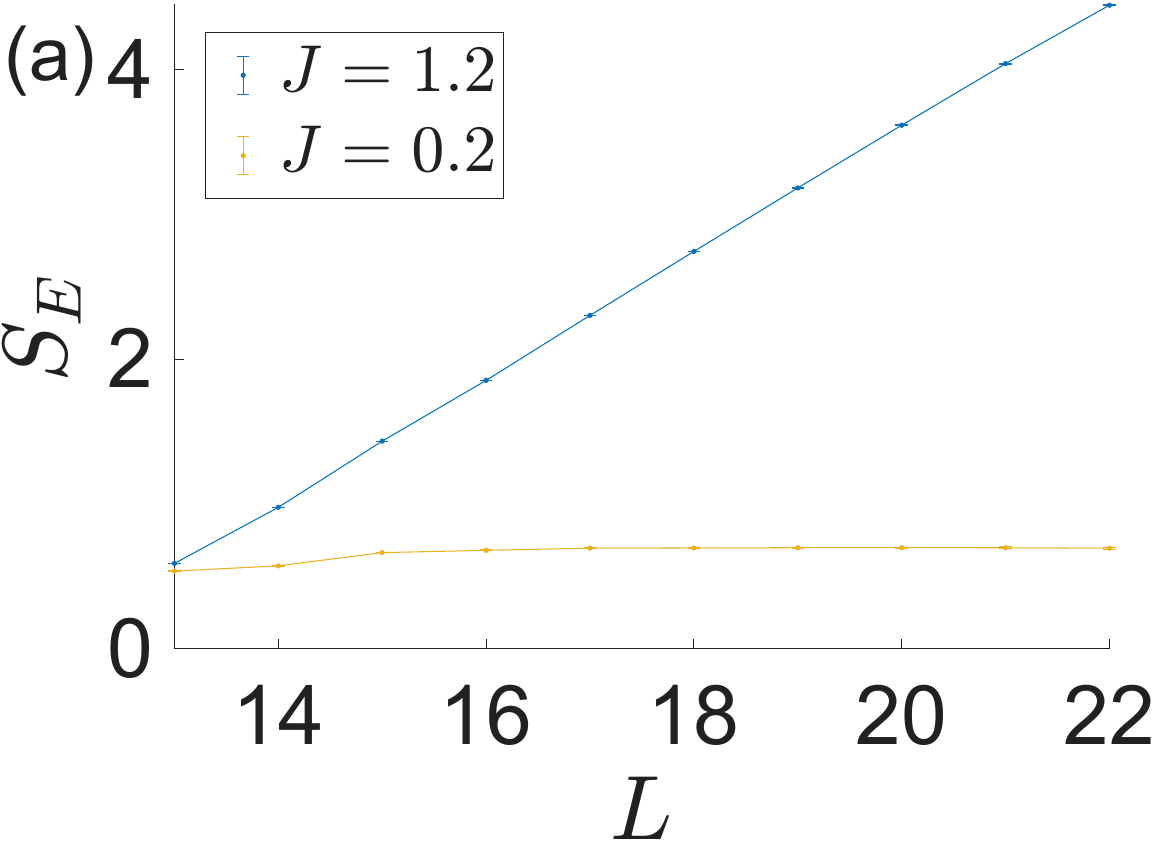}
    \includegraphics[width=0.48\linewidth]{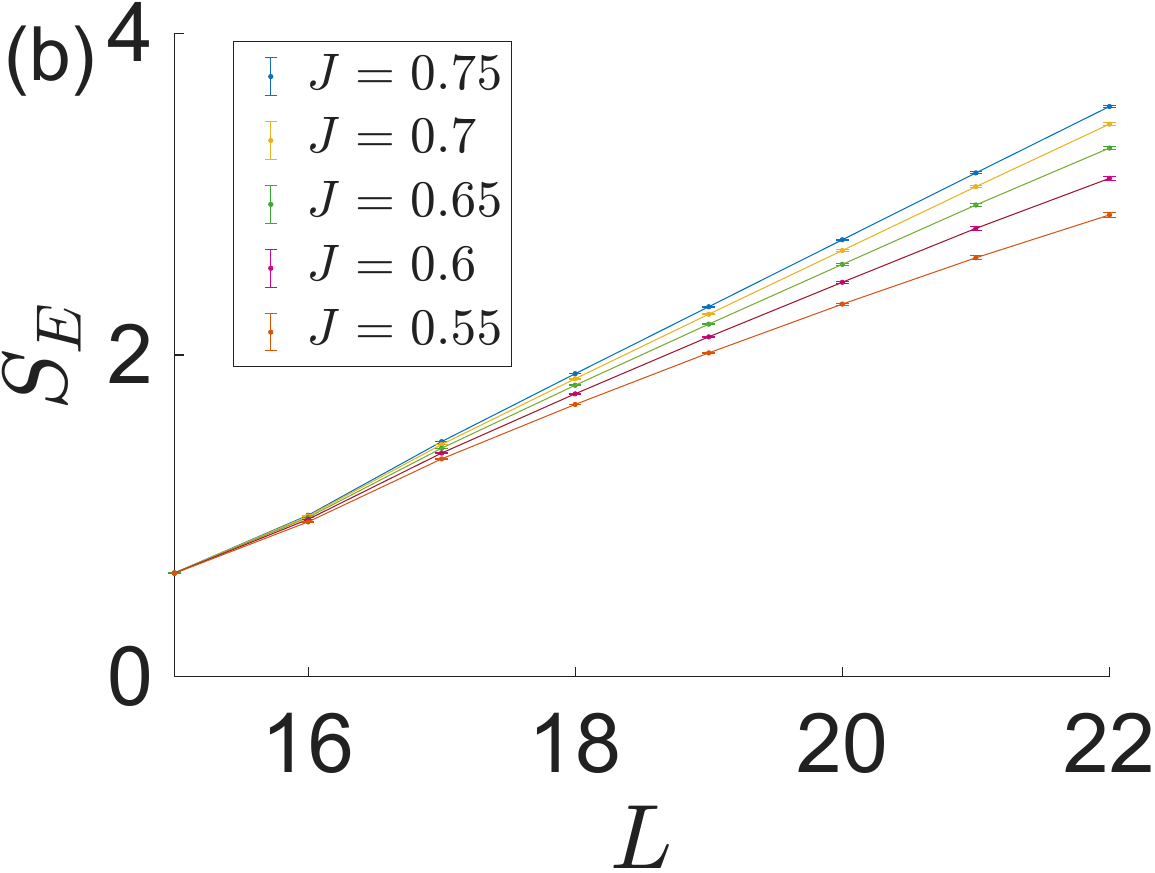}
    \caption{Entanglement entropy between disordered region and the bath. (a) $L_{thermal} = 12$. Entanglement entropy saturates in the MBL phase (J = 0.2) and grows linearly in the thermal phase $J = 0.6$. (b) ($L_{thermal} = 14$. All cases show linear increase with no tendency of slowing down, indicating that the influence of the bath on the disordered region has not disappeared.}
    \label{fig:Se}
\end{figure}

To further study the physics of avalanches, we calculate the entanglement entropy between the thermal bath and the disordered region. As shown in Fig. \ref{fig:Se} (a), if the system is deep in the localized phase, the thermal bath will only entangle with a small part of the system and the entanglement entropy will saturate as the system size increases. Conversely, if the system is in the thermal phase, the entire system will become entangled, and the entropy will increase linearly with the system size. We observe a  clear linear increase in entanglement entropy from $J = 0.55$ to $J = 0.75$, as shown in Fig. \ref{fig:Se} (b), indicating that the thermal bath successfully drives the system out of localization. 

From the numerical results of entropy and observables $s^z_L, s^x_L$, we show the influence of the thermal bath to the disordered region. Especially, when the length of the disordered region increases, the influence from the bath does not decrease to zero, but remains finite. When the length of the disordered region is much longer, even though we can not calculate the observables, we can predict that the observables for cases with and without thermal bath are still different. The difference is a way to measure the influence of the thermal bath. We can see in this case the influence penetrates deep into the disordered region. 

For MBL, the gap ratio $\ovl{r}$ serves as a general measure to characterize the extent of localization and allows for comparison across different models. Similarly, we aim to develop a general method to characterize the extent of avalanches. To this end, we define a dimensionless quantity $\phi$, referred to as the avalanche ratio, which measures the sensitivity of the disordered region to the thermal bath. For a fixed coupling and disorder strength, as well as a given thermal bath size $L_{thermal}$, we calculate the standard deviation of $\sigma^z_L$ as a function of system size $L$ both with and without the thermal bath (denoted as $\ovl{\text{std}(\sigma^z)}$ and $\ovl{\text{std}(\sigma^z)}_0$, respectively). We then identify the largest $L$ such that
\bea
\ovl{\text{std}(\sigma^z)}_0 (l_1) > \ovl{\text{std}(\sigma^z)} (l_2) \quad \quad \forall \ l_1, l_2 < L
\eea
Here, $(l_1)$ in the brackets explicitly indicates that the standard deviation is calculated on the last spin of a spin chain of size $l_1$. We refer to the largest $L$ satisfying this condition as the influenced length, denoted $L_{influ}$. The avalanche ratio $\phi$ is defined as 
\bea
\phi =(L_{influ} - L_{thermal}) / L_{thermal}
\eea
which represents the ratio of the size of the influenced disordered region to the size of the thermal bath. 
We will use $\phi$ to quantify the extent of localization and compare the models in Section  \ref{sec:compare}.

In most cases we study, the standard deviation $\ovl{\text{std}(\sigma^z_L)}_0$ decreases with $L$, while $\ovl{\text{std}(\sigma^z)}$ increases with $L$ because the influence of the bath has not yet saturated. We expect that at a sufficiently large size $L_{influ}$, these values will eventually become equal or very close, allowing us to obtain the ratio $\phi$. Therefore $L_{influ}$ can be interpreted as the length at which the influence of the bath disappears. However, if the system size is large enough, $\ovl{\text{std}(\sigma^z)}$ will also decrease with $L$, similar to $\ovl{\text{std}(\sigma^z)}_0$, which has been observed in some cases with stronger coupling parameter $J$. In such instances, the $L_{influ}$ corresponds to the point where $\ovl{\text{std}(\sigma^z)}_0$ is equal to the maximum of $\ovl{\text{std}(\sigma^z)}$, typically occurring much sooner than where the two standard deviations converge within numerical error. In this scenario, $L_{influ}$ can be interpreted as the size of the disordered region that is effectively equivalent to the thermal bath. In either case, $L_{influ}$ represents the bath-influenced range of the disordered region. This interpretation justifies the physical meaning of $\phi$ as the avalanche ratio.

Since it is not always possible to obtain numerical data of system size as large as $L_{influ}$ to calculate $\phi$, the fitting method becomes crucial for determining the value of $\phi$. In this paper, we adopt a conservative approach by using a linear function to predict the behavior of the standard deviation for sizes $L$ that are beyond our computational reach. Specifically, we fit the data without a thermal bath using the linear function
\bea
\ovl{\text{std}(\sigma^z)}_0 (L) = b_0 + k_0 L
\eea
In cases involving a thermal bath, the influence of the thermal bath leads to  nonlinear behavior in $\ovl{\text{std}(\sigma^z)}$. Therefore we employ a quadratic function to fit the data points that are smaller than but close to $L_{max}$. $L_{max}$ represents the maximum system size used in the computations. Specifically, we use a piecewise function for this fit
\bea
\notag
&\ovl{\text{std}(\sigma^z)} (L) = \\
&\begin{cases}
    b + k (L - L_{max}) + a (L-L_{max})^2 & L < L_{max} \\
    b + k (L - L_{max}) & L > L_{max}
\end{cases}
\eea
$k$ and $k^0$ represent the slope of $std(s^z)$ versus $L$ respectively for cases with and without a thermal bath. $k_0$ is always negative in the numerical results. If $k$ is positive,
\bea
\notag
L_{influ} = &\left[\ovl{\text{std}(s^z)}_0 (L_{max}) - \ovl{\text{std}(s^z)} (L_{max})\right] / (k - k_0) \\
\label{eq.std_0}
&+ L_{max}
\eea
Conversely, if $k$ is negative, 
\bea
\notag
L_{influ} = &-\left[\ovl{\text{std}(s^z)}_0 (L_{max}) - \max_L \ovl{\text{std}(s^z)} (L)\right] / k_0 \\
&+ L_{max}
\eea
This linear fitting method is not the most accurate way to calculate $\phi$. The value of $\phi$ will be underestimated when disordered region is more localized, as the actual growth of $\ovl{\text{std}(\sigma^z)}$ will slow down and eventually decrease. The advantage of this fitting method is that it is relatively stable and closely relative to the extent to which our data has saturated before reaching $L_{max}$. It focuses on the avalanches we can observe directly within the limits of our numerical capabilities, making it more meaningful for our study.

\begin{figure}[h]
    \centering
    \includegraphics[width=0.5\linewidth]{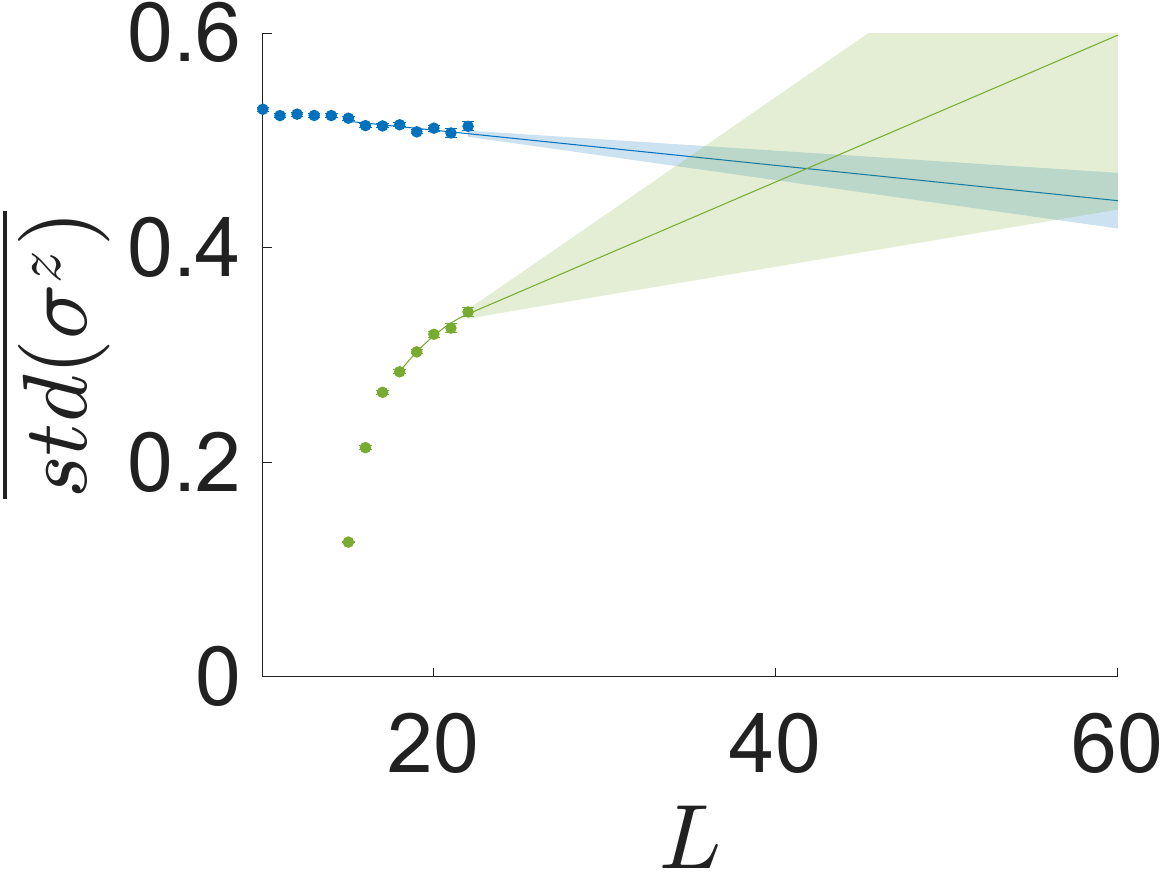}
    \caption{The standard deviation for $J = 0.6$ with no thermal bath (blue) is fitted with a linear function. Data used in the fitting range from $L = 15$ to $L = 22$. The case with $J = 0.6, L_{thermal} = 14$ (green) is fitted with a piecewise function using data from $L = 18$ to $L = 22$. The shaded areas represent the confidence intervals of the fitting. }
    \label{fig:PXP_sz_L_extended}
\end{figure}
In Fig.\ref{fig:PXP_sz_L_extended}, we present our fitting results for the case with $J = 0.6$ and $L_{thermal} = 14$. The linear prediction gives $\phi = 2.0$. When the disordered region is in the thermal phase, the value of $-k_0$ is too large, resulting in a small $\phi$. Conversely, when the disordered region is deep in localized phase, $\ovl{\text{std}(s^z)}$ does not saturate before reaching $L_{max}$, leading to a large $k$ and, consequently, a small $\phi$. The parameter region with the largest $\phi$ corresponds to the transition region and the weakly localized region. The maximum avalanche ratio is influenced by the system size $L_{bath}$ and $L$. The larger these two lengths are, the deeper we can probe into the localized phase (with $\ovl{\text{std}(\sigma^z)}$ saturated) and obtain a larger $\phi$. That implies that a more pronounced avalanche effect can be observed. We will discuss more about the avalanche ratio of the PXP model in Section \ref{sec:compare} and compare it with other models.

When $\phi$ is significantly greater than $1$, we can assert that avalanches occur, as a relatively small bath successfully influences a large disordered system. However, this is still far from the extreme version of avalanches often discussed in the literature, where a finite-size bath fully thermalizes a infinitely long disordered spin chain. Firstly, although the disordered system is deeply influenced within the region defined by $L_{influ}$, the standard deviation remains finite and decreases slowly, which differs from the signature of a thermal phase. We are unable to observe complete thermalization of the spin chains within the numerically accessible system sizes. Secondly, we can only calculate and predict the influence of the bath in a finite region and cannot make definitive statements about an infinite system. Whether a finite bath can influence a infinite disordered system remains an open question. The issue will be further discussed in Section \ref{sec:master_equation}. Thus, we are in the intermediate regime of the avalanche phenomenon, assuming the stronger version exists. In our scenario, the bath couples to a long disordered spin chain, breaks its localization and reveals the instability of LIOMs, but full thermalization has not yet been realized.

\section{General PXP-type constrained models}
To explore MBL avalanches in a broader range of systems, we generalize the PXP model to include models with more general constraints. First, we define the class of models of interest, then we use algorithms to identify the specific models that will be studied in subsequent sections.

Our general constraints are defined on spins locally to preserve the locality of the Hamiltonian. A constraint $\mathcal{C} = (n, S)$ is characterized by the set $S$ of all allowed n-spin configurations. For a list of spin-$\frac12$ particle states of length L, we use the basis of $s^z_i$ eigenstates. We say that a state in the basis satisfies the constraint if any n connected spins are allowed, i.e.
\bea
s_i \cdots s_{i+n-1} \in S \quad \forall i = 1 \cdots L - n + 1
\eea
where $s_i = 0$ or $1$, representing the state $|0\rangle$ or $|1 \rangle $. Using these allowed states as a basis, we construct a subspace of the $2^L$-dimensional Hilbert space $\mathcal{H}_{sub}$. Specifically, we define projection operator $P_i$ that projects onto the allowed states for n-spin configurations.
\bea
P^{\mathcal{C}}_i = \sum_{s_i \cdots s_{i + n - 1} \in S} | s_i \cdots s_{i + n - 1} \rangle \langle s_i \cdots s_{i + n - 1} |
\eea
Then, the total projection operator for L-spin states is given by
\bea
P^\mathcal{C} = \prod_{i=1}^{L-n+1} P_i ^\mathcal{C}
\eea
With the Hilbert subspace defined, we proceed to define the Hamiltonian. Similar to the PXP model, we can introduce terms that only flip one spin, 
\bea
H^\mathcal{C} = P^\mathcal{C} \frac{1}{2}\sum_{i=1}^L (J_i \sigma_i^x + h_i \sigma_i^z) P^\mathcal{C}
\eea
Here, $\sigma_i^x$ represents the spin flip term, while $\sigma_i^z$ denotes the disorder term. Parameters $J_i$ are uniform and $h_i$ are random, the same as those in the PXP models. Since the projection operator can be decomposed into products of local projections, and the operators $\sigma_i^x, \sigma_i^z$ have nontrivial commutation relations only with $P_j$ for $j = i - n + 1 \cdots i + n - 1$, we can express the Hamiltonian in its local form as follows:
\bea
H^\mathcal{C} = \frac{1}{2}\sum_{i=1}^L (\prod_{j = i - n + 1}^{i + n - 1} P_j^\mathcal{C}) (\sigma_i^x + h_i \sigma_i^z) (\prod_{j = i - n + 1}^{i + n - 1} P_j^\mathcal{C})
\eea
This Hamiltonian commutes with the projection operator $P$, thereby preserving the subspace $\mathcal{H}_{sub}$. The subspace is referred to as a Krylov subspace if the subspace is generated by iteratively applying the Hamiltonian to one state in the basis \cite{Moudgalya_2022, Moudgalya_2021, Herviou2021}. We focus on this subspace $\mathcal{H}_{sub}$ and  study the Hamiltonian exclusively within it.

The previously mentioned PXP model can be written as special case of this model when the constraints are $\mathcal{C}_{\text{PXP}}= (2, \{00, 01, 10\})$. Within the subspace, the PXP Hamiltonian \ref{eq.PXP} is the same as the Hamiltonian with constraints $\mathcal{C}_{\text{PXP}}$.
\bea
&P^ {\mathcal{C}_{\text{PXP}}} H_{\text{PXP}} P^{\mathcal{C}_{\text{PXP}}}\\
= &P^ {\mathcal{C}_{\text{PXP}}} \frac{1}{2}\sum_{i=1}^L (J_i \sigma_i^x + h_i \sigma_i^z) P^{\mathcal{C}_{\text{PXP}}}
\eea

Although these general models are well defined, not all constraints are suitable for studying localization transitions. In some cases, the fragmentation is so extreme that the basis states of the Hilbert subspace $\mathcal{H}_{sub}$ defined by the constraints do not form a single coherent Krylov subspace but instead is divided into many small Krylov subspaces that are not connected by the Hamiltonian. 

Therefore, we need to ensure that at least most, if not all, of the states in the basis of $\mathcal{H}_{sub}$, are connected by the Hamiltonian (The connection can also be understood as follows: states are connected in the graph that two states have an edge if one state is mapped to the other by any term in the Hamiltonian, as demonstrated in \cite{Herviou2021}). This can be easily checked by choosing random parameters and calculate eigenstates and their overlaps. If most of the overlaps are non-zero, it indicates that most states belong to a single Krylov subspace. 

Our goal is to obtain a Hilbert space that scales slowly with the system size $L$. We use a genetic algorithm to search among constraints for models that satisfy conditions mentioned above. A genetic algorithm is a type of optimization algorithm. Details about this algorithm can be found in Appendix \ref{ga} or in \cite{Katoch_2020}. We employ it to optimize the relationship between the Hilbert space dimension and the system length $L$. Specifically, we perform a linear fit for the logarithm of the Hilbert subspace dimension versus the length $L$, aiming to minimize the slope. 
\bea
\ln ( \text{dim}~ \mathcal{H}_{sub}) = k L + b
\eea

\begin{table}[h]
    \centering
    \renewcommand{\arraystretch}{1.3}
    \begin{tabular*}{\linewidth}{@{\extracolsep{\fill}} cc}
    \hline
    Model   & Asymptotic Dimension \\
    \hline
    Heisenberg & $\exp{(0.69L)}$ \\
    PXP  &$\exp{(0.48 L + 0.16)}$\\
    \Rom{1} & $\exp{(0.38 L + 0.60)}$\\
    \Rom{2} & $\exp{(0.28 L + 2.0)}$\\
    \Rom{3} & $\exp{(0.18 L + 2.9)}$\\
    \Rom{4} &$0.065 \ L^{3.0}$ \\
    \hline
    \end{tabular*}
    \caption{Formula for the asymptotic Hilbert space dimension for all cases.}
    \label{tab:dimension}
\end{table}

\begin{figure}[h]
    \centering
    \includegraphics[width=0.5\linewidth]{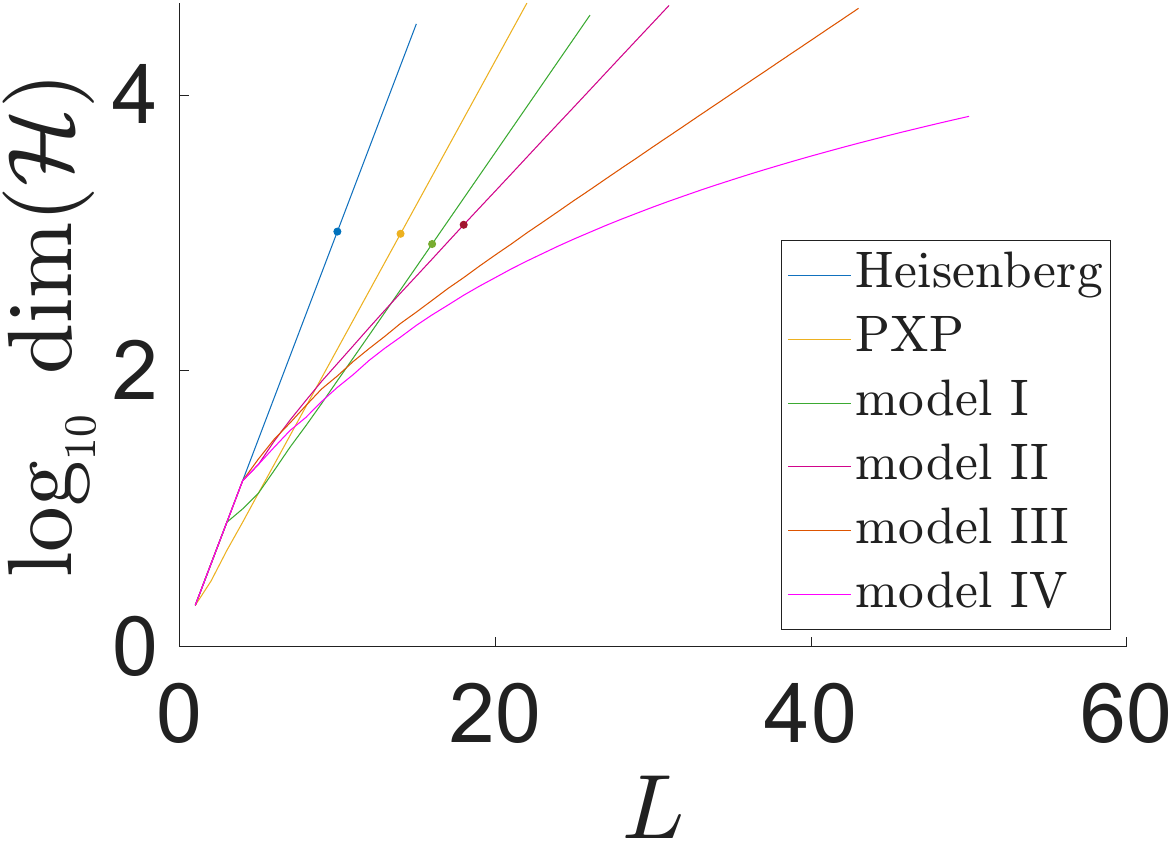}
    \caption{Dimension of the Hilbert space versus system size for all the models studied in this paper. The end points correspond to the maximum lengths reached numerically. Some lines have an extra point in the middle, representing the sizes of the thermal baths used in Section \ref{sec:compare}. Model $\Rom{4}$ does not exhibit a straight line for large lengths, indicating that the Hilbert space dimension is not an exponential function of $L$.}
    \label{fig:dimension}
\end{figure}

\begin{table*}[ht]
    \centering
    \renewcommand{\arraystretch}{1.3}
    \begin{tabular}{|>{\centering\arraybackslash}p{1cm}|>{\centering\arraybackslash}p{15cm}|}
    \hline
    Model  & Constraints (allowed spin configurations) \\
    \hline
    PXP& $\{$00, 01, 10$\}$\\
    \hline
    \Rom{1} & $\{$0000, 0001, 0010, 0100, 1000, 0011, 1001, 1010, 1011, 1111$\}$ \\
    \hline
    \Rom{2} & $\{$00000, 00001, 00010, 00100, 00011, 00101, 10001, 00110, 01010, 00111, 01011, 10011, 01100, 10100, 01101, 10101, 01110, 11000, 11001, 11010, 11100$\}$ \\
    \hline
    \Rom{3} & $\{$00001, 00010, 00100, 10000, 00011, 00101, 01001, 01010, 10010, 01011, 10011, 01100, 10100, 01101, 01111, 11000, 11001, 11010, 11011, 11100, 11101, 11110, 11111$\}$\\
    \hline
    \Rom{4} & $\{$00000, 00001, 00010, 00100, 10000, 00011, 01001, 10001, 01010, 10010, 01100, 01101, 01110, 10110, 11000, 11001, 11010, 11011, 11101, 11110, 11111$\}$\\
    \hline
    \end{tabular}
    \caption{The sets $S$ of spin configurations allowed by constraints for all the models studied in this paper.}
    \label{tab:constraints}
\end{table*}

We successfully identify several models that satisfy our requirements. Among them, we selected four models, denoted as Models \Rom{1}, \Rom{2}, \Rom{3} and \Rom{4}. the corresponding constraints are $\mathcal{C}_\text{\Rom{1}} = (4, S_\text{\Rom{1}})$,
$\mathcal{C}_\text{\Rom{2}} = (5, S_\text{\Rom{2}} )$, $\mathcal{C}_\text{\Rom{3}} = ( 5, S_\text{\Rom{3}} )$, $\mathcal{C}_\text{\Rom{4}} = ( 5, S_\text{\Rom{4}} )$. The Hilbert space dimension of these models are shown in \ref{tab:dimension}. The corresponding constraints of these models are listed in Table \ref{tab:constraints}. We also plot the Hilbert space dimensions in Fig. \ref{fig:dimension}.

We selected models with Hilbert space dimensions that scale at varying rates, covering cases from large $k$ (the PXP model) to small $k$ (Model \Rom{3}) values. This ensures a comprehensive study of different scaling behavior. As the constraints become stronger, the Hilbert space dimension even scales as a polynomial function of $L$. The fitting details of model \Rom{4} refers to Appendix. \ref{app:fitting4}. We will utilize these four models to study MBL and the influence of thermal baths on MBL systems. 

\section{simulation results for generalized constrained models}
In this section, we present detailed results of our simulation for the general constrained models we defined. Models \Rom{1} and \Rom{2} exhibit similar physics and are comparable to the PXP model discussed in Section \ref{sec:PXP}. Models \Rom{3} and \Rom{4} are similar to each other and will be discussed later in this section. 

\begin{figure}[h]
    \centering
    \includegraphics[width=0.48\linewidth]{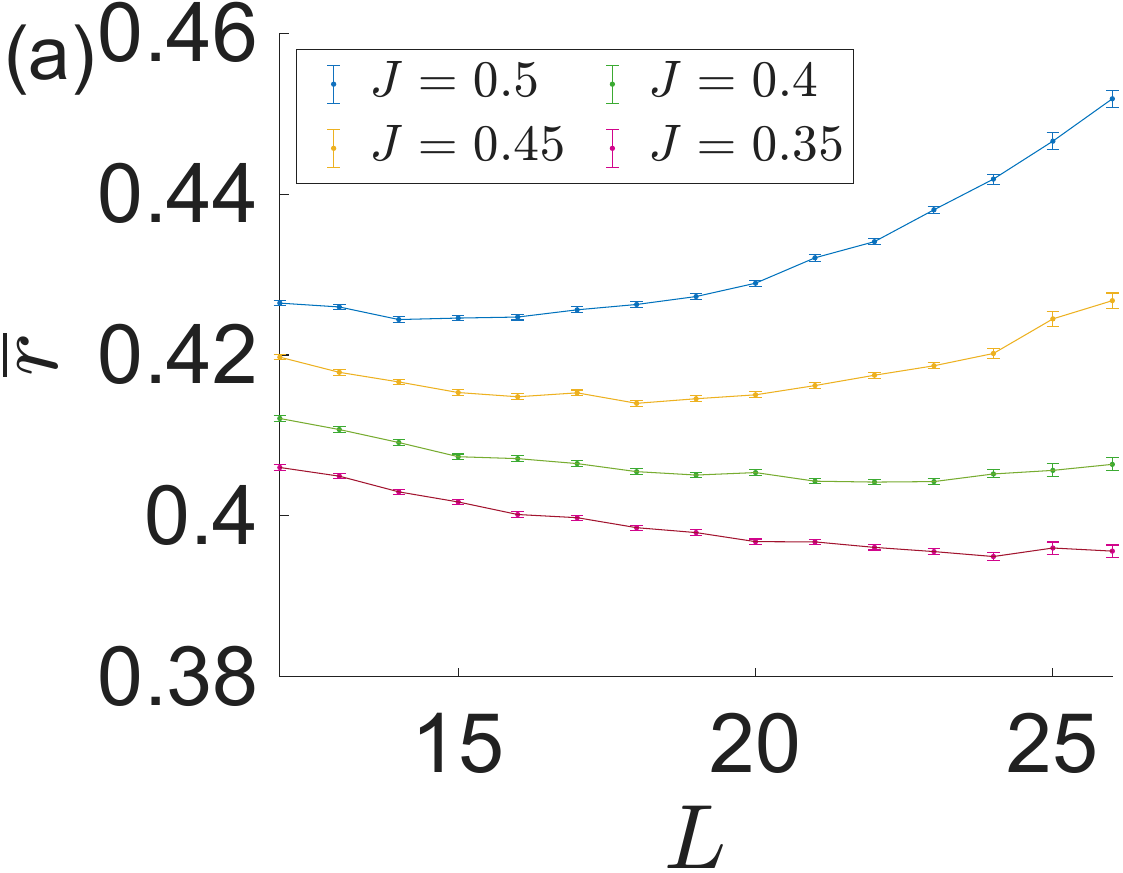}
    \includegraphics[width=0.48\linewidth]{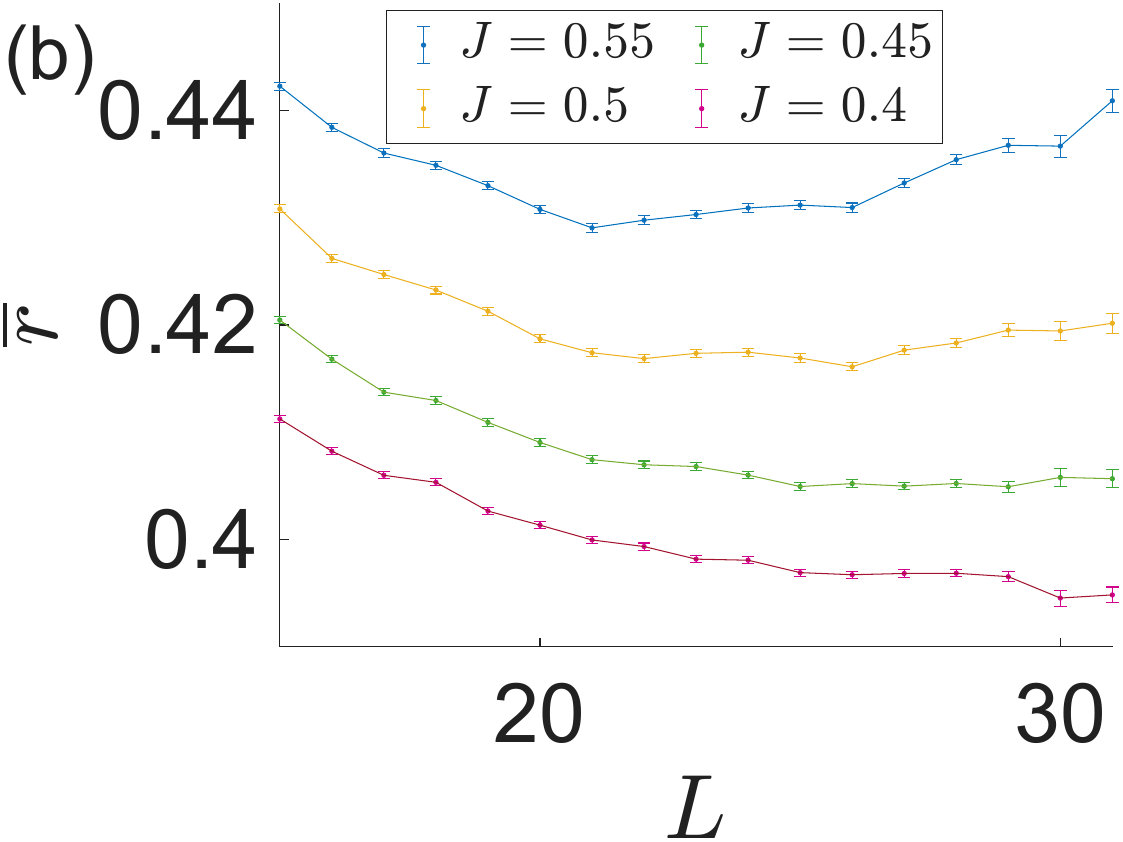}
    \caption{The average gap ratio $\ovl{r}$ versus $L$ for Models \Rom{1} (a) and \Rom{2} (b). The bottom two lines indicate that the systems are in the transition or weak localization regime.}
    \label{fig:model_14_level_L}
\end{figure}

\begin{figure}[h]
    \centering
    \includegraphics[width=0.48\linewidth]{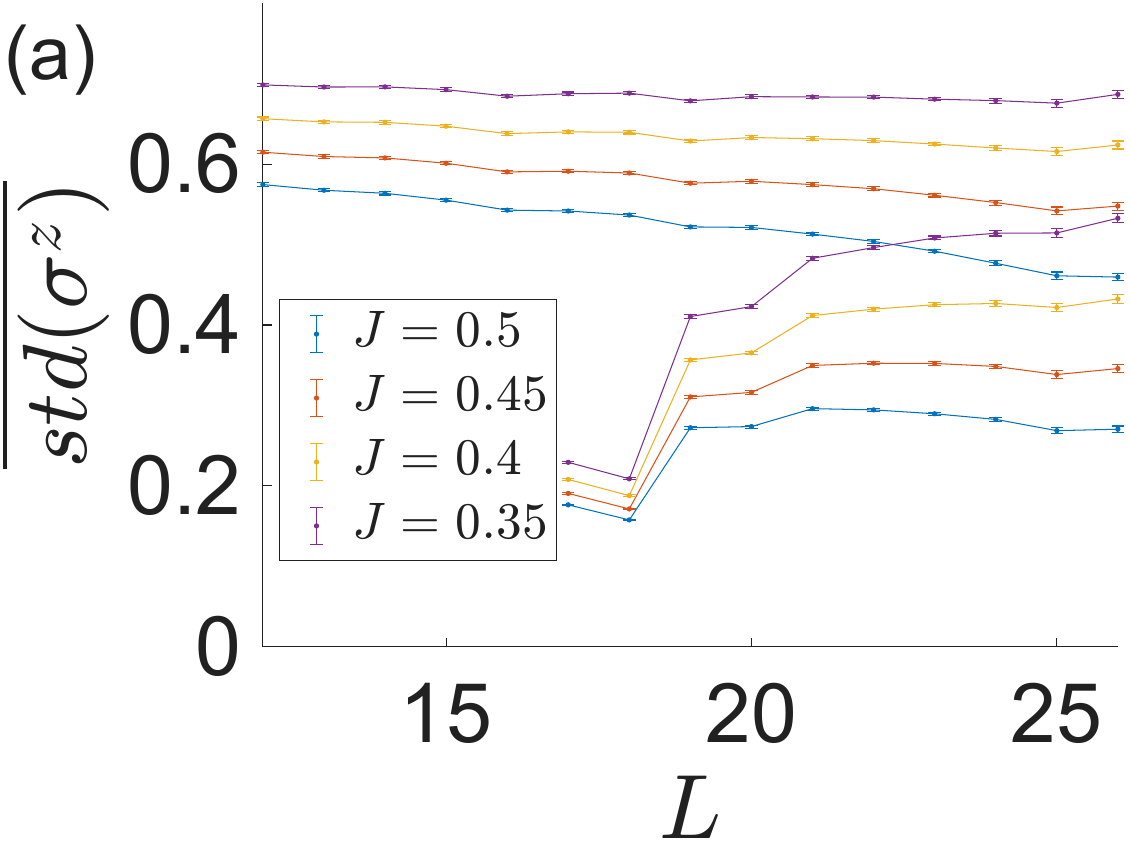}
    \includegraphics[width=0.48\linewidth]{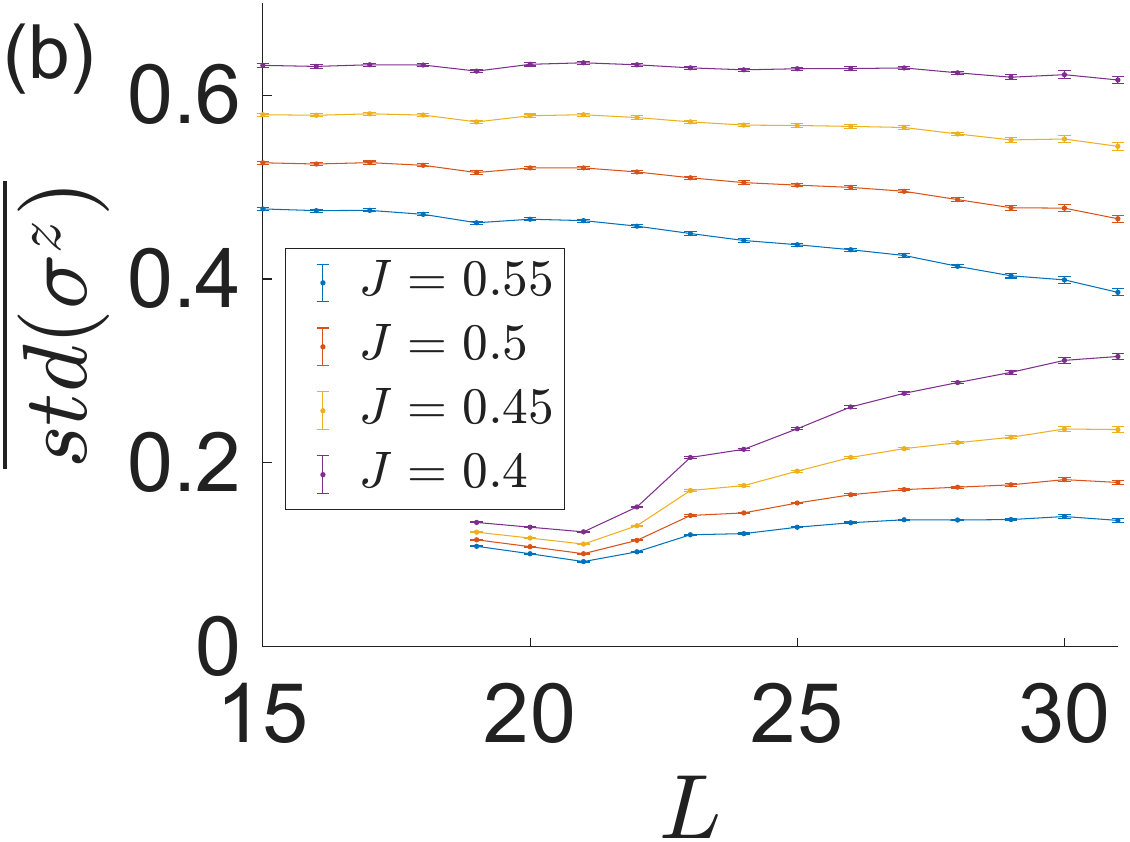}
    \caption{The standard deviation of $\sigma^z$ versus $L$ for Models \Rom{1} ($L_{thermal} = 16$) (a) and \Rom{2} ($L_{thermal} = 18$) (b). The upper lines correspond to cases without a bath, while the lower lines correspond to cases with baths. }
    \label{fig:model_14_sz_bath}
\end{figure}

We begin by studying Model \Rom{1} and Model \Rom{2}. Although the constraints themselves are somewhat abstract and do not provide much intuition, they exhibit qualitatively the same physics as the random-field Heisenberg model and the PXP model. After introducing disorder, these systems display two phases: the MBL phase and the thermal phase. Additionally, there is an intermediate parameter region where the transition occurs. This is clearly illustrated by our numerical results. Similar to the PXP model, we also plot the gap ratio $\ovl{r}$ for different system sizes $L$ and coupling constant $J$ in Fig. \ref{fig:model_14_level_L}. 
The range of parameter $J$ is centered around the transition region and weak MBL region. These parameter ranges are ideal for observing MBL avalanches. To achieve this, we can perform the same calculations as in the PXP model and examine the standard deviation of $s^z$ with and without thermal bath, as shown in Fig. \ref{fig:model_14_sz_bath}. When the system size is sufficiently large, std($s^z$) saturates and remains distinct from the case without a thermal bath.  

\begin{figure}[h]
    \centering
    \includegraphics[width=0.48\linewidth]{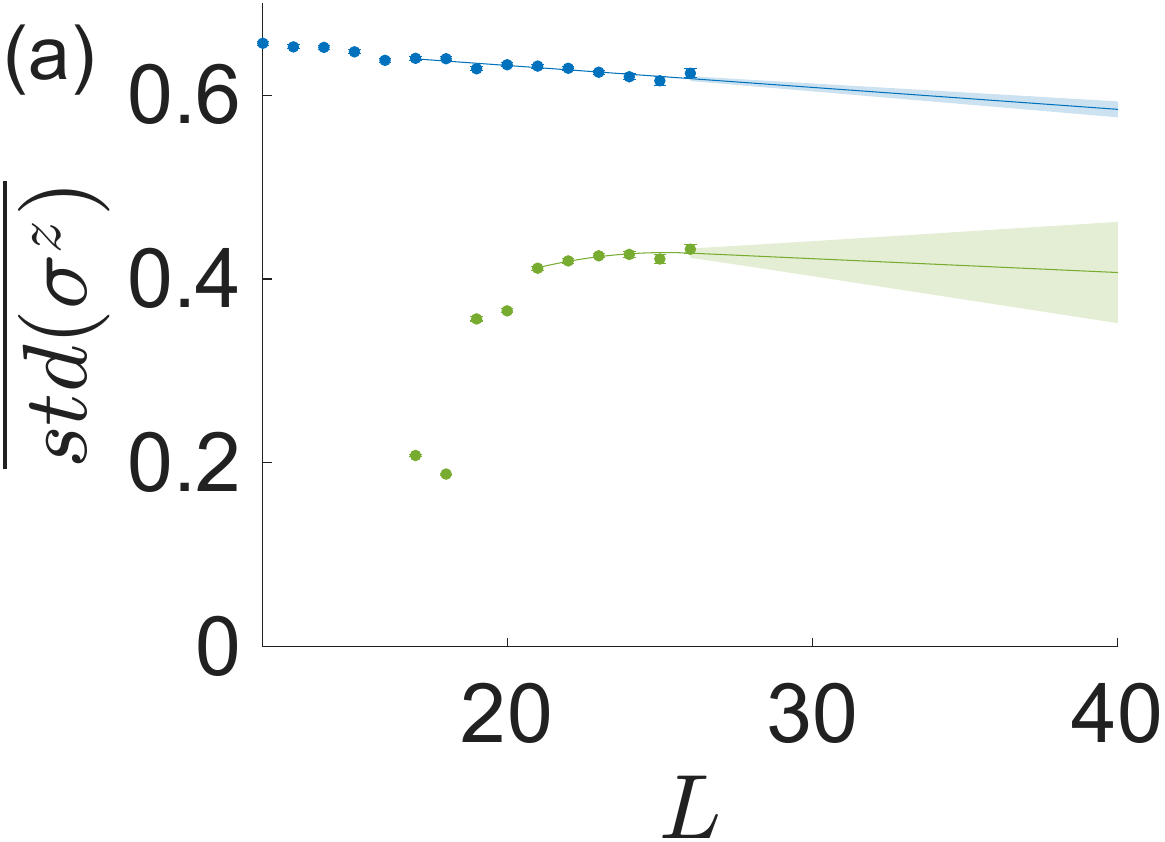}
    \includegraphics[width=0.48\linewidth]{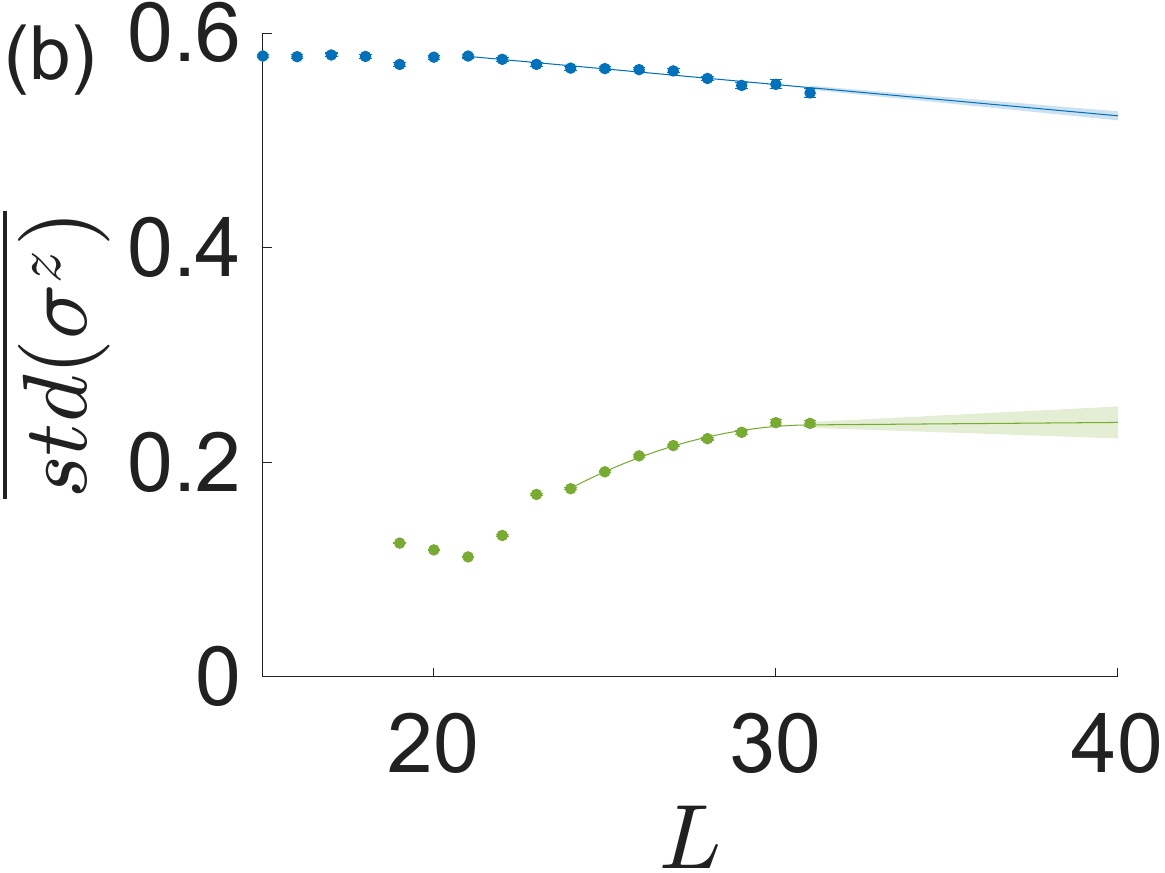}
    \caption{(a) The fitting of the standard deviation data for Model \Rom{1} with $J = 0.4$ and $L_{thermal} = 16$. The data from $L = 17$ to $L = 26$ are used for fitting in the case without the bath, and the data from $L = 21$ to $L = 26$ are used for fitting in the case with the bath. (b) The fitting of the standard deviation data for Model \Rom{2} with $J = 0.45$ and $L_{thermal} = 18$. The data from $L = 21$ to $L = 31$ are used for fitting in the case without the bath, and the data from $L = 24$ to $L = 31$ are used for fitting in the case with the bath. Figure (b) exhibits clear signatures of avalanches, with the influence of the bath persisting over very long distances.}
    \label{fig:model_14_sz_L_extended}
\end{figure}

To estimate the avalanche ratio, we use linear fitting to predict the behavior of $sz$ for much larger system length, as shown in Fig. \ref{fig:model_14_sz_L_extended}. Thanks to more data points and early saturation, the data of model \Rom{2} exhibits a stronger avalanche effect. When $J = 0.45$ and $J = 0.4$, without the bath, the disordered region is in the traditional MBL phase, and $\ovl{\text{std}(\sigma^z)}_0$ decreases very slowly with increasing system length. However, under the influence of a thermal bath, $\ovl{\text{std}(\sigma^z)}$ saturates at less than half the value of $\ovl{\text{std}(\sigma^z)}_0$. We can predict that the influence of the bath remains significant at a distance of at least five times the size of the thermal bath. This makes Model \Rom{2} an excellent candidate of studying avalanches. 

\begin{figure}[h]
    \centering
    \includegraphics[width=0.5\linewidth]{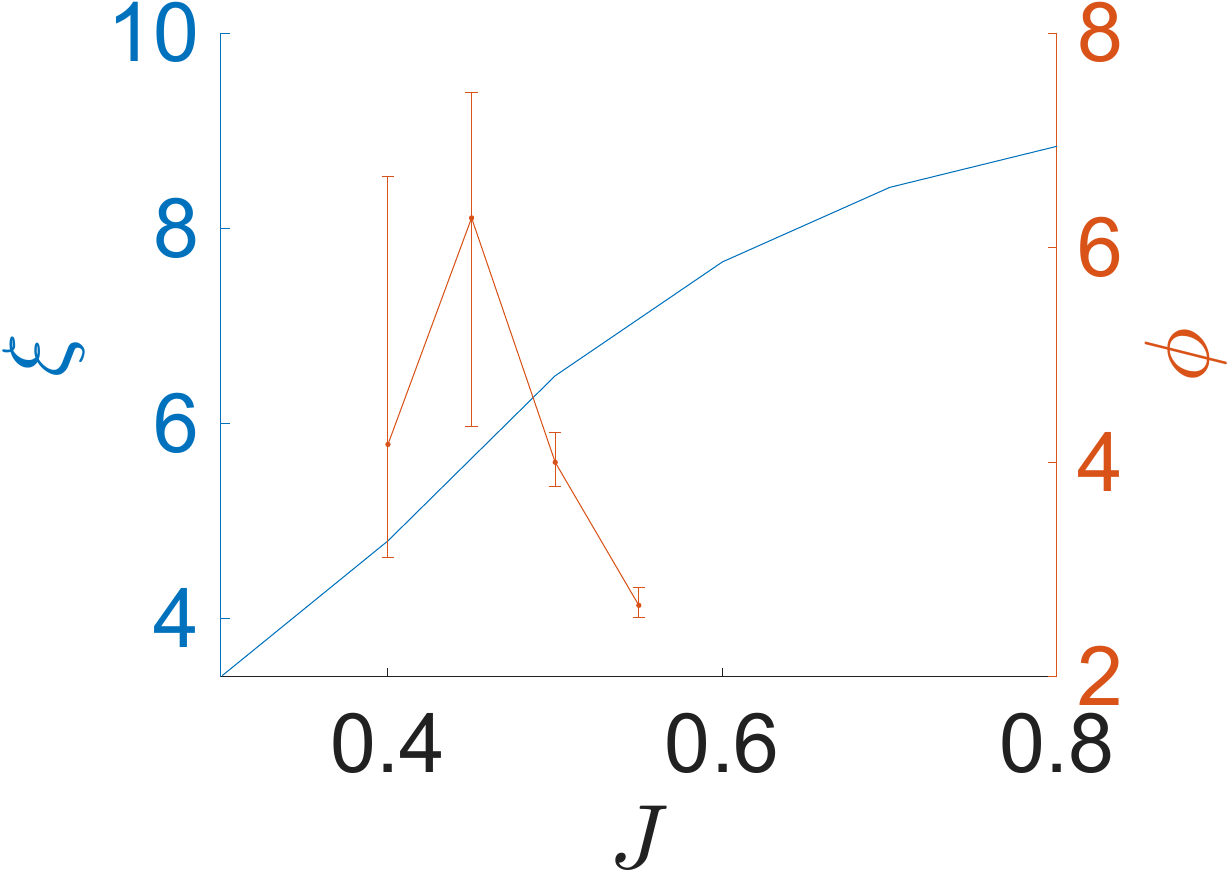}
    \caption{The correlation lengths and the avalanche ratios of Model \Rom{2} versus coupling strengths $J$. The avalanche ratio is large in the region where the correlation length is small and decreasing, indicating a traditional MBL regime.}
    \label{fig:correlation}
\end{figure}

For Model \Rom{2}, in addition to the level statistics shown in Fig. \ref{fig:model_14_level_L}, we provide further evidence that the avalanche phenomena occur within the traditional MBL phase. In Fig. \ref{fig:correlation}, we present the correlation length $\xi$ and the avalanche ratio for different coupling strengths $J$.  Comparing the trends of these two quantities, we observe that the correlation length decreases rapidly as $J$ decreases, indicating a transition into the MBL phase with a shorter localization length. But the avalanche ratio increases as $J$ decreases and becomes significantly larger when the correlation length is small.
The system size influenced by the bath is much larger than both the size of the bath and the correlation length of the disordered region. If longer system sizes were accessible computationally, the avalanche ratio would likely continue to grow as the system delves deeper in the MBL phase. This suggests that the avalanches are not merely a phenomenon of the transition regime but should be taken seriously within the small-system-size MBL regime.

If applying stronger constraints results in longer system sizes and a more pronounced avalanche effect, we could easily explore models that permit longer system sizes. However, we explain that overly strong constraints can lead to the disappearance of the thermal phase even when the disorder strength $h$ is very small compared to the coupling strength $J$. Our Models \Rom{3} and \Rom{4} fall into this category. Interestingly, the Hilbert space dimension of Model \Rom{4} is a polynomial function of the system length $L$, which is typical when constraints are particularly strong. 

\begin{figure}[h]
    \centering
    \includegraphics[width=0.48\linewidth]{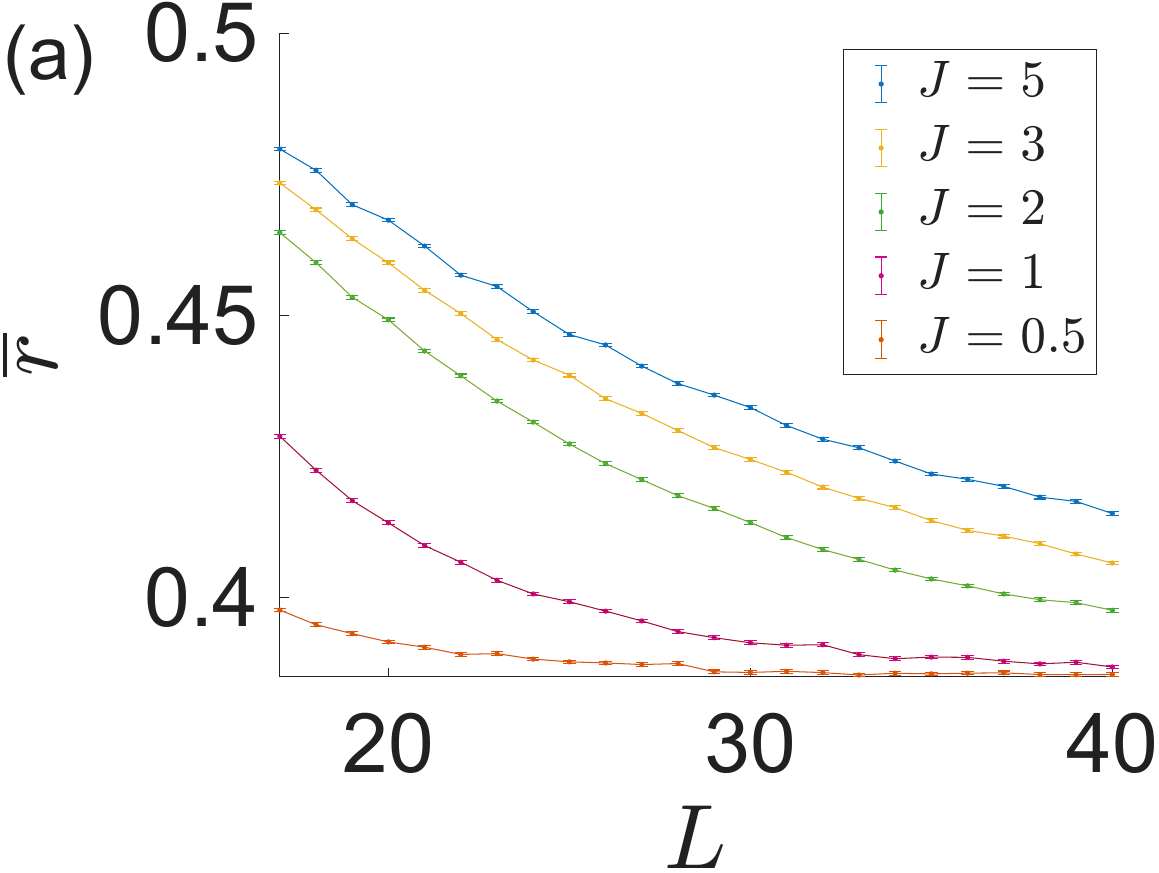}
    \includegraphics[width=0.48\linewidth]{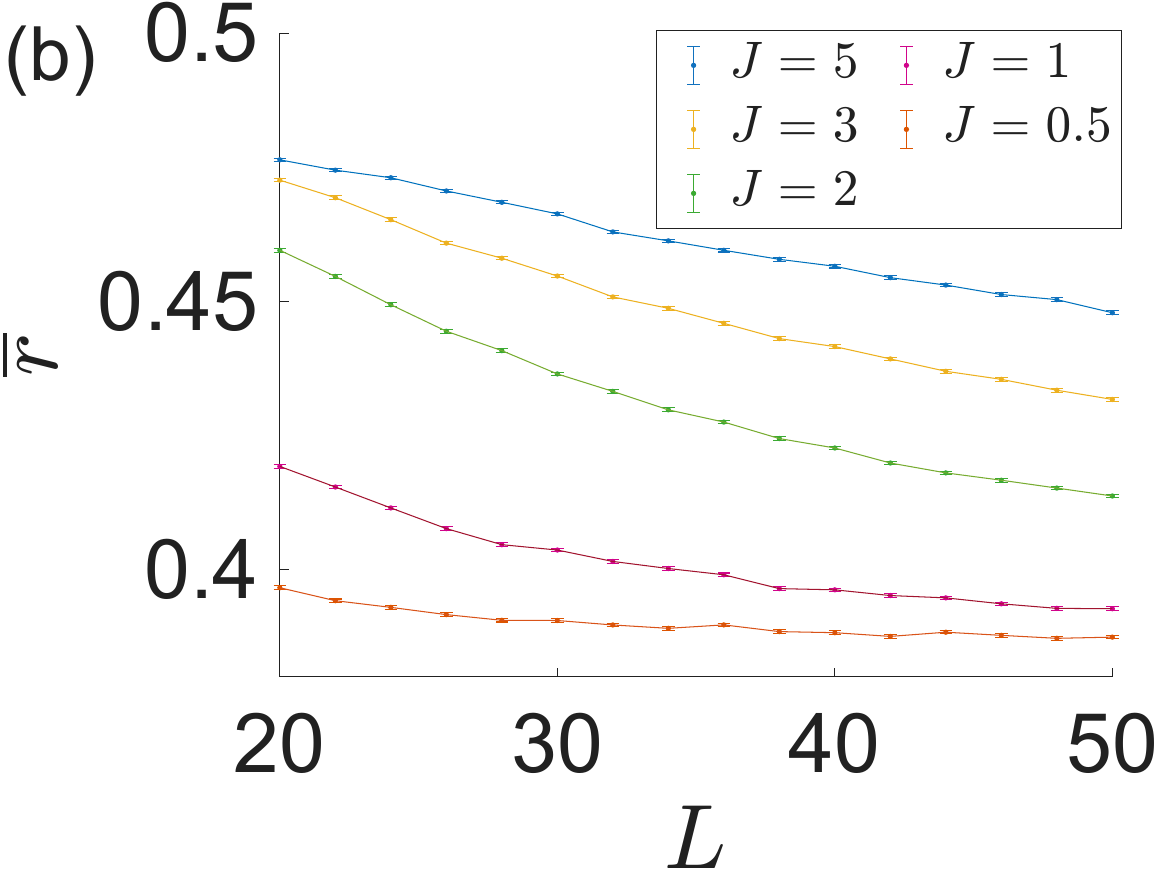}
    \caption{Average gap ratio for Models \Rom{3} (a) and \Rom{4} (b). The gap ratio decrease for all coupling strengths $J$, indicating that the thermal phase does not exist for these two models.}
    \label{fig:model_23_level_L}
\end{figure}

\begin{figure}[h]
    \centering
    \includegraphics[width=0.48\linewidth]{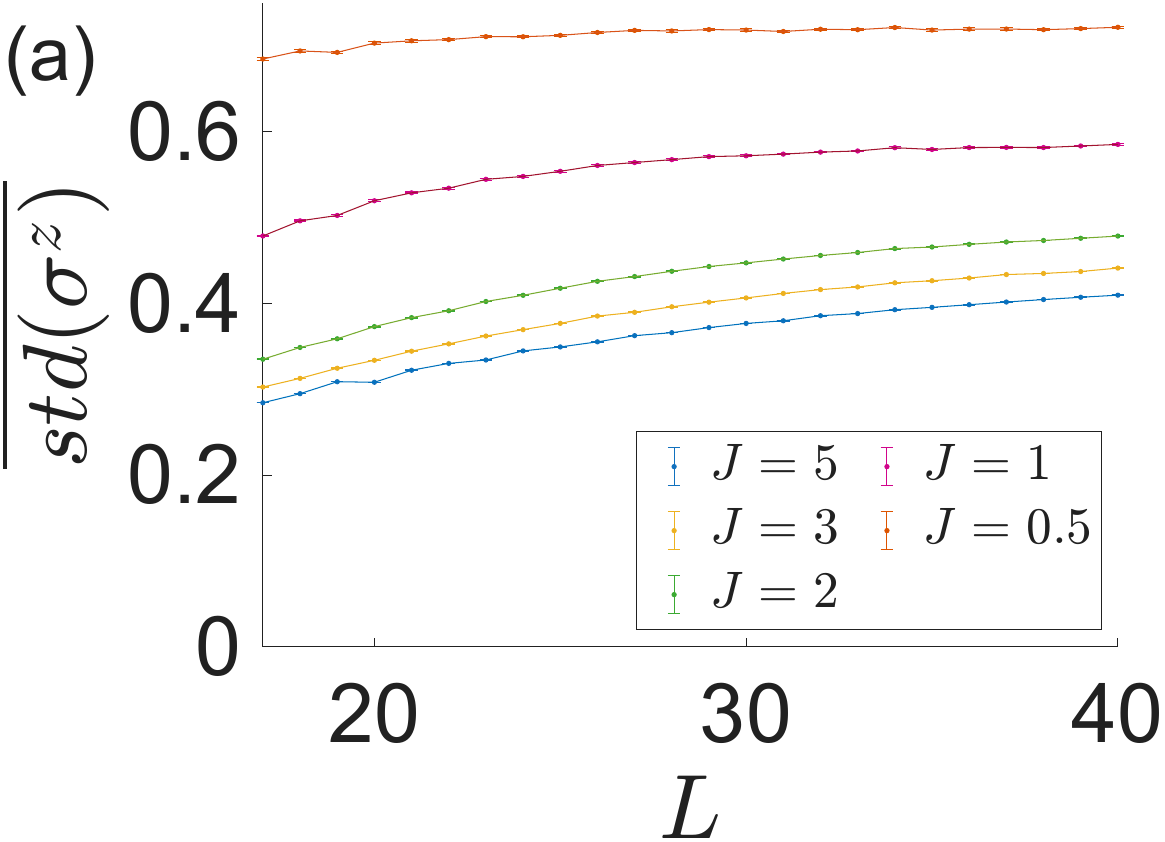}
    \includegraphics[width=0.48\linewidth]{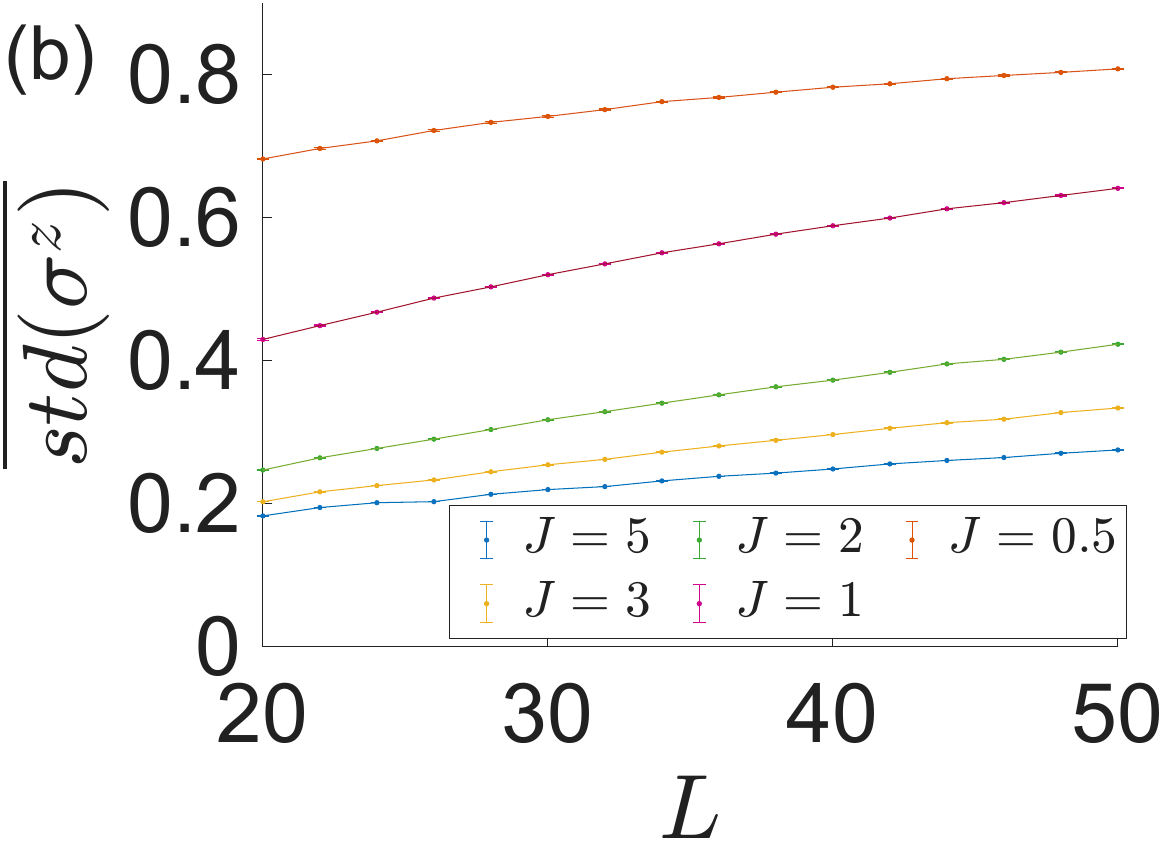}
    \caption{Standard deviation of $\sigma^z$ for Models \Rom{3} (a) and \Rom{4} (b). The data increase and then remain at a finite value, indicating that the thermal phase does not exist.}
    \label{fig:model_23_sz_L}
\end{figure}

We present the gap ratio $\ovl{r}$ and the standard deviation of $\sigma^z$ for these two models in Figs. \ref{fig:model_23_level_L} and \ref{fig:model_23_sz_L}. We observe that $\ovl{r}$ decreases continuously and $\ovl{\text{std} (\sigma^z)}$ remains constant in the large system size limit. This behavior is characteristic of MBL. However, this MBL is induced by constraints rather than disorder, as this localization occurs for any coupling strength $J$.

It is well established in the literature that strong constraints are associated with slow dynamics and localization \cite{Roy_2020, Pancotti_2020, Lan_2018, Feldmeier_2019}.
These two models highlight the importance of the extent of constraints. To study the transition effectively, it is crucial to avoid entering the region where the constraint-induced localization occurs. This consideration is significant in the model engineering process for studying avalanches.

\section{Model comparison and analysis}\label{sec:compare}
After exploring several models, we can compare them and select the one with the highest avalanche ratio under the same conditions. This comparison will guide our exploration of additional models in the future. In addition to PXP model, Models \Rom{1} and \Rom{2}, we also use a standard spin $\frac12$ model to facilitate comparisons between models with and without constraints. Since all of our constrained models lack a conserved quantity, we employ the random-field Heisenberg model and break the $s^z$ conservation by applying a uniform magnetic field in the $x$ direction. Specifically, the Hamiltonian is 
\bea\label{eq.Heisenberg}
\hat{H} = \frac{1}{2}\sum_{i=1} ^ L (h^z_i \sigma^z + h^x_i \sigma^x) + \frac{1}{4}\sum_{i=1}^{L - 1} J_i\boldsymbol{\sigma}_i \cdot \boldsymbol{\sigma}_{i + 1}
\eea
Here, $h^x_i = h^x$ and $J_i = J$ are constant while $h^z_i \in [-h^z, h^z]$ represent the random field with a uniform distribution. The value of $h^x, J, h^z$ can vary between the bath and the disordered region when studying avalanches, but they remain constant within each region. Since this model lacks constraints and is not the primary focus of this paper, we will include the detailed numerical results on MBL and avalanches in Appendix \ref{app:Heisenberg} and only present its data when comparing with other constrained models.

The central goal is to determine which model provides a stronger signature of avalanches given limited computing power. To ensure a fair comparison, we maintain the same Hilbert space dimension across the 4 models. This requires choosing the parameters $L_{max}$ and $L_{thermal}$ such that the dimensions of the bath and the entire Hilbert space are nearly identical. For the Heisenberg model, the PXP model, Model \Rom{1} and Model \Rom{2}, we use $L_{max} = 15, 22, 26, 31$ and $L_{thermal} = 10, 14, 16, 18$, respectively. The Hilbert space dimensions are shown directly in Fig. \ref{fig:dimension}. We use small dots on the lines to indicate the bath Hilbert space dimensions, all of which are close to 1000. 

\begin{figure}[h]
    \centering
    \includegraphics[width=0.5\linewidth]{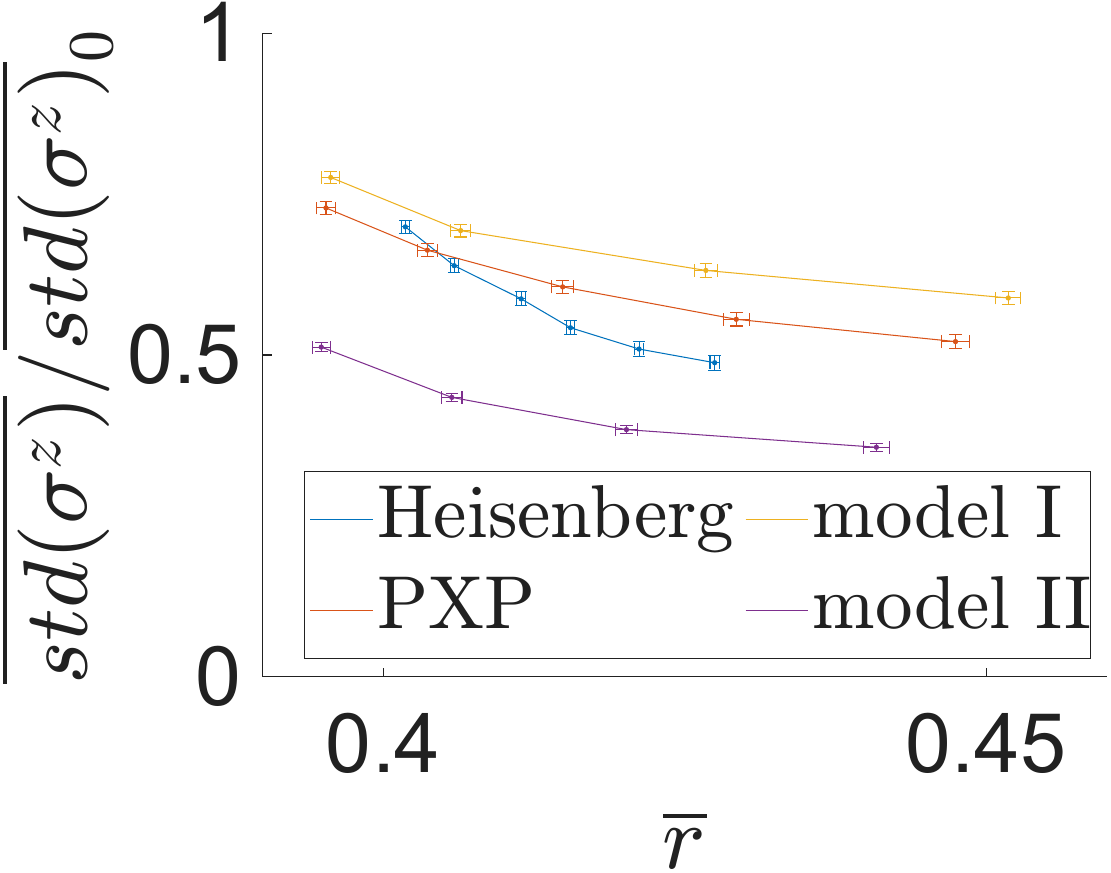}
    \caption{Comparison of the difference in standard deviation of $\sigma^z$ with and without the thermal bath. Data points corresponding to different models and coupling strengths $J$ are presented. The horizontal position is based on the gap ratio $\ovl{r}$ for a fair comparison. The model \Rom{2} exhibits stronger avalanches compared to other models.}
    \label{fig:compare_sz_r}
\end{figure}

The first comparison we can make is the difference between $\ovl{\text{std}(\sigma^z)}$ and  $\ovl{\text{std}(\sigma^z)}_0$. We present ratio of these two values at the largest system length, $L_{max}$, when changing the parameter $J$ in Fig. \ref{fig:compare_sz_r}. The x-axis represents $\ovl{r}$ for the case without a bath at length $L_{max}$. Points at similar horizontal positions correspond to the same level of localization. On the y-axis, we observe the data for Model \Rom{2} is lower than for the other models, indicating that the influence of the thermal bath on the disordered region is stronger. Since this comparison lacks sufficient accuracy, it is difficult to draw definitive conclusions about the other three models. 

\begin{figure}[h]
    \centering
    \includegraphics[width=0.5\linewidth]{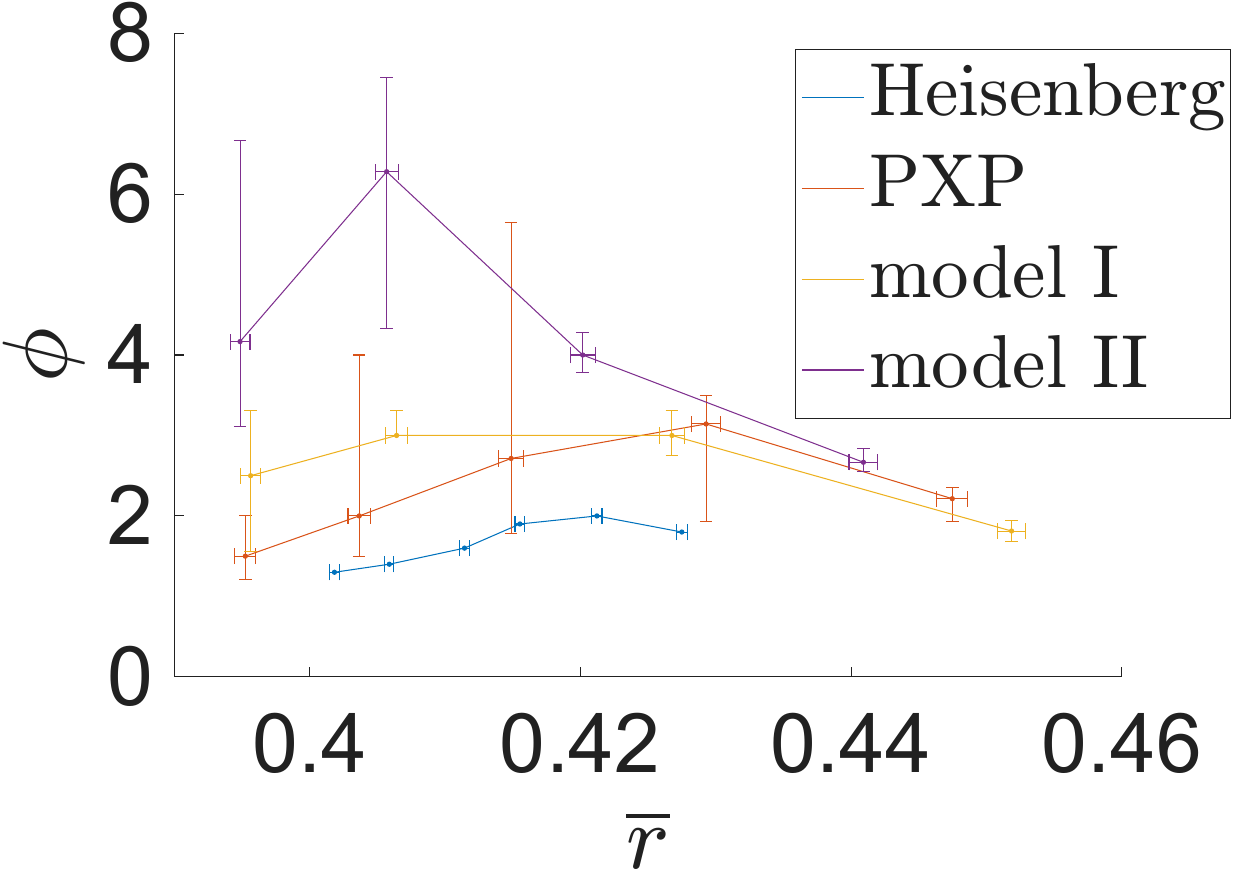}
    \caption{Avalanche ratios for different models and $J$ are presented together. Although the data is not highly accurate, the avalanche ratio of Model \Rom{2} is higher on the localized side (small $\ovl{r}$) compared to other models. Note that the number of data points for the Heisenberg model Eq. \ref{eq.Heisenberg} is insufficient to estimate the error bars accurately; therefore, they are not shown.}
    \label{fig:compare_phi_r}
\end{figure}

Next, we compare the avalanche ratio $\phi$ that we defined. We plot $\phi$ versus $\ovl{r}$ for each case in Fig. \ref{fig:compare_phi_r}. For the three constrained models, when $\ovl{r}$ is large, the differences are minimal because $\phi$ primarily depends on the data without a bath (specifically $k_0$ in Eq. \ref{eq.std_0}). As the system becomes more localized, the values of $\phi$ begin to diverge, with Model \Rom{2} outperforming the other two. This larger $\phi$ is attributed to the stability and early saturation of the data with a bath, as directly observed in Fig. \ref{fig:model_14_sz_L_extended}.
Note that we do not plot error bars for the Heisenberg model data due to the insufficient number of data points. Consequently, the values are far less accurate than those of the other cases. More details are in Appendix. \ref{app:Heisenberg}.  

From these comparisons, we can conclude that the constraints and longer system sizes enhance the stability of the data and result in a higher avalanche ratio. This is particularly valuable given the limited system sizes available. However, attributing these results solely to the system sizes would be an oversimplification. The unique characteristics of each model also play a crucial role in determining its strengths and weaknesses for studying avalanches. 

\begin{figure}[h]
    \centering
    \includegraphics[width=0.5\linewidth]{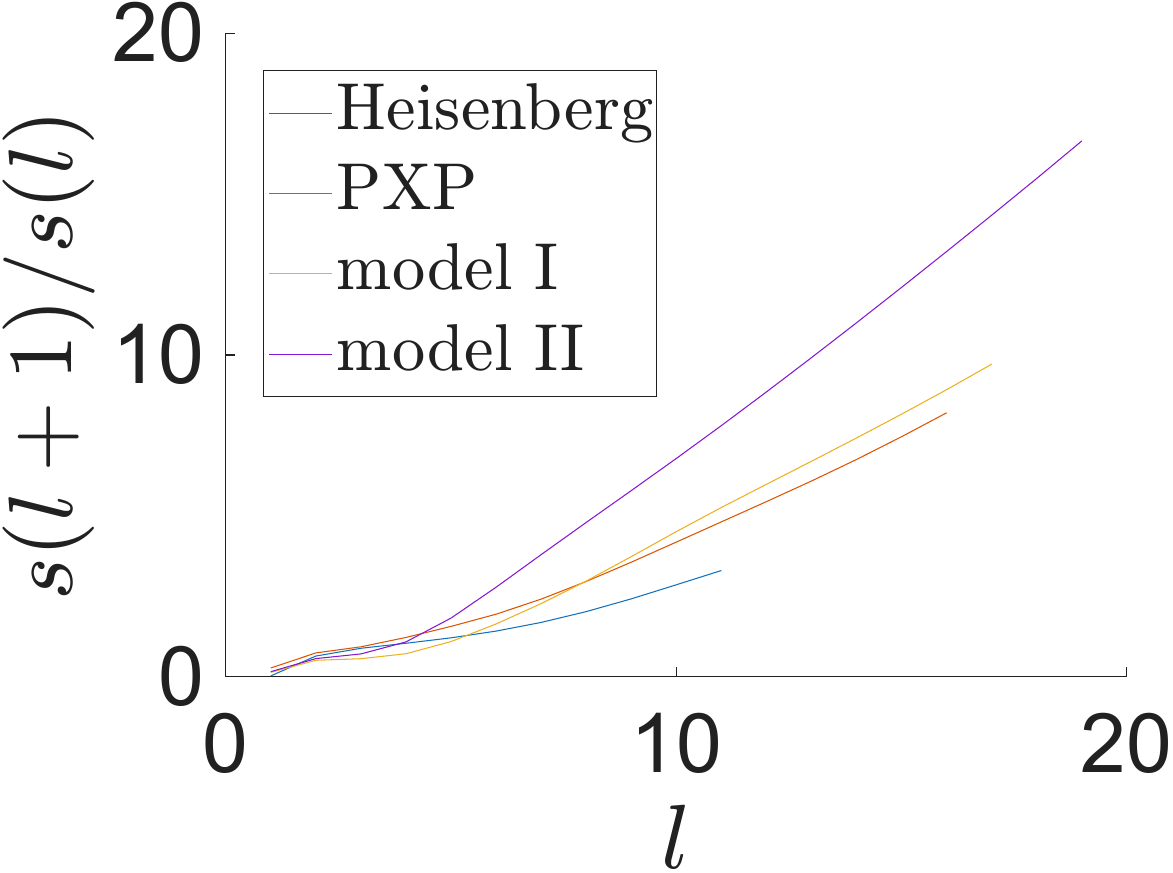}
    \caption{Operator growth data for different models. The growth rates are similar when $l$ is small, but Model \Rom{2} exhibits much faster growth at large $l$. This may be potentially connected stronger avalanche behavior in Model \Rom{2}.}
    \label{fig:operator}
\end{figure}

One theoretical approach that guides our comparison is the operator growth. In the paper \cite{Weisse_2024}, the commutator of the operator with the Hamiltonian is calculated to illustrate the divergence of the LIOMs in large system sizes. The quantity $s(l)$ is defined as the magnitude of operator $\sigma^z_L$ after being conjugated by Hamiltonian $l$ times. If the ratio $s(l+1) / s(l)$ does not saturate (as it does in AL) but instead grows to infinity with $l$, it indicates that LIOMs will eventually diverge. We present the operator growth for these models in Fig. \ref{fig:operator}. At small $l$ the differences between models are not obvious, but as $l$ increases, the data for Model \Rom{2} grows faster than Model \Rom{1} and the PXP model, which in turn grow faster than the Heisenberg model. Although it is challenging to rigorously connect these operator growth results with the avalanche ratio $\phi$ obtained in ED, we can assert that there is some correlation, as the case with the fastest operator growth (Model \Rom{2}) also shows the clearest signature of avalanches.

\section{ED versus Lindblad master equation analysis}\label{sec:master_equation}
Since exploring avalanches using ED for spin 1/2 models is not feasible for most cases, many alternative methods have been developed to calculate the coupling. One of the successful approaches is the Lindblad master equation. This theory allows for the estimation of the time scale at which a system is thermalized by a large thermal bath. After obtaining the ED results, it is insightful to compare them with the predictions from the Lindblad master equation to assess their accuracy. 

The general form of the Lindblad master equation is given by
\bea\label{eq.lme}
\mathcal{L}[\rho] = -i [ H, \rho] + \gamma \sum_\mu \left ( L_\mu \rho L_\mu^ \dagger  - \frac12 \{ L_\mu ^\dagger L_\mu , \rho \}\right )
\eea
where $L_\mu$ are the Lindblad operators. For the PXP model, we choose $L_1 = P^z_1$, $L_2 = \sigma^x_1 P^z_2$. For the random-field Heisenberg model with a transverse field, we select $L_1 = \sigma^x$, $L_2 = \sigma^y$, $L_3 = \sigma^z$. We follow the method introduced in paper \cite{Morningstar_2022, Sels_2022}: diagonalize the Hamiltonian, use the eigenstates as a basis and restrict the density matrix to its diagonal elements to focus on perturbative effects. More details about this method refer to Appendix. \ref{append.lindblad}. Although this approximation may not be fully applicable because we set $\gamma = 1$, the results will be sufficiently accurate in their order of magnitude. We obtain the difference between the largest and the second largest real part of the eigenvalues, denoted as $\Gamma$. The time scale for system thermalization is given by $1/\Gamma$.  

\begin{figure}[h]
    \centering
    \includegraphics[width=0.48\linewidth]{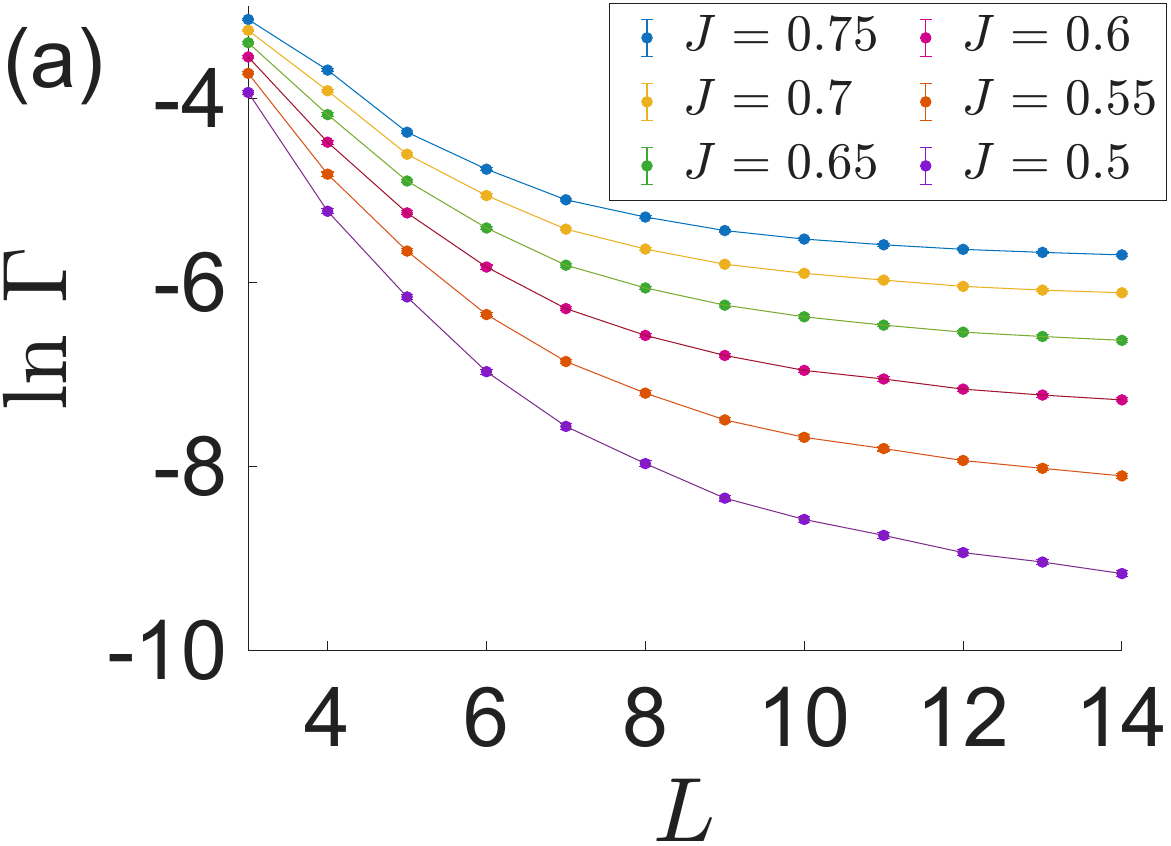}
    \includegraphics[width=0.48\linewidth]{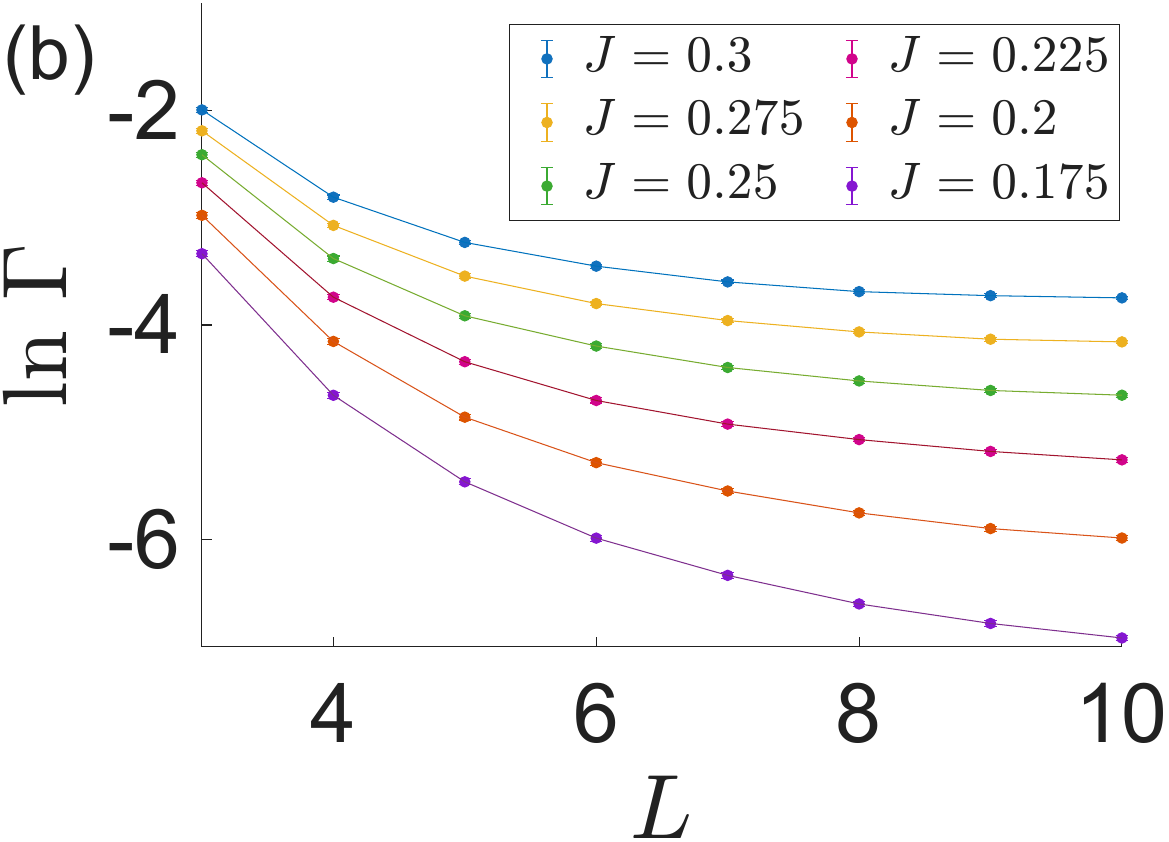}
    \caption{Relaxation energy scale of the PXP model and the Heisenberg Model \ref{eq.Heisenberg} for different $J$.}
    \label{fig:enter-label}
\end{figure}

Next, we couple this system to a thermal bath. The timescale for thermalization by the bath is related to the density of states $\rho = 1/\Delta$, where $\Delta$ is the average energy gap. We compare $\Delta$ with $\Gamma$ to determine whether the system can be thermalized by the thermal bath. The disordered region can not be thermalized if $\Gamma$ is much smaller than $\Delta$, but can be thermalized if they are of the same order of magnitude. We explicitly write the condition as
\bea\label{eq.delta_gamma}
\log \Delta \sim \log \Gamma
\eea
where $\sim$ indicates that their values are approximately equal.
With this theory from the Lindblad master equation, we can predict how large a bath is needed to thermalize the disordered system, and we compare the predictions with ED calculations. On one hand, the Lindblad master equation allows us to estimate how effectively the disordered part will be thermalized by the bath. For each disorder sample, we calculate the probability that $\Gamma_i$ is smaller than $\Delta$. We define a quantity 
\bea \label{eq.P}
P(\Gamma < \Delta) = \frac{|\{ i = 1 \cdots N_{sample} \ |\  \Gamma_i < \Delta \}|}{N_{sample}}
\eea
This represents the probability that the system remains unthermalized across different disorder realizations. On the other hand, based on ED results, we can estimate the extent to which the systems are not thermalized. We define another quantity $\ovl{\text{std}(\sigma^z)}/ \ovl{\text{std}(\sigma^z)}_0$ to indicate the degree of localization. When the ratio is $1$, the system remains unaffected and  localized, whereas if it is close to $0$, the system is thermal.

\begin{figure}[h]
    \centering
    \includegraphics[width=0.48\linewidth]{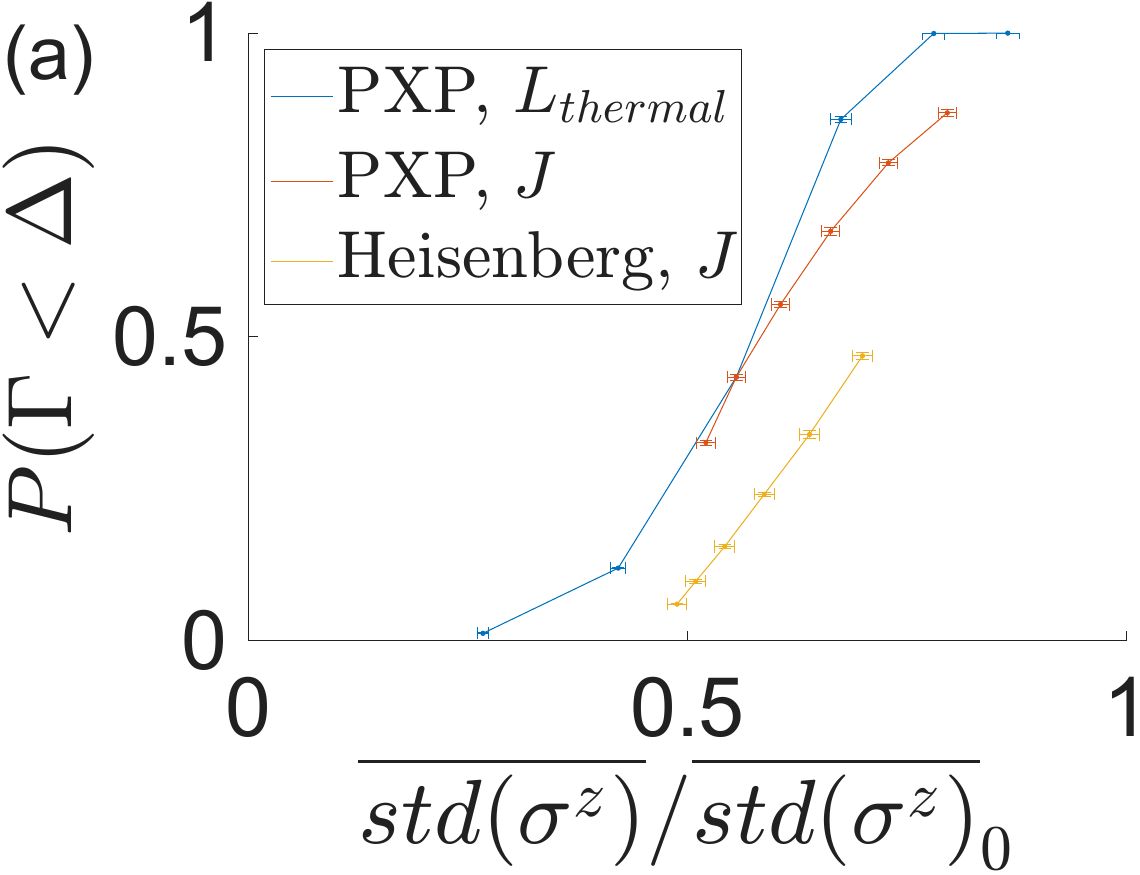}
    \includegraphics[width=0.48\linewidth]{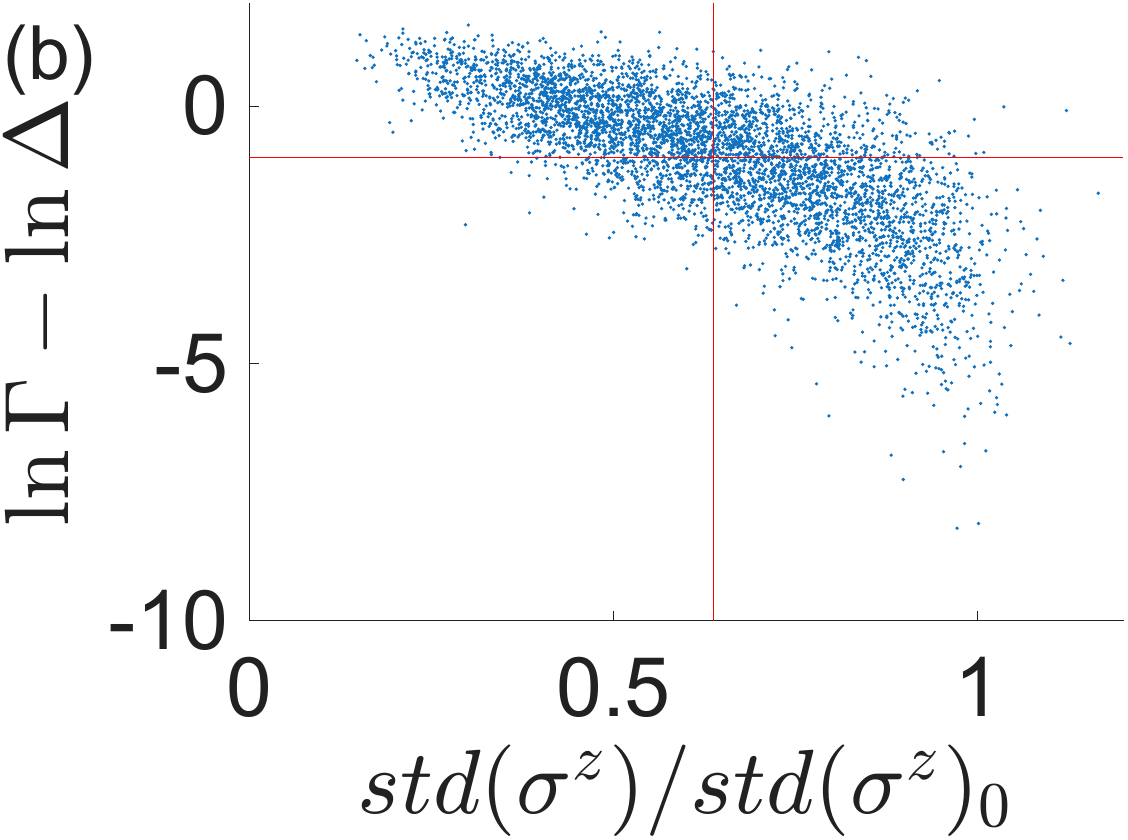}
    \caption{Comparison between the results of ED and the Lindblad master equation. (a) The probability P defined in Eq. \ref{eq.P} (from the Lindblad master equation results) versus the ratio of the standard deviation with and without the thermal bath (from ED results). Both quantities qualitatively represent the extent of remaining localized under the influence of the thermal bath. Data points pass through the central region around (0.5, 0.5), indicating the consistency between these two methods. (b) The comparison of these two quantities for each disorder configuration shows a clear correlation. There are 5000 data points corresponding to $J = 0.7$, $L_{thermal} = 12$ and $L = 20$. The red lines indicate the average.}
    \label{fig:master_ed}
\end{figure}

We present the results of these two quantities in Fig. \ref{fig:master_ed}. For this comparison, we use several data sets. The first two sets are for the PXP model. In the first set, we fix $J =0.7, L = 22$, and vary the bath size from $8$ to $18$. The size of the disordered region changes correspondingly. In the second set, we fix $L_{thermal} = 14, L = 22$, but vary $J$ from $0.75$ to $0.5$ in increments of $0.05$. The third set of data uses the random-field Heisenberg model with a transverse field, where we fix $L_{thermal} = 10, L = 15$ and vary $J$ from $0.3$ to $0.175$ in increments of $0.025$. 
\begin{figure}
    \centering
    \includegraphics[width=0.5\columnwidth]{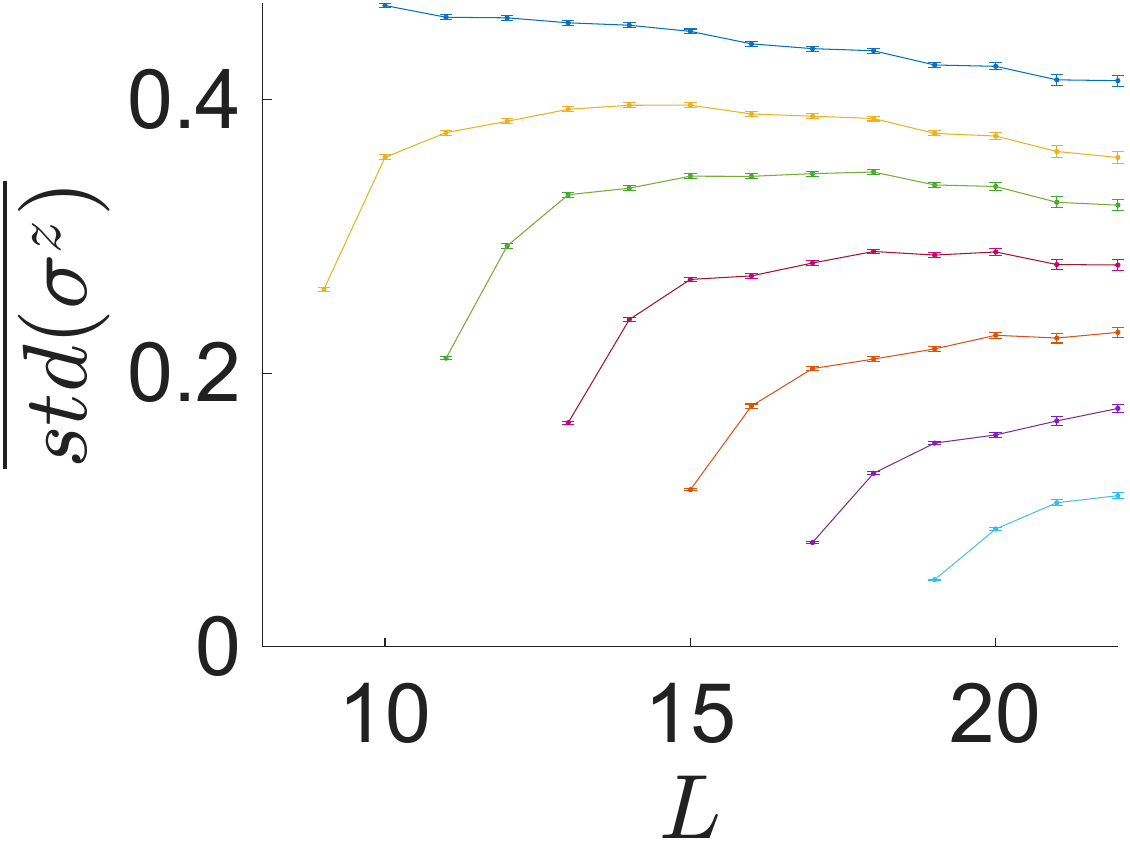}
    \caption{The standard deviation $\sigma^z$ for the PXP model with $J = 0.7$. In addition to the case without the bath, we calculate the standard deviation of $\sigma^z$ for $L_{thermal} = 8, 10, 12, 14, 16, 18$ to compare the influence of baths of different sizes. The data are stable at $L = 22$ for most cases, but the value of the standard deviation of $\sigma^z$ vary significantly. }
    \label{fig:enter-label}
\end{figure}
Since the interpretation of these two quantities  is not strictly quantitative, and given the approximations and finite size effects, we do not expect a specific functional relationship between them. However, the data points should pass through the central region (around the point (0.5, 0.5)), meaning that in the parameter region described by Eq. \ref{eq.delta_gamma}, the disordered region should be noticeably influenced by the bath. This is indeed observed for the PXP model, but the agreement is less satisfactory for the Heisenberg model \ref{eq.Heisenberg}. The results are more consistent for larger disordered regions, because the change in $\Gamma$ versus system length slows down later than the change in $\sigma^z$. Since the length of the disordered region is only $5$ for this model, this effect contributes to the off-center position of the data for the model in Eq. \ref{eq.Heisenberg}.

Overall, the results of this comparison suggest that comparing $\Gamma$ and $\Delta$ is a good criterion for determining whether the system will be thermalized. However, even though condition \ref{eq.delta_gamma} is satisfied, the disordered region may not behave like a true  thermal bath. Instead, these regions remain in the intermediate regime as discussed at the end of Section \ref{sec:PXP_avalanche}. Achieving true thermalization is more challenging than what the Lindblad master equation predicts.

Furthermore, we present the distribution of these two quantities for each disorder realization to gain deeper insight into the comparison. In Fig. \ref{fig:master_ed} (b), each point represents a specific disorder configuration. We observe a correlation between $\Gamma$ and $\ovl{\text{std}(\sigma^z)}/ \ovl{\text{std}(\sigma^z)}_0$. This correlation is expected, as both quantities are related to extent of thermalization.

Let us assume that the Lindblad master equation results accurately capture the trend in the time scale required to thermalize the disordered region. The calculation indicates that $\Gamma$ decreases as $L$ increases, resembling a logarithmic function. This suggests that the avalanche ratio would, in principle, grow exponentially with the size of the thermal bath. However, in a real disordered system, the probability of finding a large thermal bath is also exponentially small. Without further, more accurate quantitative studies, it is difficult to draw definitive conclusions about the existence of MBL. 

\section{conclusion and outlook}
We study avalanches as a mechanism to break the MBL phase in disordered spin chains by comparing the changes  in observables with and without a thermal bath. The analysis of observables in eigenstates, combined with the ED algorithm, provides an accurate and static method that reflects the influence of the thermal bath on the disordered system. This approach can be applied to relatively large systems, especially when focusing on states at a specific energy. The trends in the changes of observables due to the bath as a function of  $L$ offer predictions for even larger system sizes.

From the numerical results, we can assert that the disordered region is sensitive to a thermal bath, as illustrated by the avalanche ratio $\phi$ defined in this paper. Crucially, the sensitivity increases as the disordered system approaches the MBL phase, resulting in a larger $\phi$. Being too deep in the MBL phase requires larger systems to observe the avalanche effects, which are not accessible numerically. However, the influence of the bath does not decay or disappear within the scale of the bath size or the correlation length in the MBL phase if the bath exceeds a certain threshold, raising reasonable suspicion about the existence of MBL phase in extremely large systems.

In addition to the ED study, many other methods provide insight into the nature of avalanches. Operator growth successfully explains the breakdown of LIOMs and demonstrates that the influence of the bath on the disordered region does not decay exponentially. The operator growth calculation shows excellent consistency in quasi-periodic spin chains and the constrained systems discussed in this paper. Therefore, it can serve as a valuable tool for predicting the strength of avalanches. Further exploration of the quantitative connection between avalanches and operator growth is necessary in the theory.

Another method used to study the physics of avalanches is the Lindblad master equation. The Lindblad master equation is a well-established method for dealing with systems coupled to a thermal bath, and it provides results that are reasonably consistent with our ED findings. The interpretation of these results is crucial. Although the disordered region is significantly influenced by the thermal bath, it is not fully thermalized. Instead, the disordered region remains in an intermediate state for a long distance. The scenario in which the thermal bath absorbs the disordered region to form a larger thermal bath and accelerates the thermalization process is not observed in the ED results. It is possible that an intermediate state (different from both MBL and thermal phases) and even the MBL state itself could be stable in the thermodynamic limit.

The constrained models that we identified and studied provide a valuable platform for exploring avalanches. The physics of the intermediate state can be studied in greater detail to better understand the stability of MBL. Calculations on systems with additional sites are feasible for those with more powerful computational resources, potentially leading to a  clearer understanding of avalanches and the observation of higher avalanche ratios. Although MBL can not be studied using integrable models due to its inherent nature, some models exhibit much better properties within the limits of available computing power. Our work opens the door to using model engineering to study the physics of MBL. The genetic algorithm played a crucial role in discovering the Model \Rom{2}, which exhibits a large avalanche ratio, and other models with rich behaviors. Other search algorithms can be employed or further customized to explore other problems, including those related to localization and beyond. 

In this work, we propose new methods to diagnose MBL and thermal phases, and use ED to study them in a straightforward manner. We establish the sensitivity of disordered systems to thermal baths and distinguish intermediate states from MBL and thermalized states. Although we do not draw many decisive conclusions on the stability of MBL phase itself due to limited computational power, this work provides a new perspective and insights into the phenomenon of avalanches.

\section{ackownledgments}
 I would like to thank Yuan-Ming Lu and Kyle Kawagoe for valuable discussions during the development of this paper. This work is supported by the Center for Emergent Materials at The Ohio State University, a National Science Foundation (NSF) MRSEC, through NSF Award No. DMR-2011876.

\appendix
\section{Random-field Heisenberg model with transverse field} \label{app:Heisenberg}
In this appendix, we present detailed numerical results on MBL and avalanches for the model in Eq. \ref{eq.Heisenberg}. Similar to Models \Rom{1} and \Rom{2}, we focus on the transition parameter region. First, we calculate the gap ratio $\ovl{r}$ versus $L$ for six different coupling strengths from $J = 0.175$ to $J = 0.3$. 
\begin{figure}[h]
    \centering
    \includegraphics[width=0.5\linewidth]{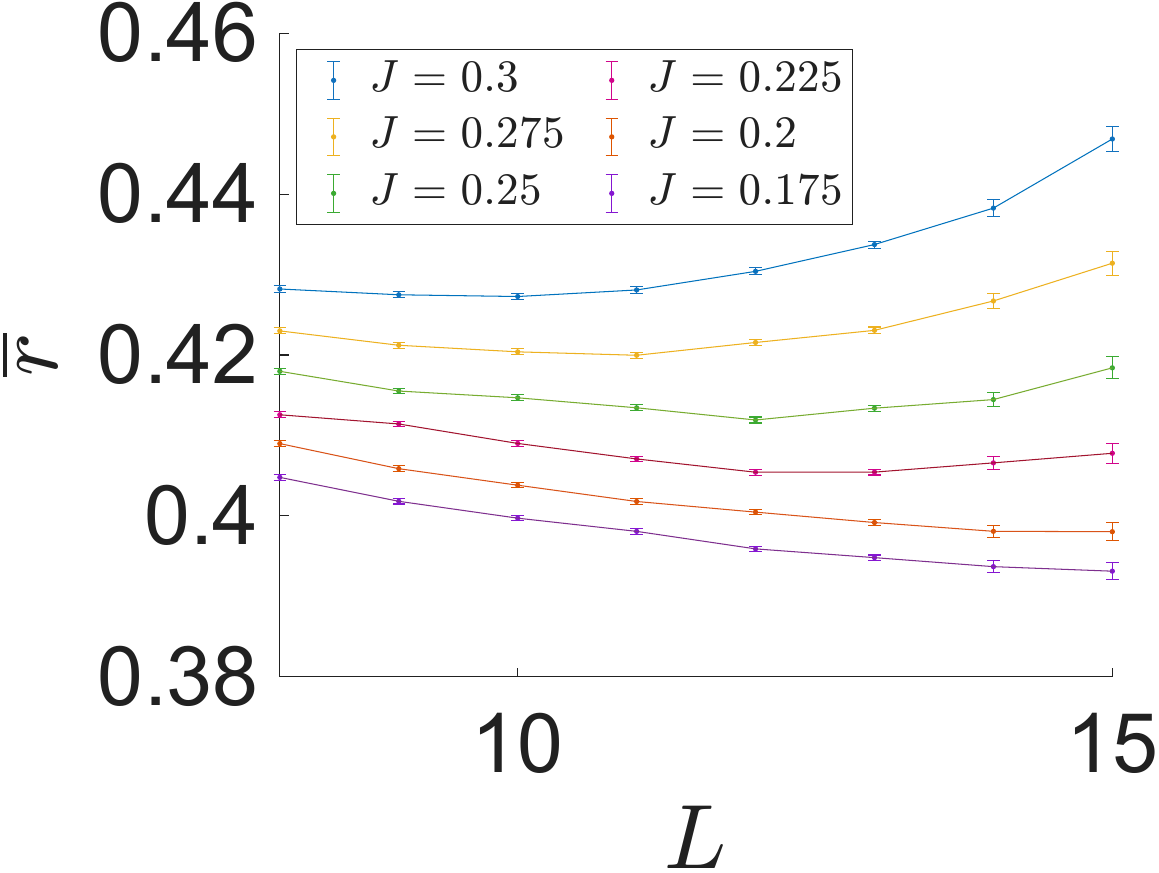}
    \caption{Gap ratio versus system size for various coupling strengths $J$.}
    \label{fig:Heisenberg_level_L}
\end{figure}

 Additionally, for each parameter $J$, with and without a thermal bath, we show the value $\ovl{\text{std}(\sigma^z)}$ versus $L$. We set $h^z = 1, h^x = 0.1$ in the disordered region and $h^z = 0.2, h^x = 0.5, J = 1$ in the thermal region. We use $L_{thermal} = 10$ for these cases to compare with those constrained models. $\ovl{\text{std}(\sigma^z)}$ begins to exhibit saturation behavior around $L = 15$, making it difficult to predict the behavior for longer lengths. Thus, at a specific parameter $J = 0.2$, we calculate  $\ovl{\text{std}(\sigma^z)}$ with $L_{thermal} = 8, 9, 10$. This additional data allows us to be more certain that the values are approaching saturation.
\begin{figure}[h]
    \centering
    \includegraphics[width=0.48\linewidth]{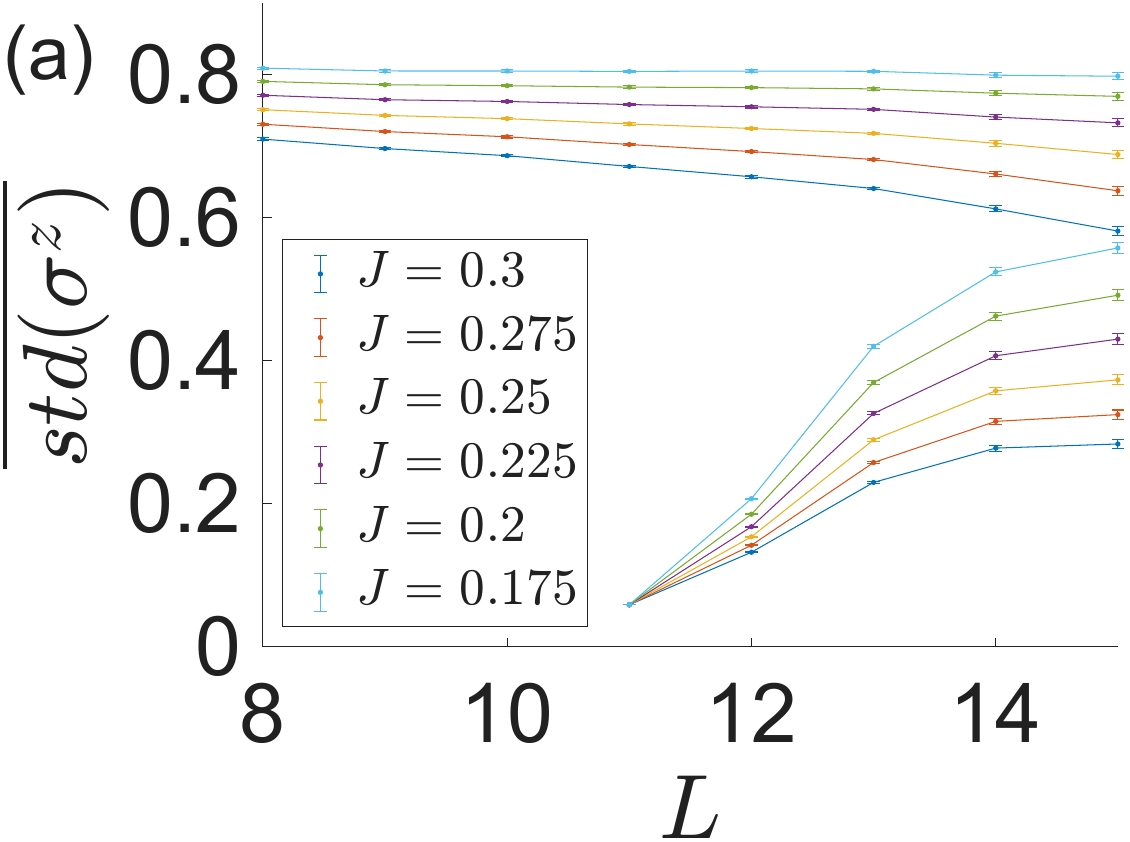}
    \includegraphics[width=0.48\linewidth]{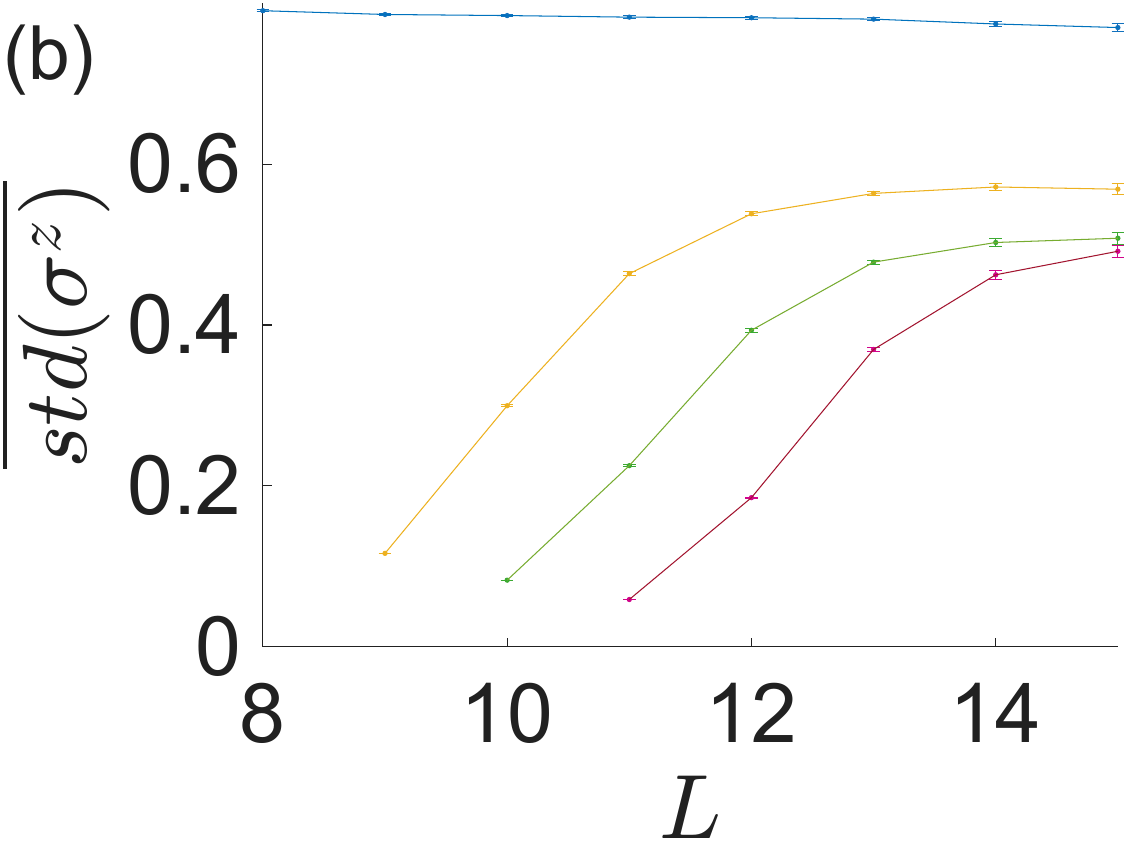}
    \caption{Standard deviation of $\sigma^z$ versus system size. (a) $\ovl{\text{std} (\sigma^z)}$ for various $J$ with and without the bath, with the bath size fixed at $L_{thermal} = 10$. (b) For $J = 0.2$, comparison between the case without the bath (the flat blue line) and with thermal bath size $L_{thermal} = 8, 9, 10$ (the data start from $L_{thermal} + 1$). We can predict that the data for $J = 0.2, L_{thermal} = 10$ will saturate with a few more sites.}
    \label{fig:Heisenberg_sz_L}
\end{figure}

\begin{figure}[h]
    \centering
    \includegraphics[width=0.5\linewidth]{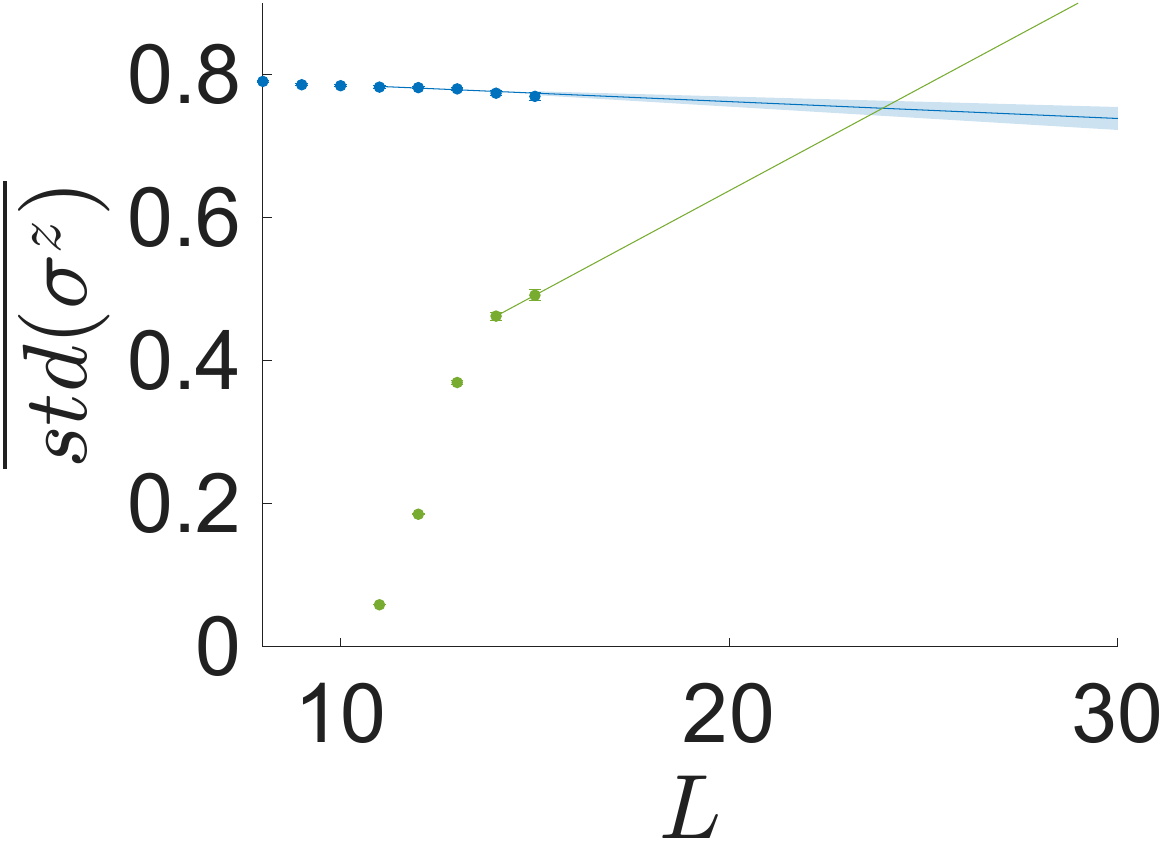}
    \caption{Linear fit for data with $J = 0.2, L_{thermal} = 10$. Since only two points are used for the data without the bath, the uncertainty of this approximation may be very large.}
    \label{fig:Heisenberg_sz_L_extended}
\end{figure}

To fit and obtain the avalanche ratio, we use a linear function for all cases. As shown in Fig. \ref{fig:Heisenberg_sz_L_extended}, for the case with a thermal bath, the data points are not sufficiently stable; only the final two points are useful for predicting the trend. We draw a straight line through these two points, making it impossible to estimate the confident interval. The avalanche ratio calculated in this way is underestimated, highlighting the importance of the number and stability of data points. If longer system lengths were accessible, the avalanche ratio would be larger, as would be expected for any model.

\section{Details of the simulation}
In this appendix, we provide more details on our simulation results, focusing solely on ED algorithms. Details relevant to the DBSCAN and GA algorithms will be discussed in Appendices \ref{dbscan} and \ref{ga}. The ED algorithm in this paper follows the paper \cite{Sierant_2020_a}.

The shift-invert method of ED (SIMED) is used in Matlab's eigs function to find eigenvalues closest to a specific value. Both SIMED and POLFED are used for calculating high-energy eigenstates of a Hamiltonian, but each has its own advantages. SIMED requires more memory but is faster, while POLFED is more time-consuming but uses significantly less memory. Therefore, we use the SIMED algorithm when the system size is small, as it is faster and memory is not yet a bottleneck. For larger system sizes, we use POLFED. Additionally, POLFED requires an approximation of the density of states, which we can estimate by linear fitting based on SIMED data to predict the density of states for larger system sizes. 

For larger system sizes, we are limited by computing power and can compute fewer samples of disorder. We list the number of samples for each model corresponding to each system size $L$.

\begin{table}[h]
    \centering
    \renewcommand{\arraystretch}{1.3}
    \begin{tabular*}{\linewidth}{@{\extracolsep{\fill}} ccccc}
    \hline
    Heisenberg & $<$13 & 13 & 14 & 15 \\
    PXP & $<$20 & 20 & 21 & 22 \\
    Model \Rom{1} & $<$24 & 24 & 25 & 26 \\
    Model \Rom{2} & $<$29 & 29 & 30 & 31 \\
    \hline
    $N_{sample}$ & 10k  & 5k & 2k & 2k \\
    \hline
    \end{tabular*}
    \caption{number of disorder samples}
    \label{tab:PXP_n_samples}
\end{table}

We always calculate $N = 100$ eigenstates with energy closest to $\left(E_{max} + E_{min}\right) / 2$. However, for certain cases where the Hilbert space dimension is smaller than $100$, we choose $N = 50$ instead.

For the Lindblad master equation, since the entire spectrum is required, we use the eig function in Matlab directly. We use 5000 disorder configuration samples for all the data involving the Lindblad master equation and only 1000 samples for the operator growth calculation due to its stability.

The algorithm is implemented in Matlab, and approximately 100,000 CPU hours were used for all the simulations. 

\section{DBSCAN algorithm} \label{dbscan}
The density-based spatial clustering of applications with noise algorithm \cite{Ester_1996, Schubert_2017} is widely used to identify clusters in data points. Since it may not be familiar to MBL community, we provide a brief introduction to this algorithm and its application in our study. Note that we use the dbscan function from Matlab's Statistics and Machine Learning Toolbox. 

Given a set of data points $X = \{p_i | i = 1 \cdots N \}$, where the position of each point $p_i$ is $(x_i, y_i)$ in two dimensions. Our task is to identify clusters of points. Specifically, we assign a label to each point $p_i$ to determine which cluster it belongs to, or if it is a noise point (i.e., does not belong to any cluster).  We define two parameters: $\epsilon$ is the distance within which two points are considered neighbors. $minPts$ is the minimum number of points required to form a cluster. The process for determining which cluster a point belongs to is as follows:

1. A point with at least $minPts$ neighbors (including itself) is identified as a core point. 

2. Neighboring core points belong to the same cluster. 

3. A point that is not a core point is identified as noise if it has no core point neighbors. If it has a neighboring core point, it belongs to the cluster of that core point. If it has multiple neighboring core points, it is assigned to the cluster of one of the neighboring core points depending on the order of the points in the input data set $X$.

\begin{figure}
    \centering
    \includegraphics[width=0.48\linewidth]{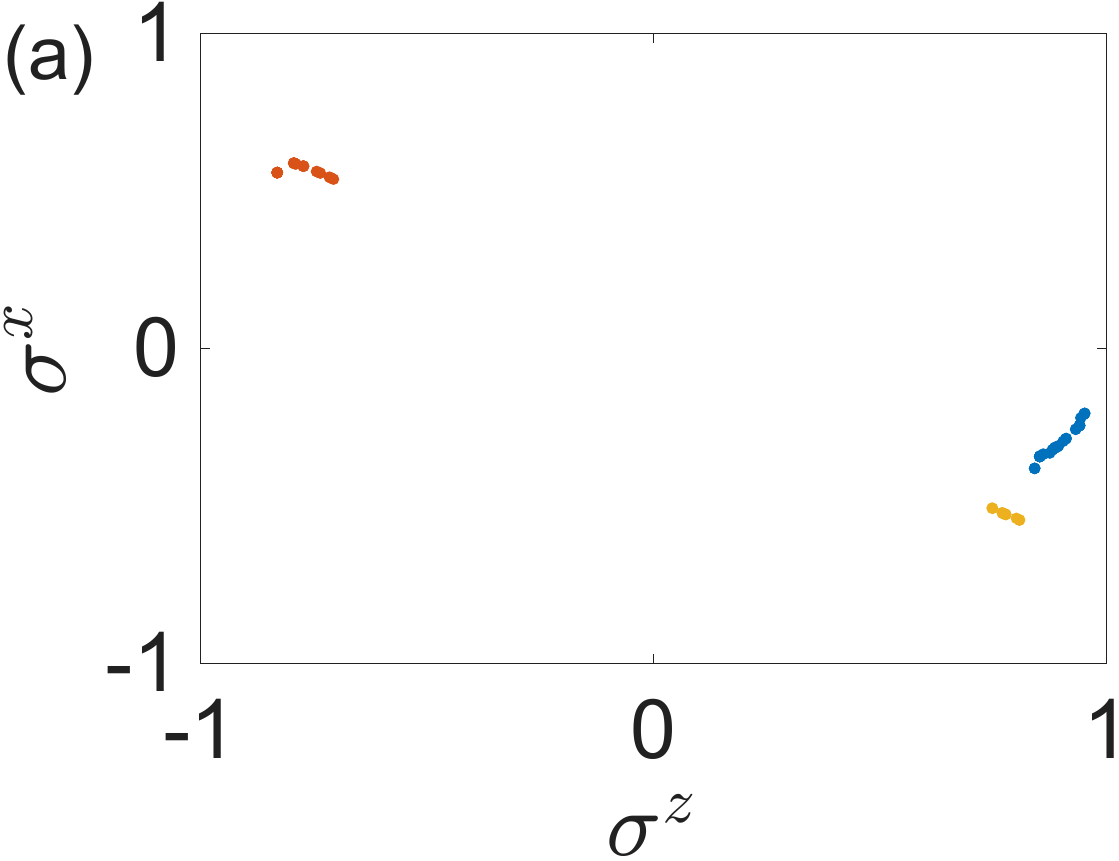}
    \includegraphics[width=0.48\linewidth]{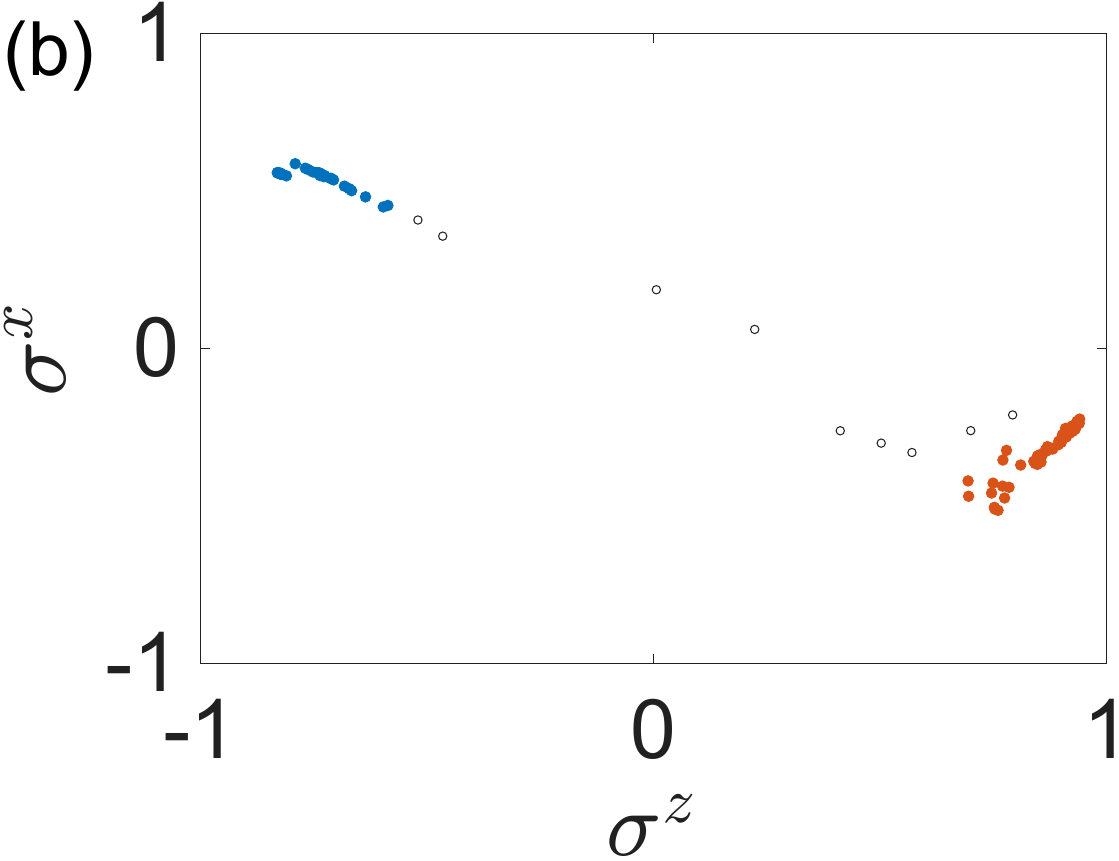}\\
    \includegraphics[width=0.48\linewidth]{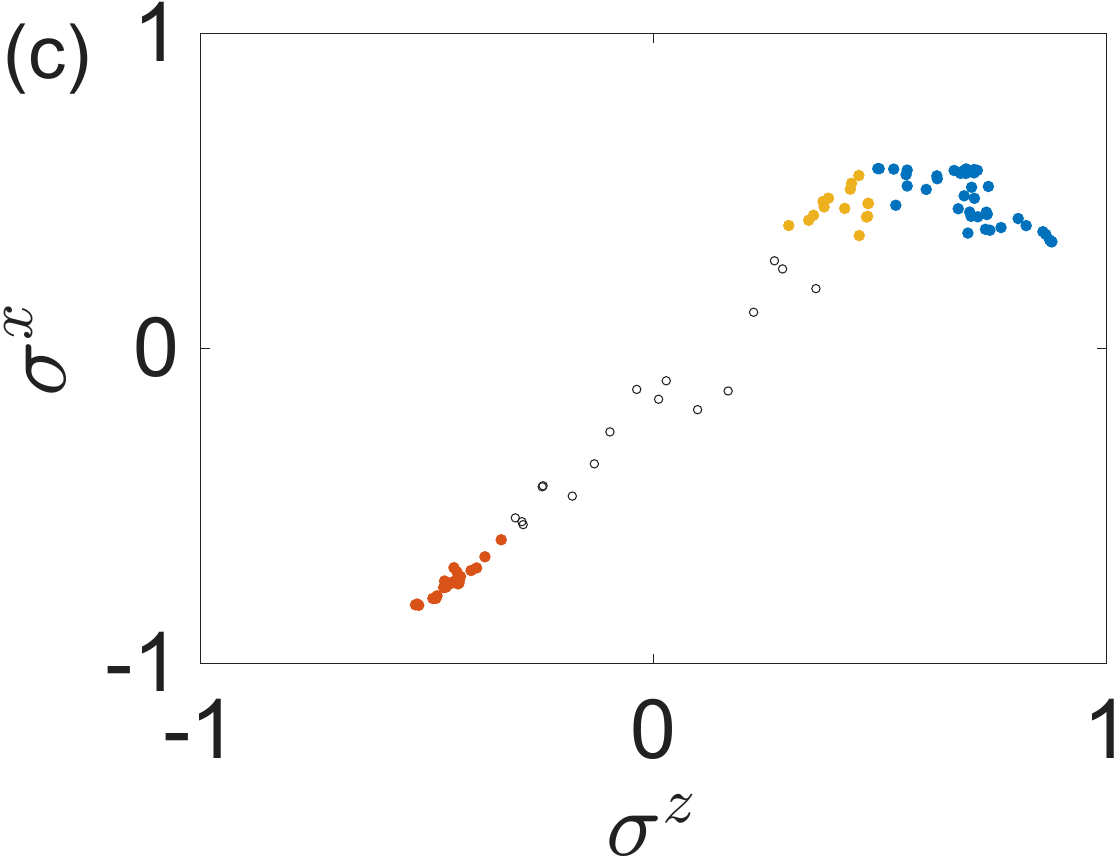}
    \includegraphics[width=0.48\linewidth]{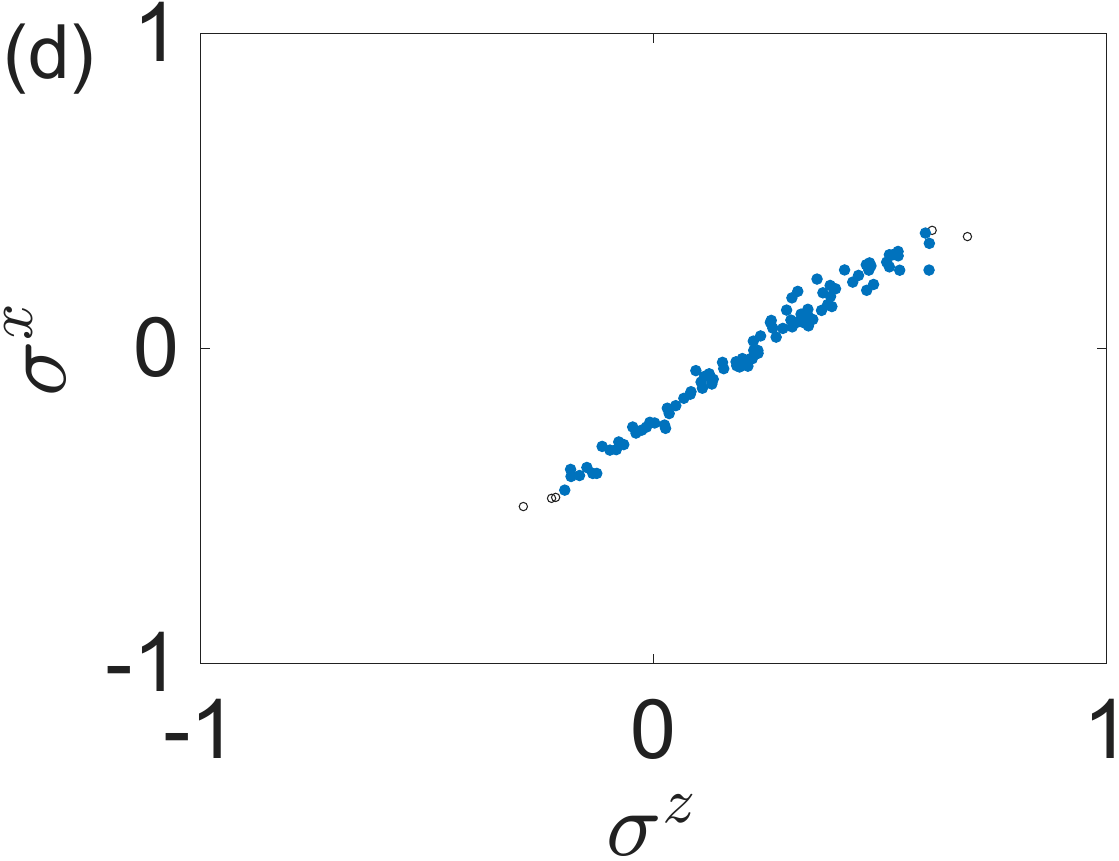}\\
    \includegraphics[width=0.48\linewidth]{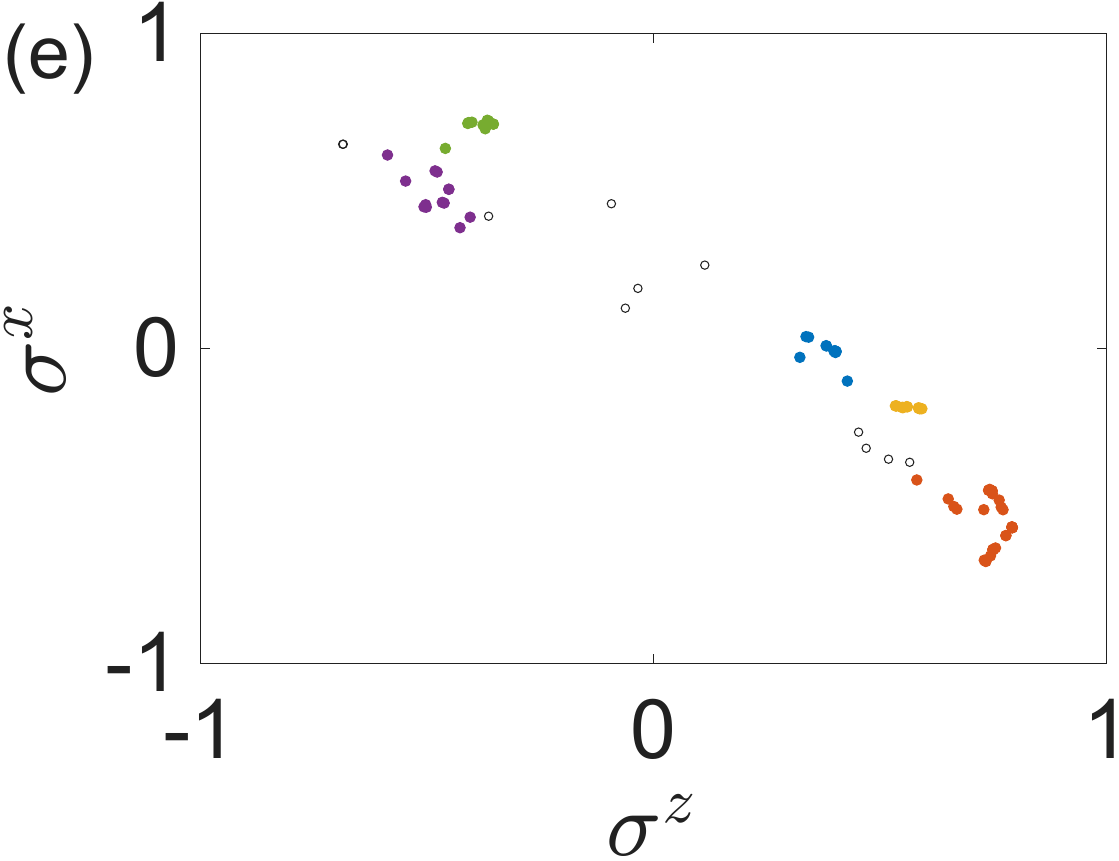}
    \includegraphics[width=0.48\linewidth]{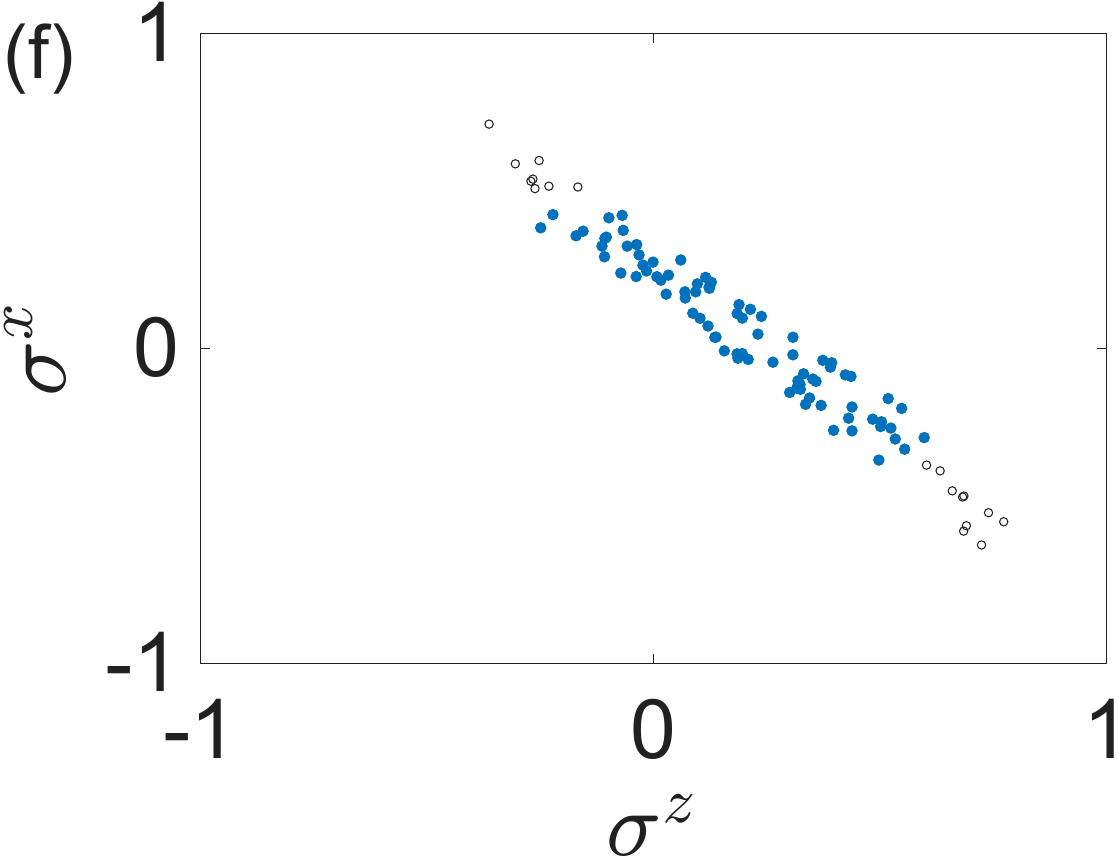}\\
    \caption{Distributions of observables and their clusters identified by the DBSCAN algorithm for $J = 0.6, L = 22$ and $L_{thermal} = 14$ in the cases with a thermal bath. (a), (c) and (e) show three examples without the bath, while (b), (d) and (f) show the corresponding results with the bath. The number of clusters is reduced by the bath, indicating the system becomes more thermal under the influence of the bath.}
    \label{fig:dbscan}
\end{figure}

In our case, we have $N = 100$ points in our set $X$ and we set $\epsilon = 0.1$ and $minPts = 10$. Typical graphs of clusters identified by the DBSCAN algorithm are shown in Fig. \ref{fig:dbscan}. Different colors represent different clusters, while black empty circles represent noise points that do not belong to any cluster. We present cluster graphs for three different disorder configurations (three rows) and compare the cases with (left) and without (right) a thermal bath. $J = 0.7, L = 22$ for all cases, with $L_{thermal} = 14$ for the case with a thermal bath. The distribution of points becomes tighter after introducing the bath, leading to the formation of fewer clusters.

\section{Genetic algorithm} \label{ga}
Genetic algorithms are widely used for optimization and finding multiple solutions. Since it is not very common in MBL community, we provide a brief introduction to the genetic algorithm and how we use it to find our models. Inspired by gene evolution, the genetic algorithm evolves a population in a specific direction to optimize fitness and identify potential solutions. For more detailed information, please refer to relevant papers, such as \cite{Katoch_2020}.

The goal of a genetic algorithm is to search for solutions $x$ within a set $S$ such that the fitness function $f(x)$ is optimized. To achieve this, we start with a population $\{x_i \in S \ | \ i= 1 \cdots N_{population}  \}$, which is a random collection of $x$ as the starting point. Then we perform $N_{evolution}$ steps of evolution on this population. Each evolution step consists of several processes, including selection, crossover and mutation, to form the population of the next step. Selection favors the variables $x_i$ with better fitness $f(x_i)$. Crossover involves combining two variables $x_i, x_j$ to create a new variable. Mutation refers to altering a variable $x_i$ to generate a new variable. After evolution, the population shifts towards variables with better overall fitness.

In our problem, the search space is consists of constraints $\mathcal{C} = ( n, S )$, where $n$ is a number and $S$ is a set of size $n$ spin configurations. For each search, we fix $n$ and a subset $S_0$ of the full set of $n$-spin configurations $F_n = \{s_1 \cdots s_i \ | \ i = 1\cdots n, s_i \in \{0, 1\} \}$. The entire search space is the set of constraints  $\Omega_{n, S_0} = \{ \{ n, S\} \ | \ S \subset S_0 \}$. This set can be encoded as a $|S_0|$-digit binary variable $x_i$, where each digit represents whether an element in $S_0$ is present or not. Each variable $x_i$ corresponds to a subset $S_i \subset S_0$ and therefore to a constraint $\mathcal{C}_i = \{ n, S_i \}$. 

The fitness function in the genetic algorithm is defined as a function of binary list $x_i$:
\vspace{2pt} 
\begin{table}[H]
\centering
\renewcommand{\arraystretch}{1.3}
\begin{tabular*}{\linewidth}{@{\extracolsep{\fill}} l}
\hline
FitnessFunction($x_i$)\\
\hline
obtain constraint $\mathcal{C}_i$ corresponding to $x_i$ \\
\textbf{if} $\mathcal{H}_{sub}$ for $L = n$ is not connected by the Hamiltonian\\ 
 \quad \textbf{return} \ 0\\
\textbf{if} $\mathcal{H}_{sub}$ for $L = 20$ is not connected by the Hamiltonian\\
 \quad \textbf{return} \ 0\\
 fit \ $\ln(|\mathcal{H}_{sub}|) = k L + b $\\
 \textbf{return} $- 1 / k $\\
\hline
\end{tabular*}
\end{table}

The dimension of the Hilbert space for $L = n$ is the same as the size of set $S_i$, making it easy to check the connectivity. If it is connected, we verify the connectivity of a larger Hilbert space with $L = 20$ by diagonalizing the Hamiltonian and calculating the overlap $\langle \psi_1 | \sigma^z_L | \psi_2 \rangle$ for any two eigenstates $A_{ij} = |\psi_i\rangle, |\psi_j \rangle$. If the Hilbert space has smaller Krylov subspaces, this overlap will be zero for the $i, j$ pairs when $i, j$ are in different Krylov subspaces because our basis are eigenstates of $s^z_i$. To ensure connectivity, we check that the ratio of $0$ in the overlap is smaller than a threshold. We define connectivity as $c = |\{A_{ij}\} \neq 0| / |\{A_{ij} \}|$. In the fitness function, we require $c> 0.5$, but in the output function, we require $c>0.99$ to ensure that the majority of the Hilbert space is connected. The remaining states will be scar states if they exist and will not statistically influence the results of our calculation.

It is worth noting that although we use linear fitting in the fitness function, the dimension of the Hilbert space is not necessarily exponential. When the scaling is slow enough, the dimension can be a polynomial function of $L$. In such cases, we may not obtain an accurate linear fit, but we can still determine the best fitting parameters $k$ and $b$. This fitness function still remains effective for identifying polynomial Hilbert spaces. 

We use Matlab's genetic algorithm, $ga()$, for the optimization. During the optimization process, we save all solutions with a fitness function value better than a specified threshold. The population size is set to 50, and the number of generations is set to 100. We do not modify the crossover and mutation processes in this function. For $S_0$, we choose the sets $F_3, F_4, F_5, F_6$ and $\{s_1 \cdots s_5  \in F_5 \ | \ \sum_{i=1} ^5 s_i< 4 \}$, and obtain the models in Table \ref{tab:constraints}. 

\section{Fitting details for the dimension of model \Rom{4}}
\begin{figure}[h]
    \centering
    \includegraphics[width=0.5\linewidth]{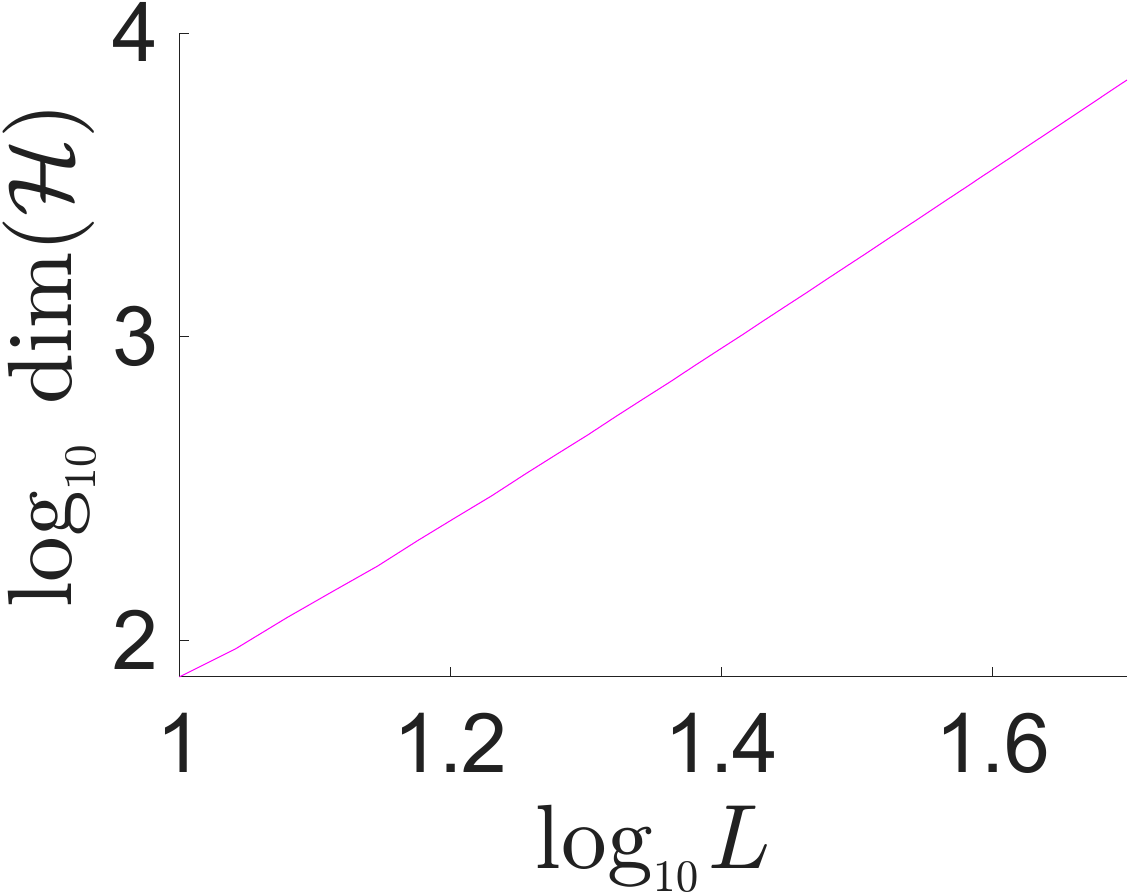}
    \caption{The same data points as model \Rom{4} in Fig. \ref{fig:dimension}. We change the horizontal axis to show a clear linear relation.}
    \label{fig:dim_4}
\end{figure}

Since the relation between log$_{10}$ dim$(\mathcal{H})$ and system size $L$ in Fig. \ref{fig:dimension} is not a straight line for model \Rom{4}, it is not possible to do fitting in that plot. We plot the relation between log$_{10}$ dim$(\mathcal{H})$ and log$_{10}L$ in Fig. \ref{fig:dim_4}. 
\label{app:fitting4}
It is clear this relation is linear. Using linear fitting, we obtain the relation as follows:
\bea
\text{log}_{10} \text{dim}(\mathcal{H}) = k \text{log}_{10} L + b
\eea
where $k, b$ are the parameters for linear fitting. Transforming this relation to abandon the logarithm, we obtain
\bea
\text{dim}(\mathcal{H}) = C L^k
\eea
where $C = 10^b$. This fitting result is presented in the table \ref{tab:dimension}.

\section{Lindblad master equation } \label{append.lindblad}
We use Lindblad master equation to study dissipation in MBL systems. This method was first introduced in Refs.~\cite{Morningstar_2022, Sels_2022}. In this appendix, we provide a detailed introduction to the approach, enabling readers to reproduce the results presented in our paper. 

We begin with Lindblad master equation Eq.~\ref{eq.lme} in the main text, where the Lindblad operators $L_i$ describe the coupling between the thermal bath and the disordered region. For the Heisenberg model, the coupling term is given by $\sigma^x_e  \sigma^x_s + \sigma^y_e  \sigma^y_s + \sigma^z_e  \sigma^z_s$, where the subscripts $e$ and $s$ denote the environment (thermal bath) and the system (disordered region), respectively. This leads to the Lindblad operators: $L_1 = \sigma^x, L_2 = \sigma^y, L_3 = \sigma^z$ and the Hamiltonian $H$ corresponds to the disordered-region Hamiltonian. 

For the PXP model, the coupling term is $P^z_{-1} \sigma^x_0 P^z_1 + P^z_{0} \sigma^x_{1} P^z_{2}$. Accordingly, the Lindblad operators are $P^z_1, \sigma^x_1 P^z_2$. The Hamiltonian $H$ is the disordered PXP Hamiltonian, excluding the first term $\sigma^x_1 P^z_2$, since it is already included in the coupling. Moreover, both the Lindblad operators and Hamiltonian of the PXP model must be projected onto the constrained subspace using the projector $P^ {\mathcal{C}_{\text{PXP}}}$. The final forms of Lindblad operators and the Hamiltonian are 
\bea
\notag
&L_1 = P^ {\mathcal{C}_{\text{PXP}}} P^z_1 P^ {\mathcal{C}_{\text{PXP}}}, \quad 
L_2 = P^ {\mathcal{C}_{\text{PXP}}} \sigma^x_1 P^z_2 P^ {\mathcal{C}_{\text{PXP}}}\\
\notag
&H = P^ {\mathcal{C}_{\text{PXP}}} H_{\text PXP} P^ {\mathcal{C}_{\text{PXP}}}
\eea

In the Lindblad master equation, the influence of the bath enters through the Lindblad operators $L_i$, which couple to one end of the disordered spin chain. All operators act on the density matrix $\rho$ of the spin chain. We treat $\rho$ as a vector using the Choi's isomorphism and denote it by $|\rho\rangle _\#$. Under this representation, the Lindbladian $\mathcal{L}$ becomes a linear operator $\mathcal{L}_\#$ acting in the doubled Hilbert space.
\bea \notag
&\mathcal{L}_\# = -i H \otimes \mathbb{I} + i \mathbb{I} \otimes H^T \\ 
&+ \gamma \sum_\mu \left( L_\mu \otimes L_\mu ^*  - \frac12 L^\dagger_\mu L_\mu \otimes \mathbb{I} - \frac12 \mathbb{I} \otimes L_\mu^ T L_\mu ^* \right) 
\eea

To extract the dissipation rate, we diagonalize $\mathcal{L}_\#$. It has a zero eigenvalue with corresponding eigenvector $|\rho_{0,\#} \rangle$, representing the steady state. All other eigenvalues $\lambda_i$ have negative real parts, and the corresponding eigenvectors are $|\rho_{i,\#} \rangle$. The time evolution of a density matrix $\rho$ governed by 
\bea
\frac{d}{dt} \rho  = \mathcal{L} [\rho]
\eea
leads to exponential decay toward the steady state $\rho_0$:
\bea
\rho(t) = \rho_0 + \alpha_i \rho_i e ^ { \lambda_i t}
\eea
where $\alpha_i$ are the projection coefficients. The slowest decay mode corresponds to the nonzero eigenvalue $\lambda_1$ with the largest real part (closest to zero), and the decay rate is 
\bea
\Gamma = - \text{real} \ \lambda_1
\eea

This decay rate $\Gamma$ assumes an infinitely large thermal bath. In realistic settings with finite bath size, we compare $\Gamma$ with the characteristic energy scale of the bath to determine whether the system thermalizes. The relevant energy scale is the average level spacing (or energy gap) of the bath, denoted by $\Delta$. The thermalization condition is thus determined by comparing $\Gamma$ and $\Delta$.

An alternative perspective is through effective coupling strength. The decay rate $\Gamma$ implies an effective coupling between the bath and the farthest spin in the disordered region of order $\sqrt{\Gamma}/\sqrt{N_{\text{bath}}}$, where $N_{\text{bath}}$ is the Hilbert space dimension of the bath. This scaling arises from two factors:

$\bullet$  On the disordered side, the transition amplitude scales as $\sqrt{\Gamma}$ since $\Gamma \sim |\langle \psi_i | V_s | \psi_f \rangle|^2$.

$\bullet$ On the bath side, for a local operator $V_e$, the average coupling between bath eigenstates satisfies $\sum_{j=1}^{N_{\text{bath}}} |\langle \phi_i | V_e | \phi_j \rangle|^2 = 1$, implying a typical matrix element $\sim 1/\sqrt{N_{\text{bath}}}$.

Meanwhile, the average energy gap scales as $\Delta \sim 1/N_{\text{bath}}$. Therefore, thermalization occurs if
\bea \notag
\sqrt{\Gamma} / \sqrt{ N_{\text{bath}}} > 1/N_{\text{bath}}
\eea
which is equivalent to comparing $\Gamma$ with $\Delta$.

\section{Random matrix as thermal bath} \label{append.random_matrix}

In this paper, we use weakly disordered spin chains as thermal bath to couple with MBL systems. However, assessing whether these spin chains genuinely serve as thermal baths requires more evidence beyond level statistics, such as the average adjacent gap ratio $\ovl{r}$, especially since constraints in such systems can introduce complexities like quantum many-body scars. In this appendix, we benchmark the spin-chain bath by comparing it to a true thermal bath constructed from a random matrix ensemble.

Specifically, we use a Hamiltonian drawn from the Gaussian Orthogonal Ensemble (GOE) as a thermal bath and couple it to a disordered PXP system. Since all Hamiltonians in our study are real, time-reversal symmetry allows for complex conjugation of the states, making GOE—comprising real symmetric matrices—a natural choice. Each GOE Hamiltonian has entries drawn from a Gaussian distribution with zero mean. The off-diagonal elements have standard deviation $\sigma = E_{\text{GOE}}$, while the diagonal elements have $\sigma = \sqrt{2}E_{\text{GOE}}$.

We set the Hilbert space dimension of the GOE matrix to $N_{\text{GOE}} = 144$, matching the PXP spin chain with length $L=10$. The energy scale $E_{\text{GOE}}$ is set to $1/8$ to ensure a similar density of states. The full Hamiltonian used in this benchmark is

\bea
\notag
&H = H_{\text{GOE}} + H_\text{PXP} \\
&+ J_{thermal} P_{-1}^z \sigma^x_0 P^z_1 + J_{disordered} P^z_0 \sigma^x_1 P^z_2
\eea
with coupling strengths $J_{\text{thermal}} = 1$ and $J_{\text{disordered}} = 0.6$. The second line contains the coupling terms: $P^z_1$, $P^z_2$, and $\sigma_1^x$ act on the disordered PXP chain, while $P^z_{-1}$, $P^z_0$, and $\sigma_0^x$ act on the random matrix bath.

The operators on the random matrix bath are defined as:
\bea
\notag
&P^z_{-1}  \sigma_0^x =\mathbb{I}_{N_{\text{GOE}} / 4} \otimes P^z \otimes \sigma^x\\
\notag
&P^z_{0} = \mathbb{I}_{N_{\text{GOE}} / 2} \otimes P^z
\eea
where the subscripts on identity operators denote their dimensions. This form of coupling is designed to mirror the structure used in spin-chain baths, allowing for a fair comparison.

\begin{figure}[h]
    \centering
    \includegraphics[width=0.5\linewidth]{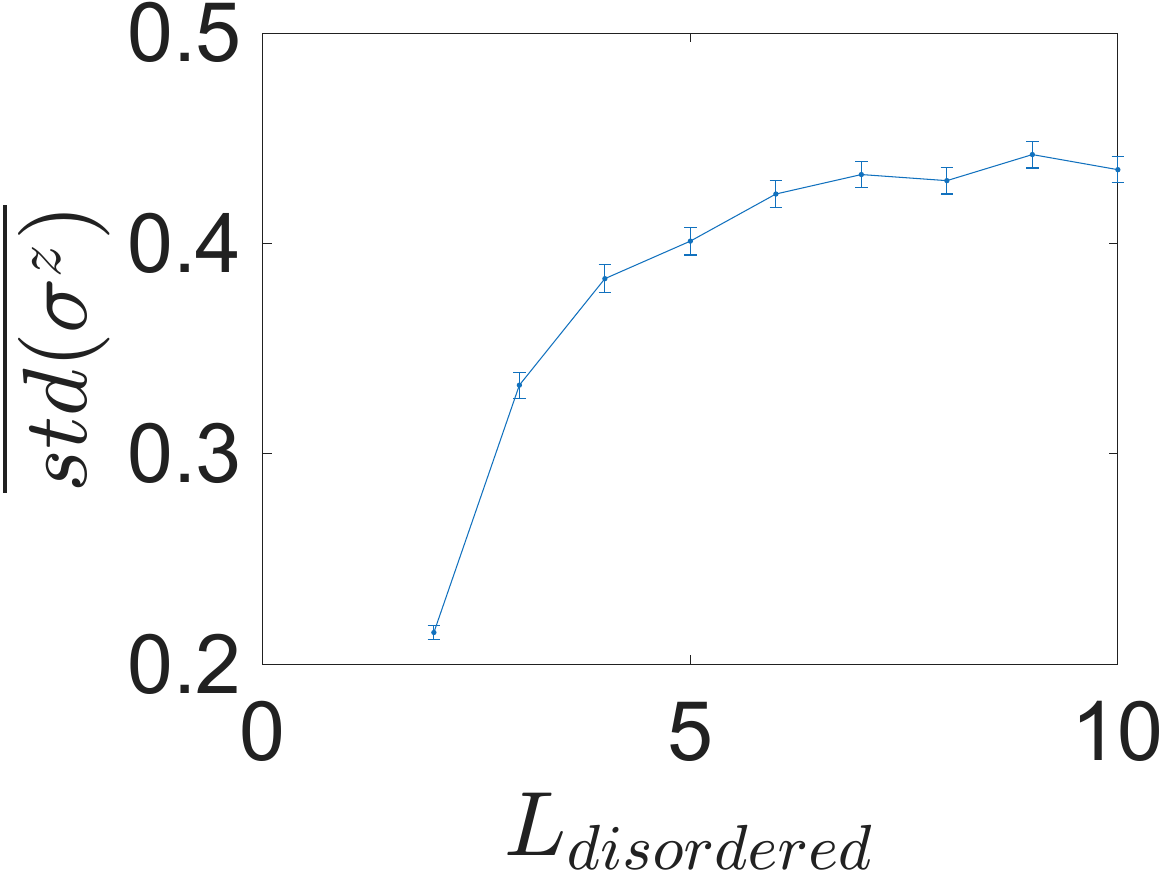}
    \caption{Standard deviation of $\sigma^z$ for the disordered PXP chain coupled to a random-matrix thermal bath.}
    \label{fig:random_matrix}
\end{figure}

We compute the standard deviation of the $\sigma^z$ expectation value on the last spin of the disordered region, following the same procedure as in the main text. Fig~\ref{fig:random_matrix} presents this standard deviation as a function of the disordered region size $L_{\text{disordered}}$.

This figure can be directly compared to the curve with $L_{\text{thermal}} = 10$ in Fig.~\ref{fig:sz_L_6}(a), where the relation of $x$ axis is $L = L_{\text{thermal}} + L_{\text{disordered}}$. We observe that the curve shape is similar and $\ovl{\text{std}(\sigma^z)}$ saturates near $0.43$ for large system sizes in both setups—using the random matrix bath and the PXP-chain bath, demonstrating consistency between different types of thermal environments.

Since the random matrix is dense rather than sparse, computing its eigenstates requires more resources. We therefore reduce the number of disorder realizations to $N=1000$ and restrict the system size to $L_{\text{disordered}} = 10$ (equivalent to $L = 20$), which is slightly smaller than the maximum $L=22$ used in the main text.

\bibliography{mbl}

\end{document}